\documentclass[fleqn,usenatbib]{mnras}

\usepackage{newtxtext,newtxmath}

\usepackage[T1]{fontenc}
\usepackage{ae,aecompl}
\usepackage{gensymb}
\usepackage{color}
\usepackage{soul}

\usepackage[]{graphicx}	
\usepackage{amsmath}	
\usepackage{enumitem}
\setlist[enumerate]{parsep=0pt}

\newcommand{\degs}{$^{\circ}$}
\newcommand{\vsini}{$v\,\textrm{sin}\,i$}
\newcommand{\kms}{km\,s$^{-1}$}
\newcommand{\ms}{m\,s$^{-1}$}

\title[Central Line Moments]{Identifying activity induced RV periodicities and correlations using Central Line Moments}

\author[J. Barnes]
{J.~R.~Barnes$^{1}$,
S.~V.~Jeffers$^{2}$,
C.~A.~Haswell$^{1}$,
M.~Damasso$^{3}$,
F.~Del~Sordo$^{4}$,
F.~Liebing$^{2}$, \newauthor
M.~Perger$^{4,5}$,
G.~Anglada-Escud\'{e}$^{4,5}$ \\
$^{1}$School of Physical Sciences, The Open University, Walton Hall, Milton Keynes. MK7 6AA. UK \\
$^{2}$Max Planck Institute for Solar System Research, Justus-von-Liebig-Weg 3, 37077 Göttingen. Germany \\
$^{3}$INAF - Osservatorio Astrofisico di Torino Strada Osservatorio 20, I-10025 Pino Torinese (TO), Italy \\
$^{4}$Institut de Ci\`{e}ncies de l’Espai (ICE, CSIC), Campus UAB, Carrer de Can Magrans s/n, 08193 Bellaterra, Spain \\
$^{5}$Institut d’Estudis Espacials de Catalunya (IEEC), c/ Gran Capit\`{a} 2-4, 08034 Barcelona, Spain \\
}

\date{Accepted for publication in MNRAS - 2024 September 10.}

\pubyear{2021/2022}

\begin{document}
\label{firstpage}
\pagerange{\pageref{firstpage}--\pageref{lastpage}}
\maketitle

\begin{abstract}
The radial velocity (RV) method of exoplanet detection requires mitigation of nuisance signals arising from stellar activity. Using analytic cool and facular spot models, we explore the use of central line moments (CLMs) for recovering and monitoring rotation induced RV variability.
{Different spot distribution patterns, photosphere-spot contrast ratios and the presence or absence of the convective blueshift lead to differences in CLM signals between M dwarfs and G dwarfs.}
Harmonics of the rotation period are often recovered with the highest power in standard periodogram analyses. {By contrast, we show the true stellar rotation may be more reliably recovered with string length minimisation}. For {solar minimum} activity levels, {recovery of the stellar rotation signal from CLMs is found to require unfeasibly high signal-to-noise observations}. The stellar rotation {period} can be recovered at solar maximum activity levels from CLMs {for reasonable} cross-correlation function (CCF) {signal-to-noise ratios $> 1000$\,-\,$5000$}. 
The CLMs can be used to recover and monitor stellar activity through their mutual correlations and correlations with RV and bisector inverse span. The skewness of a CCF, a measure of asymmetry, is described by the third CLM, $M_3$. {Our noise-free simulations indicate the linear RV vs $M_3$ correlation is {up to} 10 per cent higher than the RV vs bisector inverse span correlation. We find a corresponding $\sim$\,$5$ per cent increase in linear correlation for CARMENES observations of the M star, AU Mic. We also assess the effectiveness of the time derivative of the second CLM, $M_2$, for monitoring stellar activity.}

\end{abstract}




\section{Introduction}
\label{section:intro}
The recovery of dynamically induced radial velocity signals due to orbiting exoplanets can be severely hampered by activity signals from the host star. As a consequence, the majority of radial velocity surveys have targeted low-activity stars that are well into their main sequence lifetime. Although large scale photometric surveys searching for transiting planets do not necessarily pre-select low-activity stars, the youngest stellar clusters with ages of a few $10 - 100$~Myr are generally avoided due difficulties posed by activity-related variability. {In addition, moderately active stars that have reached the main sequence can also still exhibit activity signatures that are comparable or greater in amplitude than the dynamical signatures induced by orbiting low-mass planets.}
Demographic features found in older exoplanets depend on orbital evolution or interaction with the host star at younger ages. To fully understand the Galaxy's population of planets, we must study {both} younger {and older} populations of planets that orbit {stars with significantly different activity levels.}

The study of stellar activity and identification of spectroscopic activity signals has been established since the first exoplanet detections were made \citep{saar98}. Techniques to correct for long term magnetic variability arising from changes in absorption line shapes soon followed \citep{saar00bis}. Subsequent studies have provided detailed assessments of the stellar line bisector variability due to starspots \citep{desort07}. The signatures of more complex cool spot and hot facular regions have also been investigated by \cite{meunier10plage} in detail, using the Sun as a laboratory. Extrapolation of the solar analogue has enabled stars with different spectral types, activity features and activity levels to be modelled and studied in detail \citep{barnes11jitter,meunier13,dumusque14soap,jeffers14activity,borgniet15,herrero16,barnes17jitter,jeffers22evlac}. While direct physical modelling plays a crucial and important role in characterising and correcting activity signatures, non-parametric posterior modelling and machine learning techniques have also been successfully employed to account for variable stellar activity signals combined with planetary induced radial velocity variations \citep{haywood14,rajpaul15GP,cretignier22,debeurs22,perger23,liang24deep,zhao24deep}. These techniques rely on either prior estimates of activity related effects and timescales or on activity indicators simultaneously observed and derived from the same data used to make the radial velocity measurements.

In this paper we systematically investigate the use of central line moments (hereafter, CLMs) as a simple and consistent standardised measure of absorption line shape changes due to stellar activity. CLMs have been employed in the study of M dwarf stars, in the characterisation of instrumental and activity effects \citep{berdinas16}, and for the characterisation of activity signatures in the discovery of Proxima Centauri b \citep{anglada16proxima}. We demonstrate the expected line moment signatures from simple single spots and scaled solar spot models and simulate the expected sensitivities of the method for an M dwarf and a G dwarf star.

\section{Central line Moments}
\label{section:CLMs}

The CLMs are a basic tool for measuring a statistical distribution. Similarly, by analogy they can describe the shape of a stellar emission or absorption line. Factors such as cool or hot spots, magnetic sensitivity due to the Zeeman effect and convective blueshift affect individual lines by different degrees. Because activity induced absorption line changes are often small effects, data with good signal-to-noise ratios (SNRs) are generally required for effective CLM measurement. In radial velocity studies, the cross correlation function (CCF) is commonly calculated by considering thousands of photospheric absorption lines. The CCF, can thus be thought of as a high SNR distribution that describes the effective \textit{mean} shape of all the absorption lines considered.

For a perfectly stabilised spectrograph the CLMs arising from a single observation can be associated with different physical phenomena. From a planet hunting perspective, we are interested in the velocity shift of the star due to the reflex motion induced by an orbiting unseen exoplanet. The radial velocity measurement can alternatively be measured using the first CLM, $M_1$, which measures the centre weighted velocity of the line. Because so many absorption lines are used to calculate a typical CCF or LSD line profile, the signal-to-noise ratio can be very high at typically SNR~$\sim$\,$1000$ or more. This is a necessity for precise RV measurements with $\textrm{m\,s}^{-1}$ precision, when a spectrum is recorded with a typical detector pixel velocity increment that is $\sim$\,$1000$\,$\times$ greater.

If cool starspots or hot facular regions are present, a time series of observations may show periodic CLM variability in $M_1$ as a result of line-distorting active regions combined with stellar rotation. In other words, $M_1$, behaves in the same manner as the RV measured from fitting a symmetric function to the CCF in that it is sensitive to both dynamically induced translations in velocity due to an orbiting exoplanet \textit{and} to CCF asymmetries. In an attempt to resolve this ambiguity, a very simple measure of the shape of the line CCF, in the form of the bisector inverse span (BIS), is commonly made \citep{toner88bis,martinezfiorenzano05bis,desort07}. This measure of line asymmetry measures the difference in the mean velocity of two regions of a CCF that are representative of a deep and shallow region of the line profile. It can be used to assess whether particular radial velocity periodicities arise from stellar activity. The BIS is defined such that a cool spot signature induces a radial velocity that anti-correlates with the measured centroid of the CCF (i.e. a negative correlation). The use of BIS calculates the difference in mean bisector in two regions of the CCF, simplifying the activity induced distortions, and thus running the risk of losing information. \citep{figueira13} already demonstrated that alternative measures of the CCF shape can lead to more efficient recovery of activity induced RV variability. The skewness of the CCF, defined by the third CLM, $M_3$, can be used to measure asymmetry. One advantage of measuring skewness rather than the BIS is that more of the CCF can be used, potentially offering increased sensitivity to activity induced distortions.

While RV and BIS measurement, or similarly, $M_1$ and $M_3$,  are metrics used for investigating planetary and activity indices (e.g. see simulations in \citealt{desort07,dumusque14soap} and \citealt{herrero16}), $M_2$ is also commonly monitored as the directly related line full width at half maximum (FWHM) or differential line width \citep{zechmeister18serval}. The CCF FWHM is also routinely modelled and examined in precision RV analyses (e.g. \citealt{herrero16,haswell20dmpp,jeffers22evlac,zicher22}).

The use of CLMs thus standardises and extends common RV metrics that enable the signatures of activity to be studied. We consider the first 5 line moments, $M_0$ to $M_4$, of a continuum normalised CCF. The CLMs in velocity space, $v$, are then 

\begin{equation}
\begin{array}{lr}
    M_0 = \frac{\sum_{i}^{}{{\mathcal F}_i }}{n} & \textrm{Mean of CCF area} \\
    & \\
    M_1 = {\bar v} = \frac{\sum_i{{\mathcal F}_i v_i}}{\sum_i{{\mathcal F}_i}} & \textrm{Mean~weighted~velocity}\\
    & \\
    M_2 = \sigma_v = \sqrt{\frac{\sum_i{{\mathcal F}_i (v_i-M_1)^2}}{\sum_i{{\mathcal F}_i }}} & \textrm{Standard~deviation}\\
    & \\
    M_3 = \frac{ \sum_i{{\mathcal F}_i } \left(\frac{v_i-M_1}{M_2}\right)^3       }{\sum_i{{\mathcal F}_i }}
    & \textrm{Skewness} \\
    & \\
    M_4 = \frac{ \sum_i{{\mathcal F}_i } \left(\frac{v_i-M_1}{M_2}\right)^4       }{\sum_i{{\mathcal F}_i }} - 3 & \textrm{Kurtosis} \\
    & \\
\end{array}
\label{eqn:moments}
\end{equation}

\noindent
where ${\mathcal F}_i $ is simply the flux of CCF velocity bin $v_i$ (each with width $\Delta v_i$) in the normalised CCF (i.e. the continuum level is zero and the maximum flux, ${\mathcal F_\textrm{max}} \leq 1$, is at or close to the line centre). Hence, $M_0$ is simply the {mean of the CCF summed area, where $n$ is the number of data points. Dividing by $n$ preserves the value of $M_0$ for changes in instrument resolution and detector binning. The first even moment, $M_2$,} is the flux weighted standard deviation. Finally, the fourth CLM, $M_4$, measures the Kurtosis or tailedness of the line CCF. The CLMs, $M_0$, $M_3$ and $M_4$ are dimensionless, while $M_1$ and $M_2$ have units of velocity.

\begin{figure}
	\centering
	\includegraphics[trim=0mm 0mm 0mm 0mm, width=0.99\columnwidth]{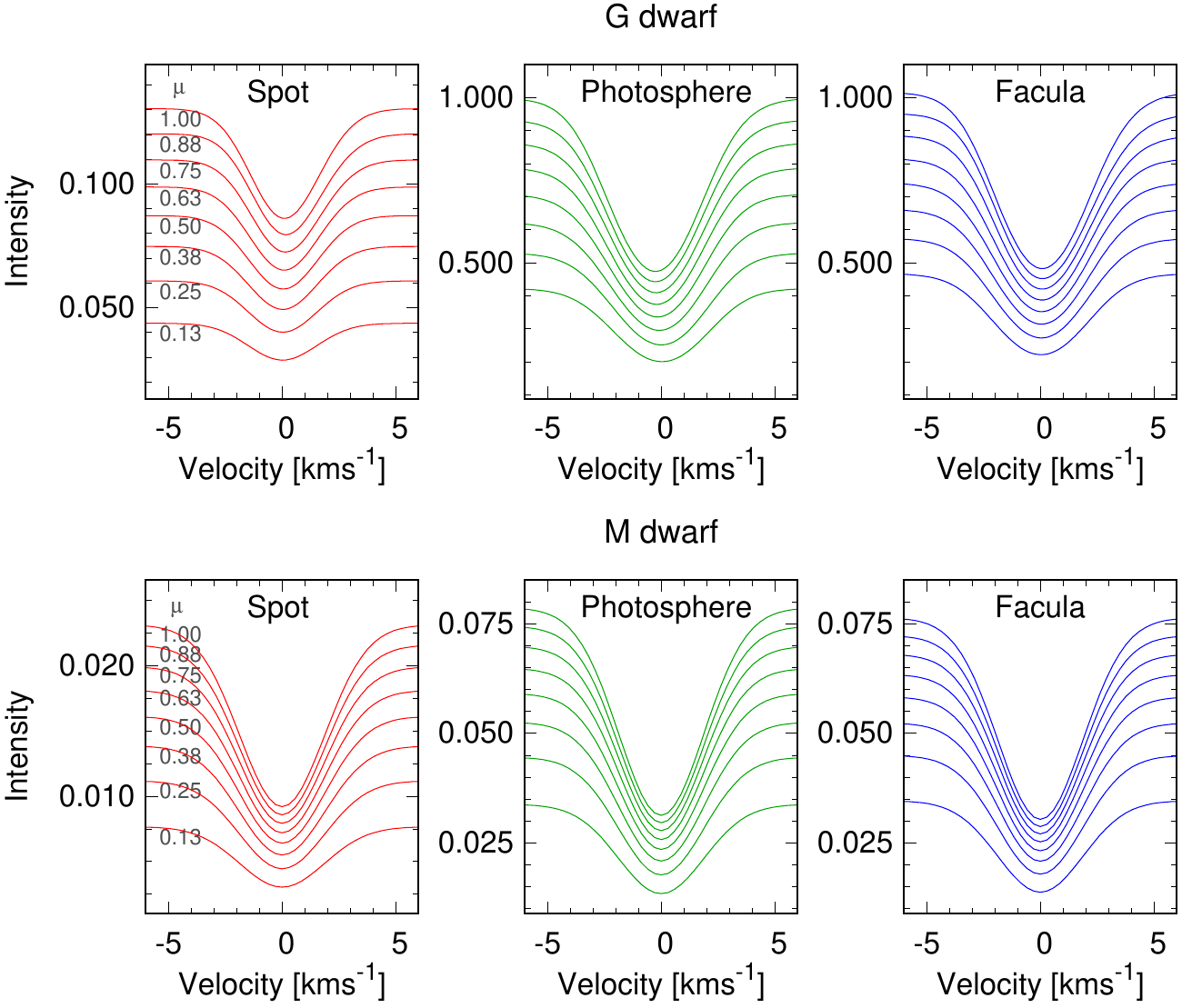}
	\caption{Model local intensity profiles representing a spot, the photosphere and a facula. Local intensities at a spectral resolution of $R = 115,000$ are shown normalised to the G dwarf photospheric continuum intensity level for 8~limb angles from $\mu = 1.0$ at disc centre to $\mu = 0.13$ near the limb.}
	\label{fig:LUTs}
\end{figure}

While the CLMs are easily obtained from CCFs that are typically calculated by precision RV pipelines, they can alternatively be calculated via
Least Squares Deconvolution (LSD) \citep{donati97zdi} of the stellar spectra. This procedure may provide less biased absolute reference estimates of higher order moments since LSD also minimises the effects of line shape distortion due to blending of absorption lines. In the ideal case, where $N_\textrm{lines}$ are considered, we can expect an effective gain in SNR by a factor of up to $\sqrt{N_\textrm{lines}}$.  For an \'{e}chelle spectrograph, several thousand absorption lines are generally considered. For a typical G star spectrum observed with mean SNR $\sim 100-200$, we can thus expect SNRs of order $5000$\,-\,$10000$ to be routinely achieved in either the CCF or least squares deconvolved profile \cite{donati97zdi,barnes98aper,barnes01aper}.

\section{Models}
\label{section:models}

We investigate the behaviour of CLMs by modelling absorption lines obtained from 3D stellar models.
We use the Doppler imaging code, DoTS \citep{cameron01mapping}, which enables line profiles to be calculated from synthetic spot models with any user-defined spot pattern \citep{jeffers14activity}. A simulated line profile is generated for each desired rotation angle with a specified SNR, emulating the CCF obtained via cross correlation. The CLM is then calculated directly from each simulated line profile. We have modified DoTS to generate absorption lines from a three temperature model and user-specified spot-filling factors. It has similar functionality to the code of \cite{dumusque14soap}. Use of DoTS for spot modelling is well documented in a number of related publications \citep{barnes11activity,jeffers14activity,barnes17jitter}. 

\subsection{Line profile construction}
\protect\label{section:lineprofiles}

\subsubsection{Cool Spots}
\protect\label{section:contrastscool}

Spot models of cool dark spots were simulated for the case of a G dwarf and a cooler M dwarf. For the G dwarf, we assumed $T_{\rm phot}$~=~5800~K and $T_{\rm spot} = 4000$~K, while for the M dwarf, $T_{\rm phot}$~=~3500~K and $T_{\rm spot}~=~3000$~K \citep{berdyugina05starspots,panja20starspotsims,jeffers23crx}. We adopted typical observation wavelengths in each case, {performing simulations with resolution $R=115,000$ at} V band wavelengths for the G dwarf (e.g. assuming HARPS-like observations) and at R band wavelengths for the M dwarf (e.g. the visible arm of CARMENES, {but also assuming $R=115,000$}).

{The cool spot and photospheric intensities as a function of limb-angle, $\mu = \textrm{cos}(\theta)$, were obtained from the limb-dependent model intensity spectra provided by} the G\"{o}ttingen Spectral Library\footnote{http://phoenix.astro.physik.uni-goettingen.de} \citep{husser13atlas} at $0.5\,-\,0.6$~\micron~(G dwarf) and $0.6\,-\,0.7$~\micron~(M dwarf) wavelengths. The corresponding photospheric intensity to cool spot intensity ratios at disc centre for the chosen temperatures and wavelengths are $I_{\rm phot}/I_{\rm spot} = 7.66$ for the G dwarf and $I_{\rm phot}/I_{\rm spot} = 3.40$ for the M dwarf. A cool spot is thus expected to induce a larger amplitude signature on a G dwarf compared with an M dwarf. 
{Simulated limb-dependent local intensity line profiles for cool spots and photosphere are shown in Fig. \ref{fig:LUTs} (left and middle panels) for 8 limb angles. Further details are given below in \S \ref{section:contrastsfacular} - \ref{section:CB}.}

\subsubsection{Facular contrasts}
\protect\label{section:contrastsfacular}

Simple models have been used to describe the observed hot facular intensity as a function of limb angle. \cite{lawrence88faculae} and \cite{lawrence88faculaemulticolour} investigated facular contrast using solar images. \cite{lawrence88faculaemulticolour} proposed the contrast between quiet photosphere and faculae as a function of limb angle, $\mu$, takes the form $(I_{\rm fac}-I_{\rm phot})/I_{\rm phot} = b(1/\mu-a)$.  \cite{ahern00} found the variation in facular contrast at 6723\,\AA{} shows very similar behaviour to bluer wavelengths (4706\,\AA{}), but with reduced contrast at the limb. \cite{unruh99} also computed models of facular contrasts as a function of limb angle and wavelength, finding general agreement with observations, but higher contrasts.
More recent studies have examined the dependence of this trend on magnetic field strength and spectral type (e.g. \citealt{yeo13faculae,norris17faculae}). {Simulations using the MURaM code \citep{vogler05muram} were used by \cite{johnson21} to obtain limb-dependent variation of facular features of different magnetic field strengths for a range of spectral type. These were adopted by \cite{zhao23soap} in their modelling of stellar activity.}

{
\subsubsection{M dwarf Facular contrasts}
\protect\label{section:contrastsfacularMdwarf}

The simulations of \cite{beeck15} showed that for active M dwarfs, faculae are not significant. We adopted the M2 dwarf model with 500G vertical magnetic field strength, simulated for {\em Kepler} band wavelengths (Fig.~3 of \citealt{johnson21}). For this model, faculae remain dark (like cool spots) with a very small maximum contrast of $(I_{\rm fac}-I_{\rm phot})/I_{\rm phot} = -0.03$) until they become indistinguishable (i.e. zero contrast) from the photosphere at limb angle, $\mu = 0.34$. As faculae approach the limb further, they exhibit very small positive contrasts of around~$0.015$ at the maximum simulated limb angle of $\mu = 0.2$. We fit the quadratic law found by \cite{borgniet15} for the Sun (see \S \ref{section:contrastsfacularGdwarf} below) to obtain contrast at any limb angle. The model tends to a contrast of $0.04$ at $\mu = 0.0$. The subtle change from negative to positive ratios in the photospheric vs facular intensity contrast can be discerned by close examination of the corresponding continuum intensities at $\mu = 0.12$ and $\mu = 1.00$ in Fig. \ref{fig:LUTs} (bottom right panel).

We also investigated the effect of M dwarf faculae by assuming that the solar law can be applied to the M dwarf case at R band wavelengths, following the observations and modelling of  \cite{lawrence88faculae}, \cite{ahern00} and \cite{unruh99}. In this case, we used the same \cite{borgniet15} solar facular contrast law as for the G2 dwarf model described below.  

\subsubsection{G dwarf Facular contrasts}
\protect\label{section:contrastsfacularGdwarf}

For simplicity}, for the G2 dwarf model, we adopted the law found by \cite{borgniet15}, based on the solar observation modelling work of \citep{meunier10plage}, where $(I_{\rm fac}-I_{\rm phot})/I_{\rm phot} = 0.131618 - 0.218744\mu + 0.104757\mu^2$. {Again, although faculae have a relatively low contrast with the quiet photospheric level at disc centre (see Fig. \ref{fig:LUTs}, top left panel)}, the contrast rises significantly near the limb: for the V band, at limb angles of $\mu = 1.0, 0.5$ and $0.1$ (limb angles $\theta = $ $0$\degs, $60$\degs~and $84$\degs), $(I_{\rm fac}-I_{\rm phot})/I_{\rm phot} = 0.018, 0.048$ and $0.11$. {The contrast tends to $0.13$ at  $\mu = 0.0$.}

\subsubsection{Convective Blueshift}
\protect\label{section:CB}

At the solar photospheric level, the balance between the hotter upwelling central regions of granular cells and the sinking cooler intergranular regions results in a net effective blueshift of spectral lines. This results in skewed absorption line bisectors with a distinctive C-shape \citep{dravins81}. Inside a sunspot {or facular region}, where convection is suppressed, the degree of blueshift is reduced leading to both an apparent relative redshift of spectral lines \textit{and} a change in line bisector shape. Moreover, \cite{gray05shapes} {showed} that changes in convective motions can be probed as a function of spectral type and class by measuring the bisector shape changes.

The convective blueshift (CB) effect contributes the highest degree of activity induced solar RV variability as the average number of active regions fluctuates on solar magnetic activity cycle timescales \citep{meunier10plage}. Consequently, it can have an important systematic effect on precision RV studies aiming to detect exoplanets and precisely measure their masses. We investigated CB in our modelling of RV signatures in active stars \citep{jeffers14activity}. The effects of CB have also been reviewed, modelled and examined in detail by \cite{meunier10plage}, \cite {meunier10recon}, \cite{borgniet15}, \cite{dumusque14soap} and \cite{meunier17blueshift}. 

{

\cite{meunier17blueshift}, examined the convective blueshift as a function of spectral type using a sample of late-F, G and K stars. \cite{liebing21} extended the sample size and included M stars and found that there is no convective blueshift or redshift for stars with $T_{\rm eff} < 4000$\,K. In the following sections, we {consider models}
without CB for all the M dwarf simulations. 


\cite{dumusque14soap} showed that $\Delta v \sim 330$~\ms{} on average for G dwarf photospheric regions compared with active regions where convection is suppressed. Moreover the different line bisector shapes in the photosphere and spot regions have subtle and important effects on the mean line profile of an active star \citep{dumusque14soap}. For our G dwarf model, we have followed a more elaborate prescription by incorporating limb-dependent line shape changes and shifts for both active regions and the quiet photosphere. These are informed by the high resolution limb-dependent solar observations of Fe{\sc{i}} lines by \cite{cavallini85} and \cite{lohner-bottcher19}. Fig. \ref{fig:LUTs} shows the local intensity profiles with the bisectors and shifts that were derived and used by \cite{zhao23soap} (see their Fig. 8). 

}


\begin{figure*}
    \centering
    \includegraphics[trim=5mm 0mm 5mm 0mm, width=2.0\columnwidth]{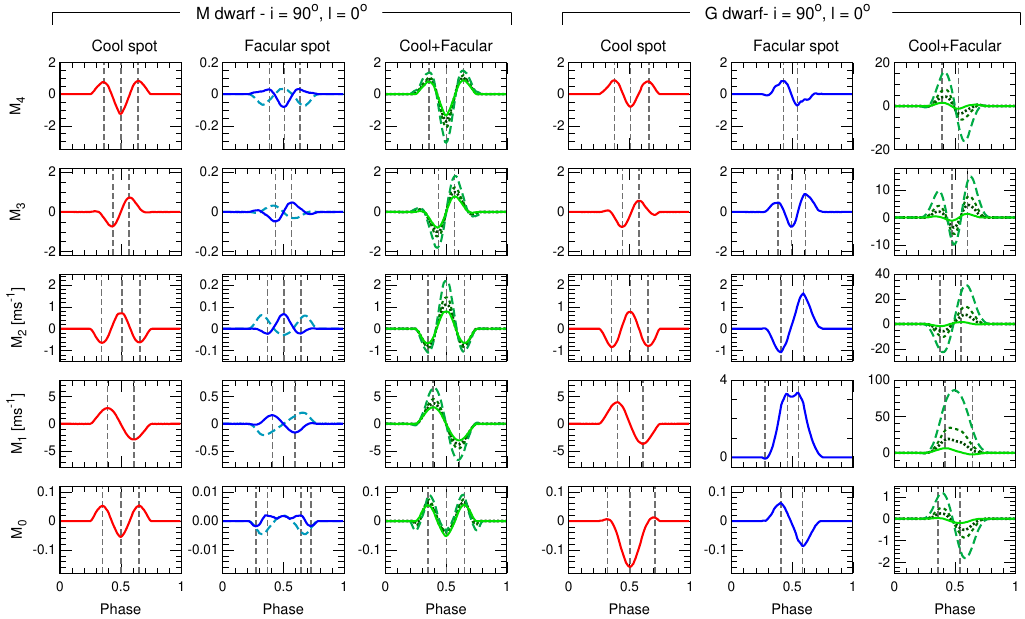}
    \caption{Central Line Moments as a function of rotation phase for an M dwarf star (columns 1-3) and a G dwarf star (columns 4-6) with $v$\,sin\,$i = 2$~\kms{} and axial inclination, $i =90$\degs. The central line moments as defined in Equation \ref{eqn:moments} are shown for an equatorial (latitude, $l=0^{\circ}$)  single cool dark spot with respective umbral and penumbral radii of $r_\textrm{u} = 2$\degs{} and $r_\textrm{p} = 4.47$\degs{} (columns 1 and 4), a bright facular spot with  $r_\textrm{f} = 4.47$\degs{} (columns 2 and 5).
    {For the cool\,+\,facular models in columns 3 and 6, $r_\textrm{u}$ and $r_\textrm{p}$ are the same as in columns 1 and 4, while the ratios of facular area to combined umbral and penumbral areas are $A_\textrm{f}/(A_\textrm{u}+A_\textrm{p}) = 1, 5.29, 10.3$~and~$26.9$ (solid, dotted, short-dashed and long-dashed curves). 
    The solid line in column 2 shows the CLMs using the 
    \citet{johnson21} 
    facular limb-dependent contrasts, while the long dashed curve shows the corresponding CLM using the 
    \citet{borgniet15} 
    limb-dependent contrast used for the G dwarf model. 
    The vertical dashed lines indicate the phases of minimum and maximum deviation for each CLM. N.B. the y-axis extent in columns 1, 3, 4 and 5 are the same (with the exception of $M_1$ for the G dwarf facular spot) and 1/10th this range in column 2 while the column 6 y-axis has a much larger extent.}}
    \label{fig:moments_single}
\end{figure*}

\begin{figure*}
    \centering
    \includegraphics[trim=4mm 0mm 3mm 0mm, width=2.0\columnwidth]{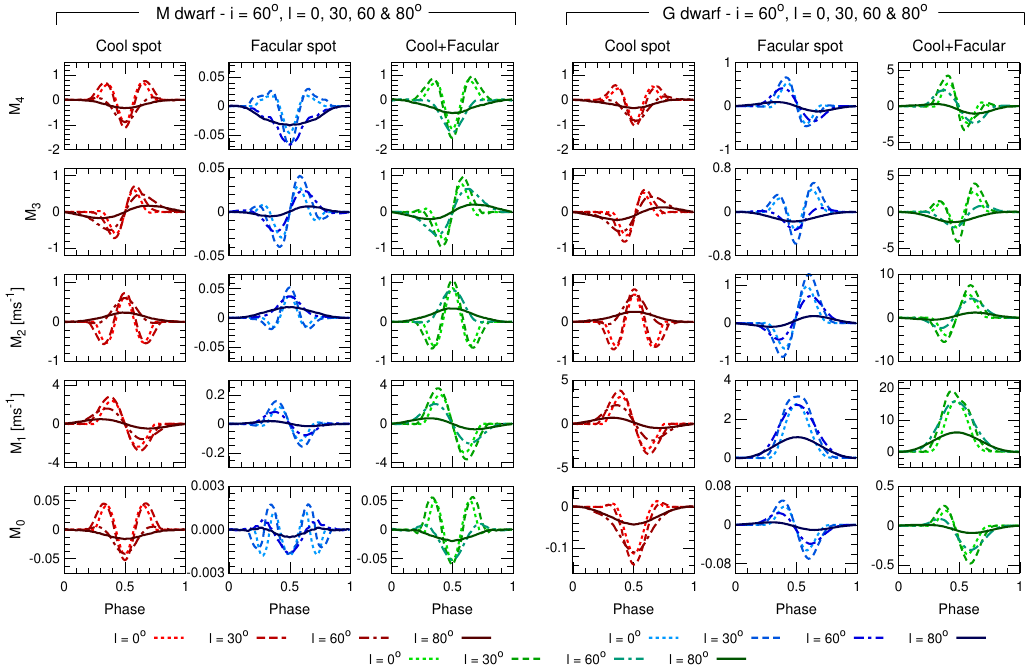}
    \caption{Central Line Moment signatures as a function of rotation phase for an M dwarf and G dwarf star with axial inclination $i = 60$\degs{} and spots {at respective equatorial, low, high and circumpolar latitudes of $l=0,~30,~60$~\&~$80^{\circ}$.} The cool, facular and cool\,+\,facular models are plotted in the same columns as Fig.~\ref{fig:moments_single}. The cool\,+\,facular models are plotted for $A_\textrm{f}/(A_\textrm{u}+A_\textrm{p}) = 5.29$.}
    \label{fig:moments_single2}
\end{figure*}


\subsection{Individual spot construction}
\protect\label{section:spotconstruction}

Line profiles can be simulated from our 3D stellar model for any spot distribution pattern. Each visible 0.5\degs\ pixel was assigned a local intensity profile appropriate for either a cool spot, photosphere or facular region. For cool spots, we simulated both umbral regions at the spot intensity and penumbral regions with an intensity that is half of the difference of the photospheric and spot continuum intensities. The continuum intensity of each visible pixel is obtained by interpolating from the appropriate limb-dependent spectra \citep{husser13atlas}, or in the case of the faculae, scaled relative to the photospheric continuum as described above in \S \ref{section:lineprofiles}. We investigated cases for a cool spot, facular spot and cool\,+\,facular spots. The latter comprises a cool spot surrounded by a facular annulus. The ratio of facular coverage to cool spot area is known to be a function of cool spot coverage: we estimated facular areas from equations 1 \& 2 of \citep{shapiro14}, which give combined cool spot umbra and penumbra areas ($A_\textrm{u}$+$A_\textrm{p}$), and facular areas ($A_\textrm{f}$). The three single spot cases are:

\medskip
\noindent
(1) A single cool spot with umbral radius, $r_\textrm{u} = 2$\degs.  A penumbral and outer umbral annulus is modelled for each spot, with $A_\textrm{p}/A_\textrm{u} = 4$, following \citet{solanki04}. The area ratio typically varies between 3 and 5 \citep{brandt90,steinegger90,beck93}. The corresponding penumbral/umbral radius ratio for $A_\textrm{p}/A_\textrm{u} = 4$ is $r_\textrm{p} = \sqrt{(A_\textrm{p}/A_\textrm{u}+1)}r_\textrm{u}$ yielding $\sqrt{5} \times 2$\degs{} $= 4.47$\degs. 

\medskip
\noindent
(2) A facular spot. For the purposes of illustration, we simulated a spot with the same total area as a cool spot. i.e. $A_\textrm{f} = A_\textrm{p} = 4.47$\degs.

\medskip
\noindent
(3) A cool spot surrounded by a facular ring. For illustration, we simulated two cases: (a) $A_\textrm{f} = A_\textrm{u}+A_\textrm{p}$ and (b) facular rings with $A_\textrm{f}$/($A_\textrm{u}+A_\textrm{p}$) =  26.9, 10.3 and 5.29 appropriate for solar minimum, maximum and high activity cases following the relationships of \cite{shapiro14}. i.e. the facular contribution is greatest for the solar minimum case and lowest for the high activity case. 

\medskip

Our $r_\textrm{u}=2$\degs~spots have penumbral radius, $r_\textrm{p}=4.47$\degs{} and facular outer radii of \hbox{$r_\textrm{f} = \sqrt{5}\sqrt{A_\textrm{f}/(A_\textrm{u}+A_\textrm{p})+1} \times 2$\degs}\ $= 6.32$\degs{} for case (a) with $A_\textrm{f} = A_\textrm{u} + A_\textrm{p}$. For case (b), $r_\textrm{f} = 11.2$\degs, $15.1$\degs{} and $23.6$\degs{} respectively for the high activity, solar maximum and solar minimum cases.  

\section{Central line moments from single spot models}
\protect\label{section:single_spots}

We characterise the expected behaviour of CLMs by first simulating equatorial single starspots during a complete stellar rotation. We then investigate CLM signatures for single spots located at different latitudes before attempting to recover periodicities from the CLMs. Figure \ref{fig:moments_single} shows the noise-free CLM signatures (calculated from absorption lines with SNR~=~$\infty$) for the {M} dwarf and {G} dwarf models for each of the three cases described above in \S \ref{section:spotconstruction} with an axial inclination of $i=90^{\circ}$ and a 
spot located on the equator. 

\subsection{CLM signatures for an M dwarf}
\protect\label{section:single_Mdwarf}

The CLM signatures for the cool, facular and cool\,+\,facular single spots are shown respectively in Figure \ref{fig:moments_single} columns 1,2 and 3. Apart from differences in amplitude, a given CLM shows very similar behaviour in each of the three spot cases.

\subsubsection{M dwarf with single cool spot}
\protect\label{section:Mdwarfcool}

{For approximately half of the rotation cycle, when spots are not visible, all CLM signatures remain flat.} A reduction in local intensity due to a cool spot results in a localised apparent bump in the line profile at the velocity of the spot. The total stellar continuum flux is also reduced since the intensity of a cool spot is lower than the photospheric intensity. This leads to a reduction in line contrast relative to the normalised continuum. A minimum $M_0$ {(Fig.~\ref{fig:moments_single}, column~$1$, row~$1$), occurs when the spot is on the meridian, at the centre of the line profile. At this phase, the spot is travelling transverse to the observer with no Doppler shift.} Relative limb darkening and foreshortening, as the cool spot is seen at different limb angles, are also important and lead to maxima in $M_0$ when the phase angle subtended by the spot relative to the stellar meridian is close, {but not necessarily exactly,} 45\degs. The {phases of maximum and minimum $M_0$} are shown in Fig. \ref{fig:moments_single} by the vertical dashed lines for each CLM.
The amplitude of variability in line area measured by $M_0$ is {over an order of magnitude smaller} than the velocity amplitudes measured by $M_1$ and $M_2$ and the {dimensionless} morphology changes measured by $M_3$ and $M_4$ (Fig.~\ref{fig:moments_single},~column~$1$, rows~$1-5$). 

{The line centroid, $M_1$, is analogous to measuring a flux weighted} RV signature due to an active region. {With no CB in the M dwarf model,} a cool equatorial spot {traces} a symmetric sine-wave signature for the half of the rotation period over which it is visible. {This signature results primarily as a consequence of the localised relative local intensity deficit due to the presence of the cool spot.}
The maximum and minimum values of $M_1$ occur when the line distortion due to the starspot is at a maximum. As with $M_0$, the exact phases are related to the foreshortening limb angle of the spot, the relative limb darkening effect between spot and photosphere and the spot size. When the spot signature is in the blueshifted portion of the line profile, the profile appears deeper in the unperturbed redshifted side, leading to a positive shift of the line centroid. The reverse signature occurs when the spot signature is in the redshifted portion of the line profile. 

For the M dwarf behaviour of $M_2$, the presence of a cool spot located at a phase angle between the limb and the meridian acts to slightly reduce the width of the profile when the maximum spot signature is present. The maximum occurs when the spot is on the meridian because the profile is effectively more U-shaped compared with the spot-free profile, thereby giving relatively more weight to regions away from the profile centre.  The cool spot $M_3$ signature behaves {in a roughly inverse sense} to $M_1$. It is defined such that a spot in the blueshifted portion of the line effectively creates a longer, skewed, tail on the redshifted side of the profile (negative skewness). {This behaviour is exactly as seen with the anti-correlation behaviour between RV and BIS.} {Finally, } $M_4$ measures the kurtosis or tailedness of the line profile. Here the maximum spot signature amplitude in $M_1$ will not only skew the profile (as seen in $M_3$), but will lead to a more tailed profile (positive Kurtosis) on average. Negative Kurtosis, corresponding to the minimum in $M_4$, is seen when the profile appears more U-shaped due to a spot at the centre of the line. 

{The overall behaviour of the even numbered moments for a single spot is similar, with $M_0$ and $M_4$ showing the same general phase-dependent behaviour, while $M_2$ behaves in an inverse manner, but with the same general morphology. This is similar to the inverse behaviour of the odd numbered moments, $M_1$ and $M_3$. However, close inspection of the CLMs in Fig. ~\ref{fig:moments_single} reveals the relative phases of the maxima and minima (vertical dashed lines) are different for each CLM. This feature is important when considering the degree of linearity in the correlation between $M_1$ and $M_3$, which we examine in detail in \S \ref{section:CLMactivity}.}

\subsubsection{M dwarf with single facular spot}
\protect\label{section:Mdwarffac}

{A facular spot with the same total area as a cool spot ($A_\textrm{f}$ = $A_\textrm{u}$+$A_\textrm{p}$) yields a signal that is more than an order of magnitude lower owing to the much lower contrast with the photosphere (Fig.~\ref{fig:moments_single},~column~$2$, with $1/10^{\rm th}$ the scale in Fig.~\ref{fig:moments_single} column~$1$). Because the adopted limb-dependent facular contrast with the photosphere is small and negative for limb angles in the range $0.34 < \mu < 1$ (\citealt{johnson21}; see \S \ref{section:contrastsfacularMdwarf}), the general morphology of M dwarf faculae is the same as for a cool spot as shown by the solid curves in Fig.~\ref{fig:moments_single},~column~$2$. Once spots approach the stellar limb, with $\mu < 0.34$, the facular contrast becomes positive, though foreshortening results in smaller intensity contributions near the stellar limb, leading to relatively low amplitude effects. 
	
In the solar-like case, where the facular contrast is always positive and increases towards the limb (\citealt{borgniet15}; see \S \ref{section:contrastsfacular} - \S \ref{section:contrastsfacularGdwarf}), behaviour that is broadly inverse to the cool spot CLM behaviour is seen (Fig.~\ref{fig:moments_single},~column~$2$, long-dashed curve).
However, since the facular intensity contrast with the photosphere is greatest at the limb, the maximum deviation for a facular spot occurs $\sim 0.05$ earlier in phase (before phase 0.5), or later (after phase 0.5) than for a cool spot for our model setup (as shown by the vertical dashed lines in columns 1 and 2 of Fig.~\ref{fig:moments_single}). For subsequent simulations, we use the facular contrast law described by \citet{johnson21} and assume that CB effects are not present.}

\subsubsection{M dwarf with single cool\,+\,facular spot}
\protect\label{section:Mdwarfcoolfac}

{The third column in Fig. \ref{fig:moments_single} demonstrates how a facular region around a cool spot reinforces the cool spot signature, leading to an increased amplitude. This is a consequence of the behaviour noted in \S \ref{section:Mdwarffac} for the \citet{johnson21} facular contrast law (the \citet{borgniet15} solar law would lead to the opposite effect: a partial damping of the dominant cool spot component). The solid {curve represents} case (a) where the facular region has the same area as the cool spot. For case (b) with the highest facular area of $A_\textrm{f}/(A_\textrm{u}+A_\textrm{p}) = 26.9$ (see \S \ref{section:spotconstruction}), the signature is only augmented by a factor of up to around $\sim 2$ in CLMs $M_1$ to $M_4$, because of the low facular contrast.}

\begin{figure*}
    \centering
    \includegraphics[width=0.995\columnwidth]{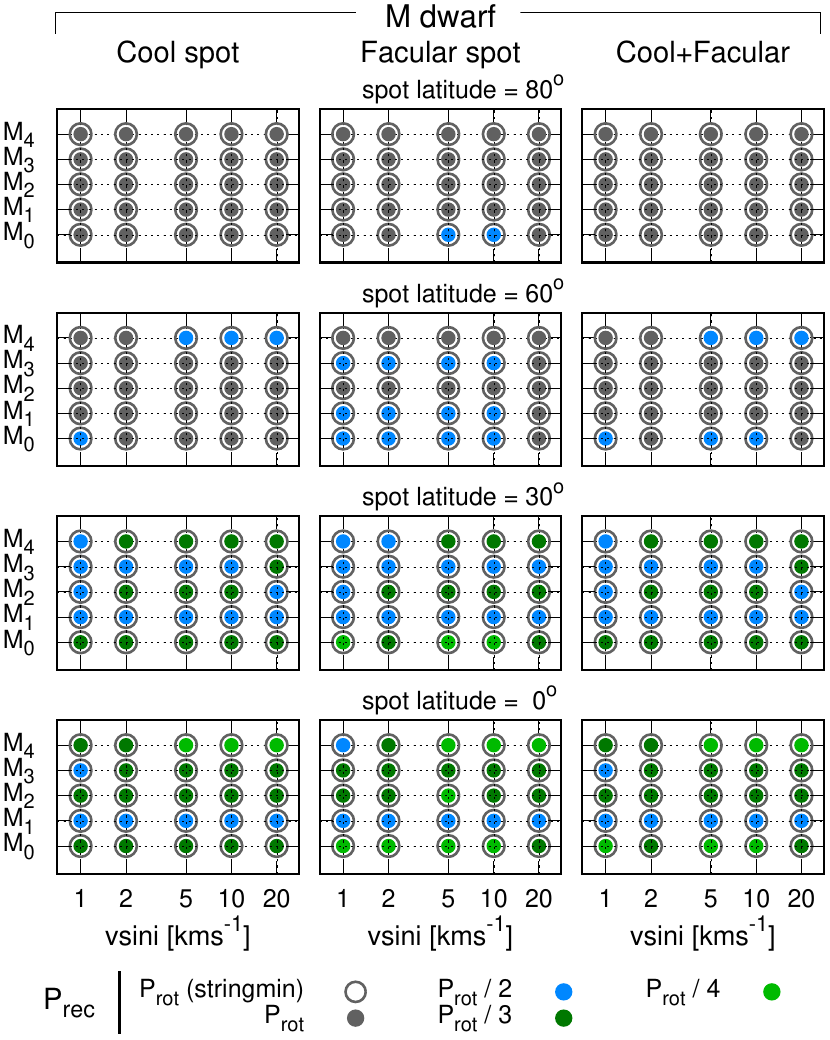}
    \hspace{4mm}
    \includegraphics[width=0.995\columnwidth]{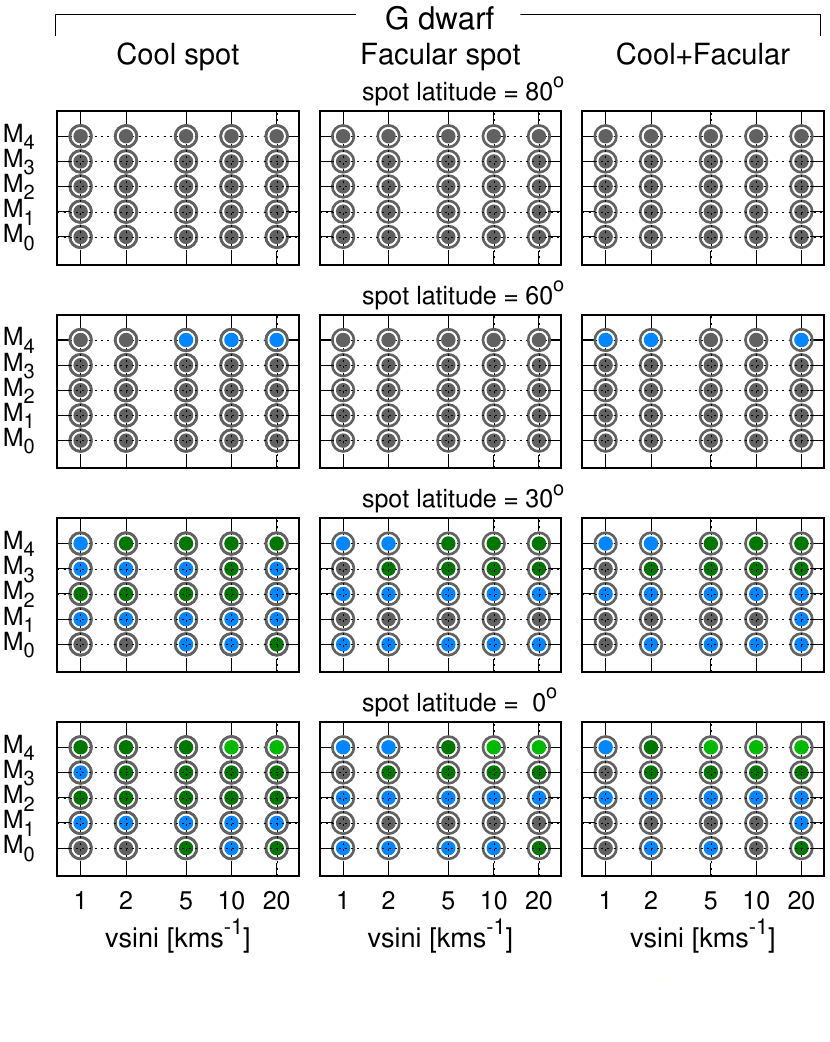}
    
    \caption{
        Period recovery matrices showing the dominant recovered period, $P_\textrm{rec}$, as a fraction of the simulated period, $P_\textrm{rot}$, for M dwarf (left) and G dwarf (right) models with $i=60^{\circ}$. $P_\textrm{rec}$ is colour coded and shown for CLM vs simulated \vsini{} value in each sub-panel. Sub-panel rows 1-4 show respective results for spot latitudes $l = 0^{\circ}, 30^{\circ}, 60^{\circ}, 80^{\circ}$. Sub-panel columns 1-3 for each stellar model show results for the cool spot, facular spot and combined {cool\,}+\,facular spot models.}
    \label{fig:sensitivity_onespot}
\end{figure*}

\subsection{CLM signatures for a G dwarf}
\protect\label{section:single_Gdwarf}

{The G dwarf CLM signatures for the cool, facular and cool\,+\,facular single spots are shown respectively in Figure \ref{fig:moments_single} columns 4, 5 and 6. Here, with a fixed spot size, the $340$\,ms$^{-1}$ convective blueshift yields a facular signature for a given CLM that is different from the corresponding cool spot signature in terms of morphology, but similar in terms of amplitude. As a consequence, the facular signature dominates the morphology of the cool\,+\,facular CLMs for the simulated $A_\textrm{f}/(A_\textrm{u}+A_\textrm{p})$ ratios, leading to relatively large CLM amplitudes.
}



\subsubsection{G dwarf with single cool spot}
\protect\label{section:Gdwarfcool}

The cool spot CLMs, $M_0$ - $M_2$, in the G dwarf model (Fig.~\ref{fig:moments_single}, column~4) have slightly larger amplitudes compared with the M dwarf model owing to the higher contrast between photosphere and spots. The cool spot distortion in an absorption line, and consequently, the CLM signature, is dominated by the relatively large local intensity continuum contrast between the spot and photosphere. At {this contrast}, the local intensity line shape, or any shift in the local intensity line profile (i.e. from CB effects) have only secondary effects. {However, the relative shift between the local intensity line profiles of the photosphere and spot can be seen as slight asymmetries. For example, in $M_1$, the positive deviation before phase 0.5 is $\sim 5$\% greater than the negative deviation after phase 0.5. Similarly, the pre-phase 0.5 minimum in $M_2$ is $\sim 6$\% greater than the post-phase 0.5 minimum. The additional complexity induced by adopting limb-dependent local intensity line shapes and CB blueshift velocity offset between photosphere and active regions (see \citealt{zhao23soap}, Fig. 8) dilutes the amplitude of the higher order moments with the effect that the $M_3$ and $M_4$ amplitudes are {\em less} than those seen for the M dwarf cool spot model.} $M_0$ shows higher amplitude than for the M dwarf model, with a lower degree of complexity (i.e. it passes through a minimum when the spot is on the stellar meridian) but is still an order of magnitude lower than the other CLMs.

\subsubsection{G dwarf with single facular spot}
\protect\label{section:Gdwarffac}

{The presence of a facular region results in a net apparent redshift of the line profile, illustrated by the behaviour of $M_1$ (Fig.~\ref{fig:moments_single}, column~5). A very small, almost negligible, blueshift (negative velocity) at phase 0.28 is seen. To first order, the shift is symmetric. The maximum redshift values occur when the facular region is viewed at phases $\sim 0.45$ and $\sim 0.55$ (the peak at $\sim 0.55$ is slightly higher than the peak at  $\sim 0.45$) as a result of the limb angle dependency of the line bisector shapes and the relative convective blueshift values as a function of limb angle (see \cite{zhao23soap}, Fig. 8). Compared with the other CLMs, $M_1$ is unique in its morphology when a facular spot or dominant facular region (see \S \ref{section:Gdwarfcoolfac} below) is present. This behaviour also contrasts with the M dwarf model where clear correlations or anti-correlations are seen between all odd moments or even moments.
}

\subsubsection{G dwarf with single cool\,+\,facular spot}
\protect\label{section:Gdwarfcoolfac}

The cool\,+\,facular case {(Fig. \ref{fig:moments_single}, column~6)} shows how the convective blueshift of the facular region dominates the morphology for {all $A_\textrm{f}/(A_\textrm{u}+A_\textrm{p}) > 1$ models (i.e. case (b) in \S \ref{section:spotconstruction});} that is, the cool\,+\, facular CLM morphologies more closely resemble the facular morphologies than the cool morphologies. {Only for $A_\textrm{f}/(A_\textrm{u}+A_\textrm{p}) = 1$ (i.e. case (a) in \S \ref{section:spotconstruction}), are more equal contributions seen from individual cool and facular components, as would be expected from Fig. \ref{fig:moments_single} columns 4 and 5.} For the solar minimum case, when $A_\textrm{f}/(A_\textrm{u}+A_\textrm{p}) = 26.9$ (long-dashed lines), the convective blueshift effect due to the large facular area results in an order of magnitude increase in the CLMs. With an average spot radius of 1\degs, as seen on the Sun \citep{takalo20sunspots}, the cool\,+\,facular variability from our models is {of order 20~\ms, which is larger than the solar RVs derived by \cite{meunier10plage}. In reality facular regions are not confined to an annulus around a cool spot, but are more distributed, which would lead to lower amplitude signatures.}

\subsection{The effect of instrumental resolution}
\protect\label{section:resolution}
{
We investigated the effect of increasing the resolution from the standard $R=115,000$ for the single spot G dwarf models with \vsini~=~2\,kms$^{-1}$ discussed in the preceding sections. The CLM signatures for a $2$\degs~equatorial spot, assuming $i=90$\degs are shown in columns 4-6 of Fig.~\ref{fig:moments_single_resolution}. Specifically, we looked at the higher resolutions of $R=140,000$ and $R=190,000$ of ESPRESSO \citep{pepe21espresso} in addition to a very high resolution of $R=500,000$. Because the rotational broadening is a significant fraction of the line broadening at \vsini~=~2\,kms$^{-1}$, we find that the effects on lower order moments is fairly subtle. The large convective blueshift that dominates the $M_1$ signature in a G dwarf (see equivalent RV signature contributions in \cite{zhao23soap}) is already well resolved at $R=115,000$ so that further increasing the resolution only increases the amplitude of $M_1$ by $1.8$\%, $1.3$\% and $0.5$\% respectively for the cool, facular and cool\,+\,facular (solar maximum ratio with $A_\textrm{f}/(A_\textrm{u}+A_\textrm{p}) = 10.3$) models at $R=500,000$. By contrast, the equivalent $M_3$ amplitude increases for the cool, facular and cool\,+\,facular models are respectively $36$\%, $63$\% and $60$\%. However, for the G dwarf model, the degree of correlation, which we investigate below, {\em decreases}, with increasing resolution. Although the amplitude changes are more pronounced in the higher order moments, we note that $M_4$ notably resolves more structure, which arises from the convective blueshift and limb-dependent bisector changes. For completeness, we also show CLM signatures for a very low \vsini~=~1\,kms$^{-1}$ in Fig.~\ref{fig:moments_single_resolution} (columns 1-3). Although the CLM amplitudes are predominantly lower compared with  \vsini~=~2\,kms$^{-1}$, there is a visibly more dramatic change in amplitude in $M_3$, while the cool\,+\,facular $M_4$ signature at $R=500,000$ shows a very large amplitude signature.
}

\subsection{Dependence of CLM signature and spot latitude }
\protect\label{section:single_latitude}

Figure \ref{fig:moments_single} demonstrates that single spots located on the stellar equator do not produce purely sinusoidal signatures in the CLMs. This was noted by \cite{boisse11} (see their Fig. 7), who also showed that fractions of the true stellar rotation period, typically $P_\textrm{rot}/2$ or $P_\textrm{rot}/3$, may be recovered with greater power than $P_\textrm{rot}$ when measuring CCF RVs. The recovered period with dominant power, {$P_\textrm{rec}$,} thus depends on the form of the CLM when it is visible, but also on the duration of visibility of a spot throughout a complete rotation, which is determined by the stellar axial inclination, $i$, and the latitude,~$l$, of the spot. We investigated the behaviour of the CLMs for a star with $i = 60^{\circ}$, {with spots at latitudes,} $l=0^{\circ}$, $30^{\circ}$, $60^{\circ}$ and $80^{\circ}$. We again considered cool spot, facular spot and cool\,+\,facular spot models. For the cool\,+\,facular spot case, we used $A_\textrm{f} / (A_\textrm{u}+A_\textrm{p}) = 5.29$. 

{The CLM signatures for a single spot at different latitudes are shown in Figure \ref{fig:moments_single2}. Equatorial and low latitude (i.e. $l=0^{\circ}$ and $30^{\circ}$) spot CLM signatures replicate the CLM signatures seen in the $i = 90^{\circ}$ simulation. In contrast, the circumpolar spots at $l=80^{\circ}$ induce signals that more closely resemble pure sinusoids in almost all CLMs because they are always visible, but with varying intensity contribution modulated on the $P_\textrm{rot}$ timescale to first order. }
As expected, the CLM amplitudes are largest for spots at the sub-observer's latitude ($90^\textrm{o} - i = 30$\degs{}). {The} signatures for $l=60^{\circ}$ are intermediate between those of low and circumpolar spots in both amplitude and morphology. 

\subsection{Recovering periodicities from single spot models}
\protect\label{section:single_periods}


\begin{table}
	\begin{center}
		\begin{tabular}{c|c|c}
			\hline
			&   M dwarf           &     G dwarf        \\
			
			\vsini{} [\kms]  &   {$P_\textrm{rot}$ [d]} &  {$P_\textrm{rot}$ [d]} \\
			\hline
			1                       &        25.25        &          50        \\
			2                       &        11.11        &          25.25      \\
			5                       &         4.45       &           11.11      \\
			10                      &         2.25       &           4.45      \\
			20                      &         1.16       &           2.25      \\
			\hline
		\end{tabular}
	\end{center}
	\caption{Sensitivity simulation \vsini{} and $P_\textrm{rot}$ combinations for the G and M dwarf models with respective $R_* = 1$R$_\odot$ and $0.5$R$_\odot$.}
	\label{tab:vsini_p}
\end{table}

{Using the single spot models, with $i=60$\degs{} and $l=0^{\circ}$, $30^{\circ}$, $60^{\circ}$ and $80^{\circ}$}, we investigated the recovered periodicities of CLM signatures for an expanded parameter space by including
the effect of rotation velocity. We considered $v\,\textrm{sin}\,i~=$ $1$, $2$, $5$, $10$ and $20$~$\textrm{km\,s}^{-1}$ for both the M dwarf and G dwarf models. Corresponding periods, based on the assumed stellar radii of $R_* = 0.5$~$\textrm{R}_\odot$ and $1$~$\textrm{R}_\odot$ were estimated and modified slightly for the shorter periods to minimise integer day-aliasing effects. The \vsini\ and rotation period combinations are shown in Table \ref{tab:vsini_p}, with periods chosen to further minimise integer day rotation period aliasing.
{For each spot latitude, $l$, and \vsini{} vs $P_\textrm{rot}$ combination, 120 daily noise free (SNR =~$\infty$) absorption profiles spanning $120$~days were synthesised with observation time uncertainties of $3$\,hrs to further minimise $1$\,d period aliasing. The effect of adding finite noise is explored with more complex models in \S \ref{section:spotmodels}.}

\subsubsection{Periodicities in Generalised Lomb-Scargle periodograms}
\protect\label{section:single_period_gls}

Period searches were performed on each derived CLM timeseries using the generalised Lomb-Scargle (GLS) periodogram analysis\citep{zechmeister18}. In each case, the \textit{dominant} recovered period, $P_{\rm rec}$ {(i.e. the periodogram peak with the highest power)} was recorded by searching for the most significant peak power at $P_\textrm{rot}$ and its fundamental harmonics at $P_\textrm{rot}/2$, $P_\textrm{rot}/3$, $P_\textrm{rot}/4$, $P_\textrm{rot}/5$ and $P_\textrm{rot}/6$. 
A {tolerance of 10 per cent} in {$P_{\rm rec}$} was required for identification as either $P_\textrm{rot}$ or one of its harmonics. Fig. \ref{fig:sensitivity_onespot} shows the recovered periodicities for the M dwarf and G dwarf models with an axial inclination of $i=60^{\circ}$ for cases with a single spot placed at latitudes $0^{\circ}$, $30^{\circ}$, $60^{\circ}$ and $80^{\circ}$. The {recovered period, $P_{\rm rec}$,} for combinations of SNR and $v\,\textrm{sin}\,i$ are colour coded according to the dominant GLS peak and plotted as solid circles. Broadly, the results can be summarised as follows:

\begin{enumerate}[wide, labelwidth=!,itemindent=!,labelindent=0em, leftmargin=0.5cm, label=(\roman*), itemsep=0.1cm, parsep=0pt]
{\bf
\item Single spots induce dominant CLM periodicities related to, but not necessarily at $P_{\rm rot}$
\item Dominant periodicities at $P_{\rm rot}/2$, $P_{\rm rot}/3$ or $P_{\rm rot}/4$ are found, depending on spot latitude and type
\item High latitude spots ($l=80$\degs): the spots are always visible, predominantly resulting in sinusoidal behaviour at $P_\textrm{rot}$
\item As a single spot is simulated at successively lower latitudes, progressively higher harmonics of $P_{\rm rot}$ tend to be recovered
\item For the G dwarf and M dwarf cool spot models, $P_{\rm rec}$, is very similar, since spot contrast is the predominant contributor to CLM variation
\item The G dwarf facular and cool\,+\,facular models recover $P_{\rm rec}$ at lower harmonics of $P_{\rm rot}$ compared with the M dwarf in $M_0$, $M_1$ and $M_2$ because the convective blueshift effect dominates and simplifies the phase dependent signature
}

\end{enumerate}

The simulations demonstrate that recovery of the true rotation period for even single spot configurations {with a GLS periodogram} should not always be expected, particularly for spots that appear at low latitudes. When facular regions are present and more sinusoidal behaviour is seen in the CLMs (e.g. G dwarf, $M_1$ models), $P_\textrm{rot}$ is recovered. {The lack of convective blueshift in M dwarfs means that the CLM signatures are more prone to showing higher harmonics of  $P_\textrm{rot}$. Nevertheless, the contribution from a facular region is much smaller on an M dwarf compared with a G dwarf because the facular contrasts are small. They would thus be relatively difficult to detect in data with finite SNR.}

A single spot (or dominant {compact } spot group) will only {potentially} yield $P_\textrm{rot}$ as the dominant periodicity in all CLMs if the spot is always visible at high latitude. This is most likely for stars with significant axial tilt relative to the observer (i.e. $i << 90^{\circ}$), and is dependent on the expected spot distribution. Under the solar paradigm, where spots generally only emerge at low-intermediate latitudes, with $l < 40^{\circ}$, a star would need to be significantly inclined to yield this behaviour. Nevertheless, there is a large body of evidence that suggests active stars possess polar or circumpolar spots of significant size (e.g. see \cite{strassmeier09starspots} for a review of stars with Doppler images and \cite{almenara22toi3884} who have recently invoked a polar spot to explain unusual planet transit lightcurves). 

\subsubsection{Periodicities from String Length Minimisation}
\protect\label{section:single_period_sl}

{
Alternative approaches have been used for recovering periodicities in data, particularly where signals are not strictly sinuosoidal. \citet{mcquillan13} demonstrated that the auto-correlation function (ACF) is able to recover the correct rotation period for simulated cases where there are amplitude changes, discontinuities in the data, or where the variability isn't strictly sinusoidal. This method works well for data that are well sampled, and was used by \citet{mcquillan13} to recover rotation periods from high cadence Kepler photometry. However, the ACF method is not effective when there are fewer data with more phase gaps. Instead, we opted to use the string length minimisation method (\citealt{burke70string}, \citealt{sworetsky83string}). This method was applied successfully to the planet hosting DMPP-3 system \citep{barnes20dmpp3}, which shows highly eccentric RV variability from the DMPP-3AB binary orbit at $507$\,d, which standard periodogram methods were not initially able to identify. A String Length (SL) periodogram is obtained simply by phase-folding the data using a sequence of trial periods. The phase-folded data for each trial period are ordered by phase (i.e. from 0 to 1) and the string length, defined by the total summed distance between neighbouring points, is calculated. An SL periodogram is thus constructed, in which the period with the {\em minimum} string length indicates the optimal phase-fold. This nonparametric approach is difficult to assess when data become noisy. For the purposes of our period recovery, we have chosen to identify $P_\textrm{rec}$ by searching for the minimum peak that lies at $P_\textrm{rot}$ or its harmonics (as described above in \S \ref{section:single_period_gls}) by searching for SL minima that lie at $\geq 99$\% or $2.5$-$\sigma$ below the SL periodogram mean.

Fig. \ref{fig:sensitivity_onespot} shows $P_\textrm{rec}$ identified from the SL minimum for for CLM vs \vsini{} at each spot latitude. The SL recovered $P_\textrm{rec}$ is indicated by the {\em open circle} symbol, which surrounds the GLS recovered $P_\textrm{rec}$ indicated by the {\em closed circles}. Importantly, the SL method recovers the simulated rotation in all cases: all open circles are grey, indicating $P_\textrm{rec}$ = $P_\textrm{rot}$, unlike the recovered GLS periodicities. In \S \ref{section:spotmodels}, we next investigate CLM signature morphology and recovery of rotation periods in the CLMs with GLS and SL searches using more realistic spot distributions of stars with solar-like activity.
}

\begin{figure*}
   \begin{center}
   \begin{tabular}{ccccc}
   \includegraphics[trim=0 -5mm 0 0mm, width=0.59\columnwidth]{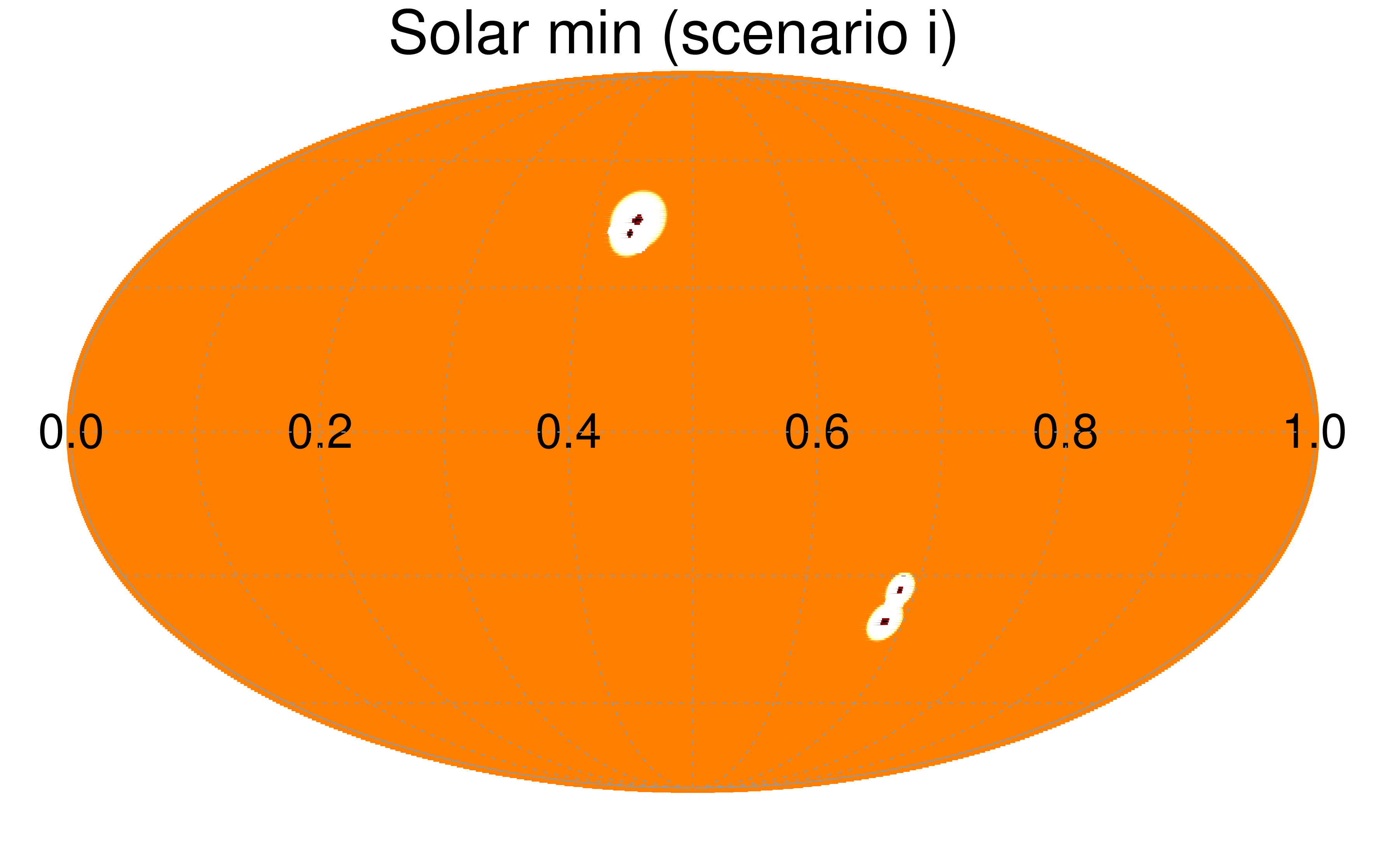} &
   \includegraphics[width=0.34\columnwidth]{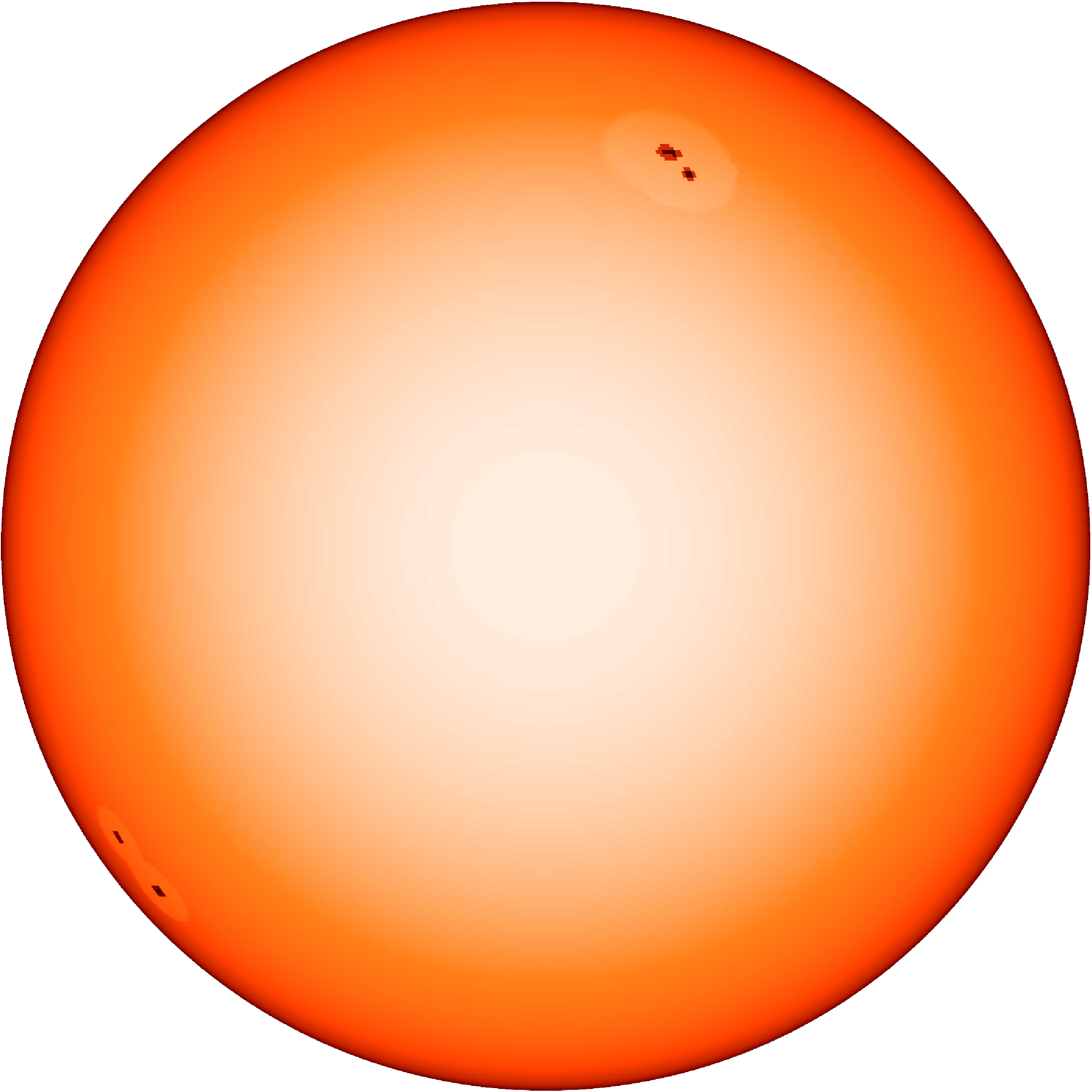} &
   \hspace{0mm} &
   \includegraphics[trim=0 -5mm 0 0mm, width=0.59\columnwidth]{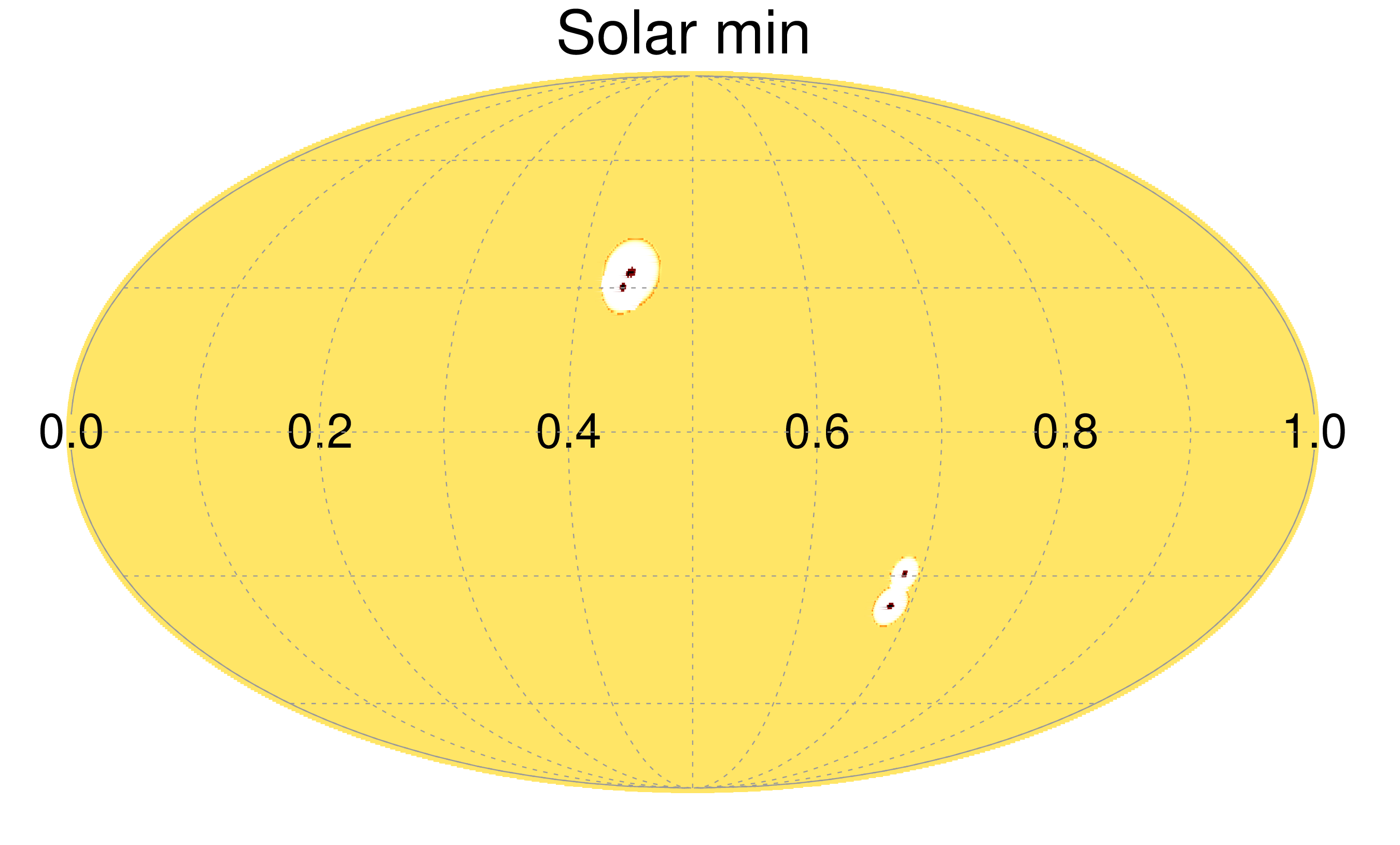} &
   \includegraphics[width=0.34\columnwidth]{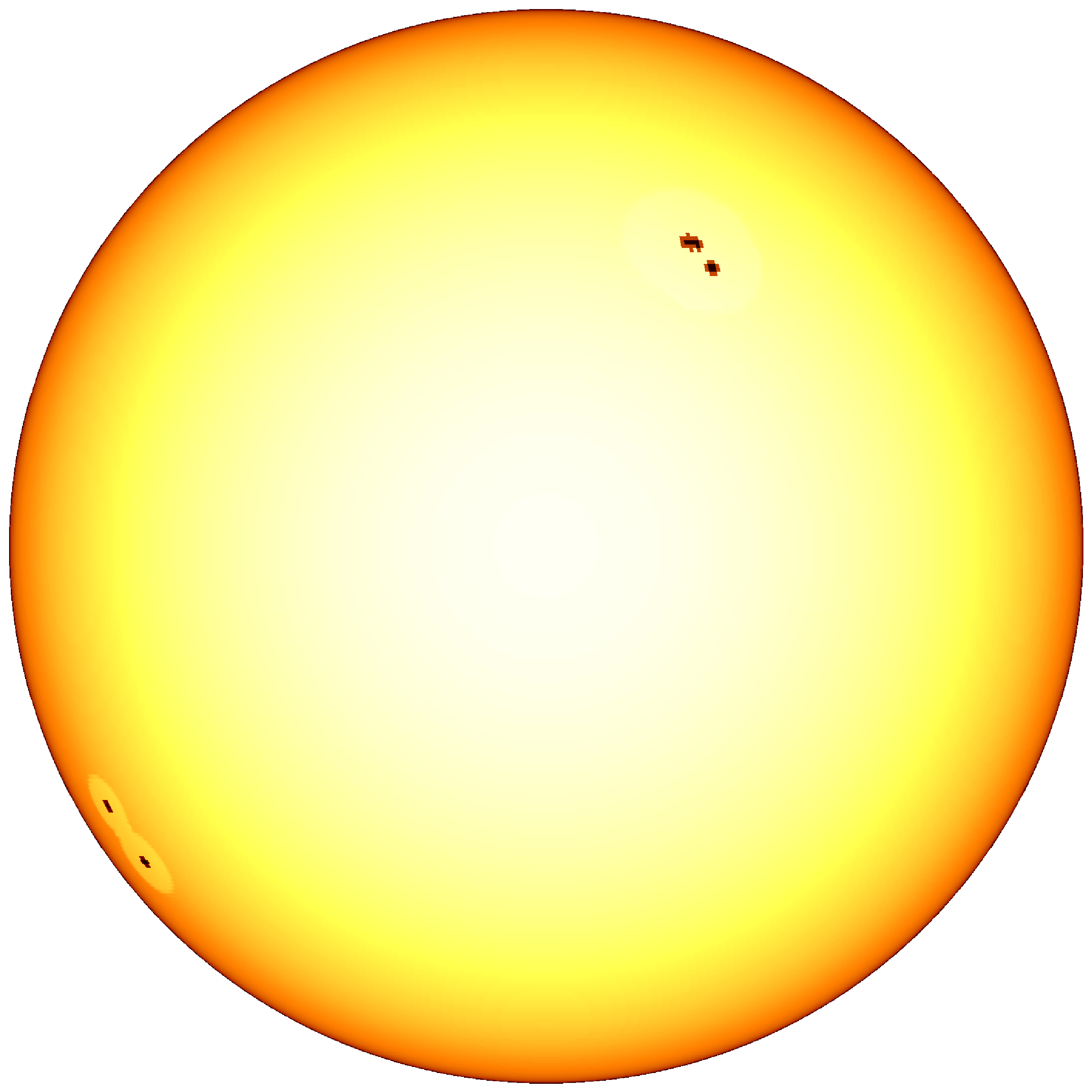} \\
   
   \includegraphics[trim=0 -5mm 0 0mm, width=0.59\columnwidth]{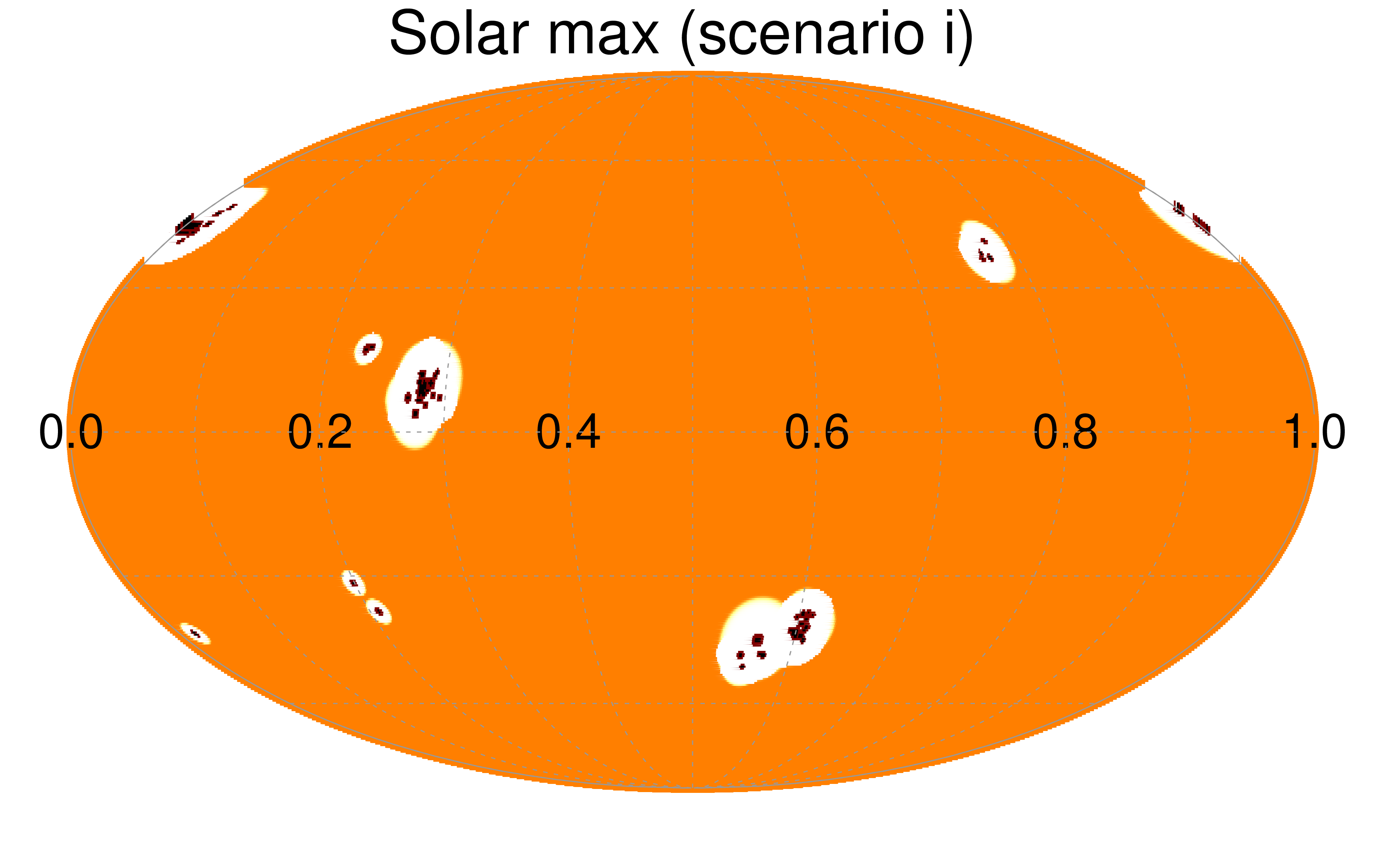} &
   \includegraphics[width=0.34\columnwidth]{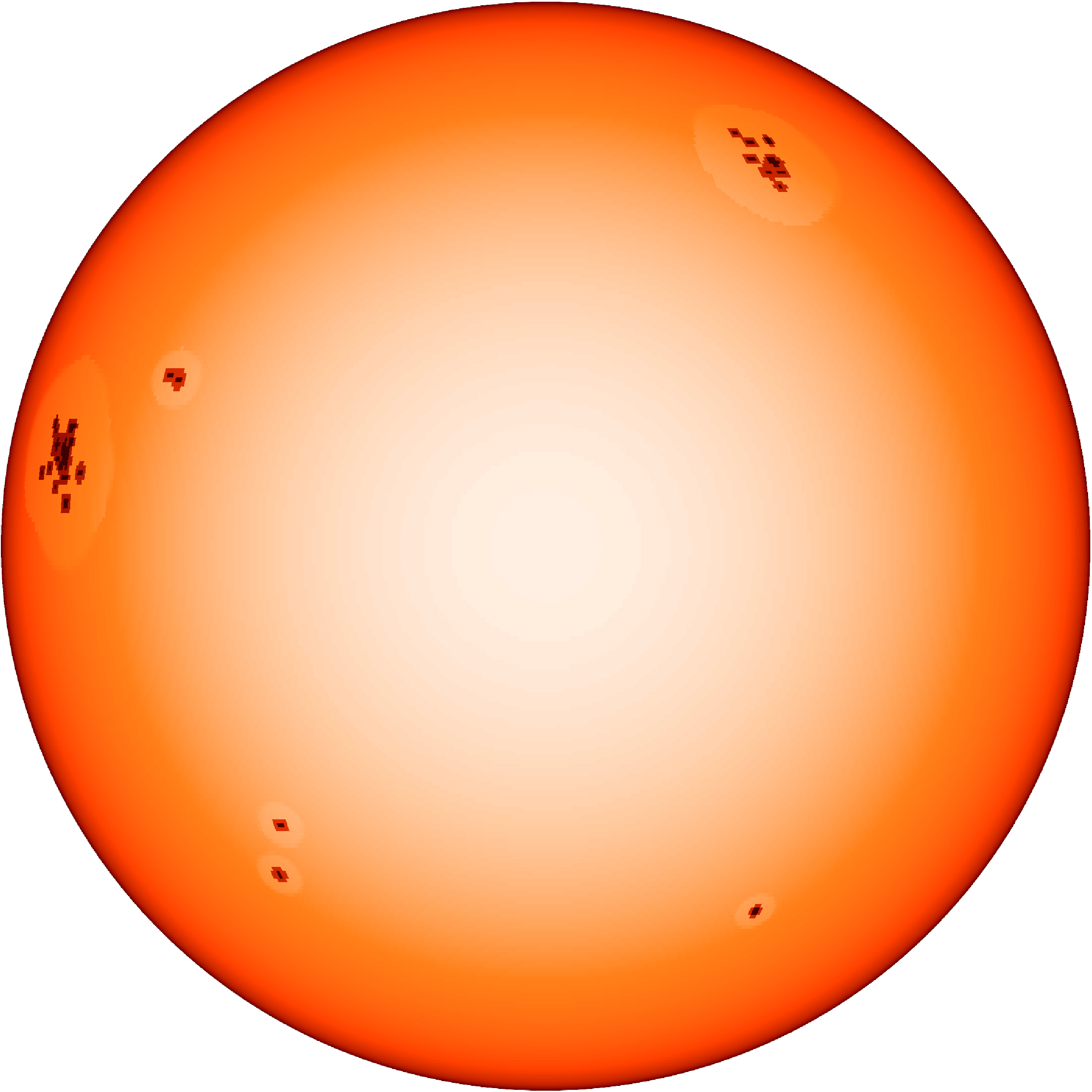} &
   \hspace{0mm} &
   \includegraphics[trim=0 -5mm 0 0mm, width=0.59\columnwidth]{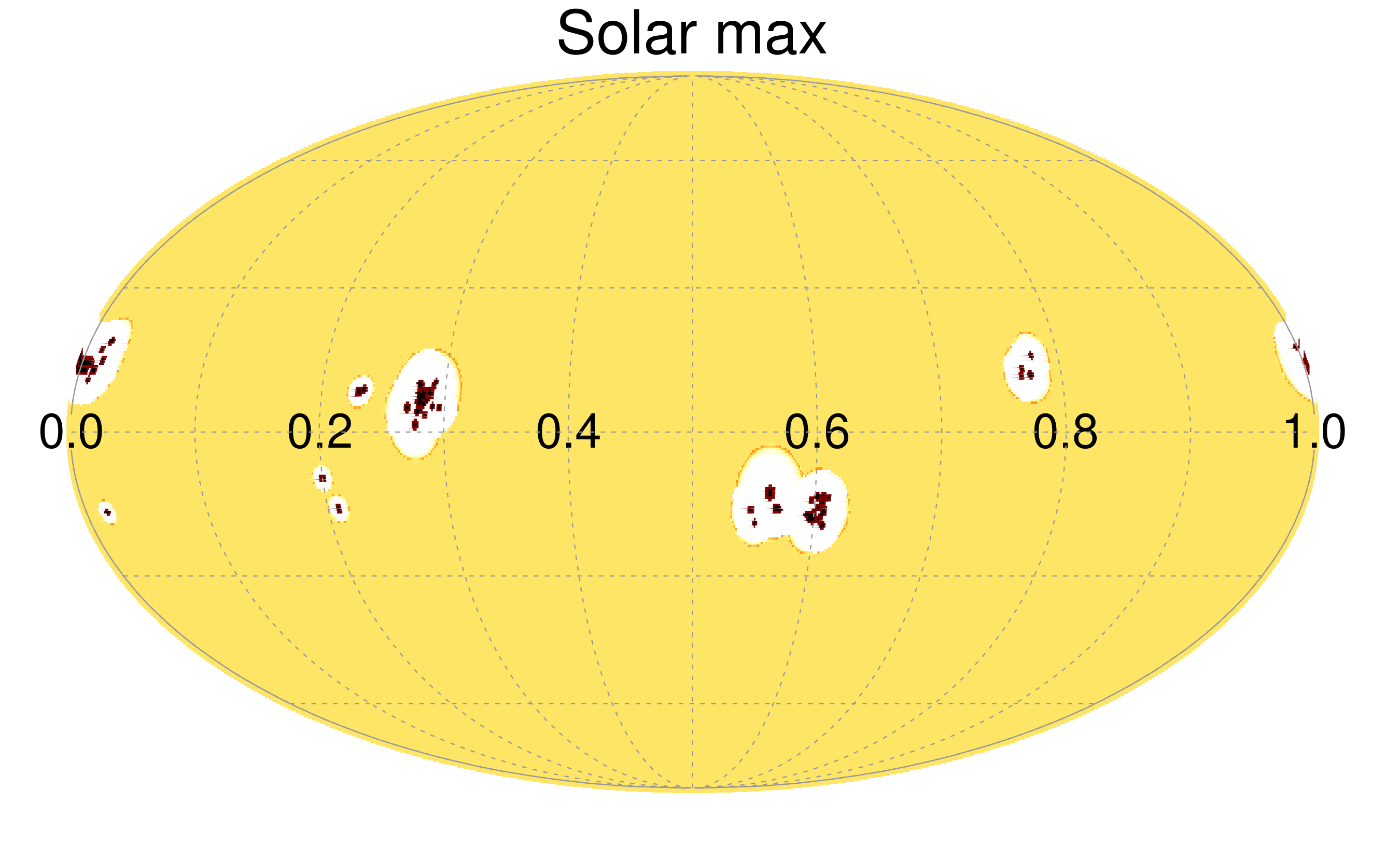} &
   \includegraphics[width=0.34\columnwidth]{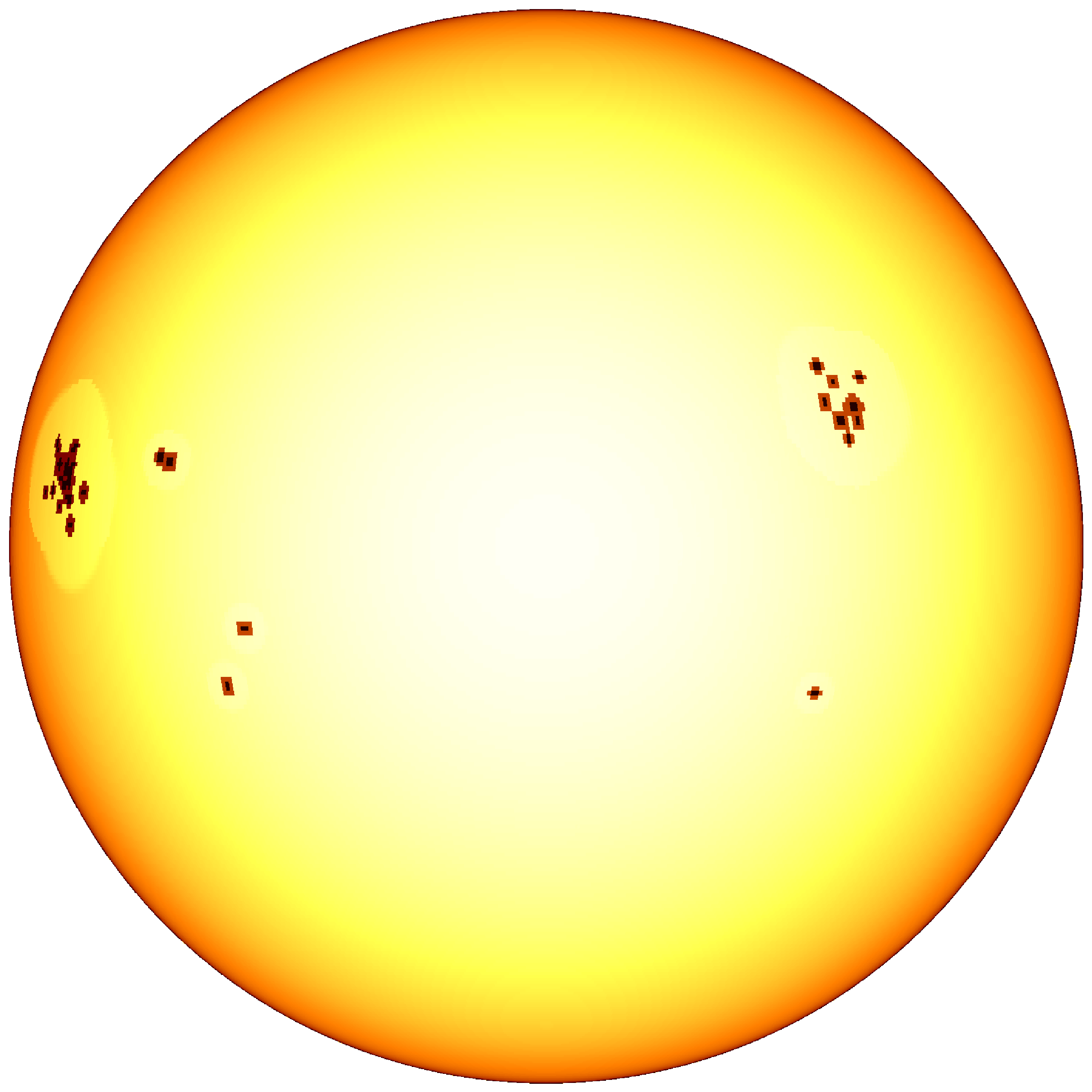} \\

   \includegraphics[trim=0 -5mm 0 0mm, width=0.59\columnwidth]{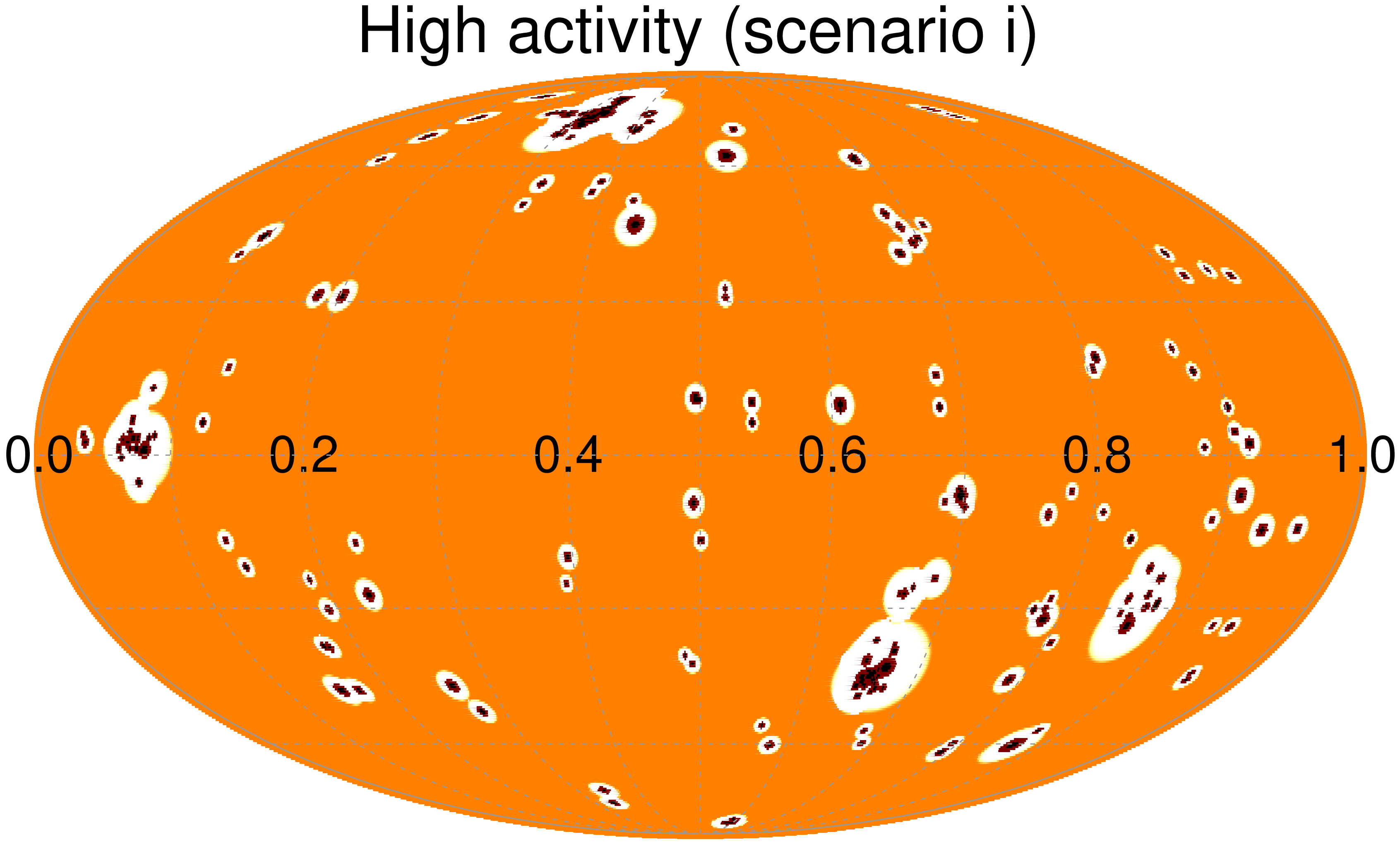} &
   \includegraphics[width=0.34\columnwidth]{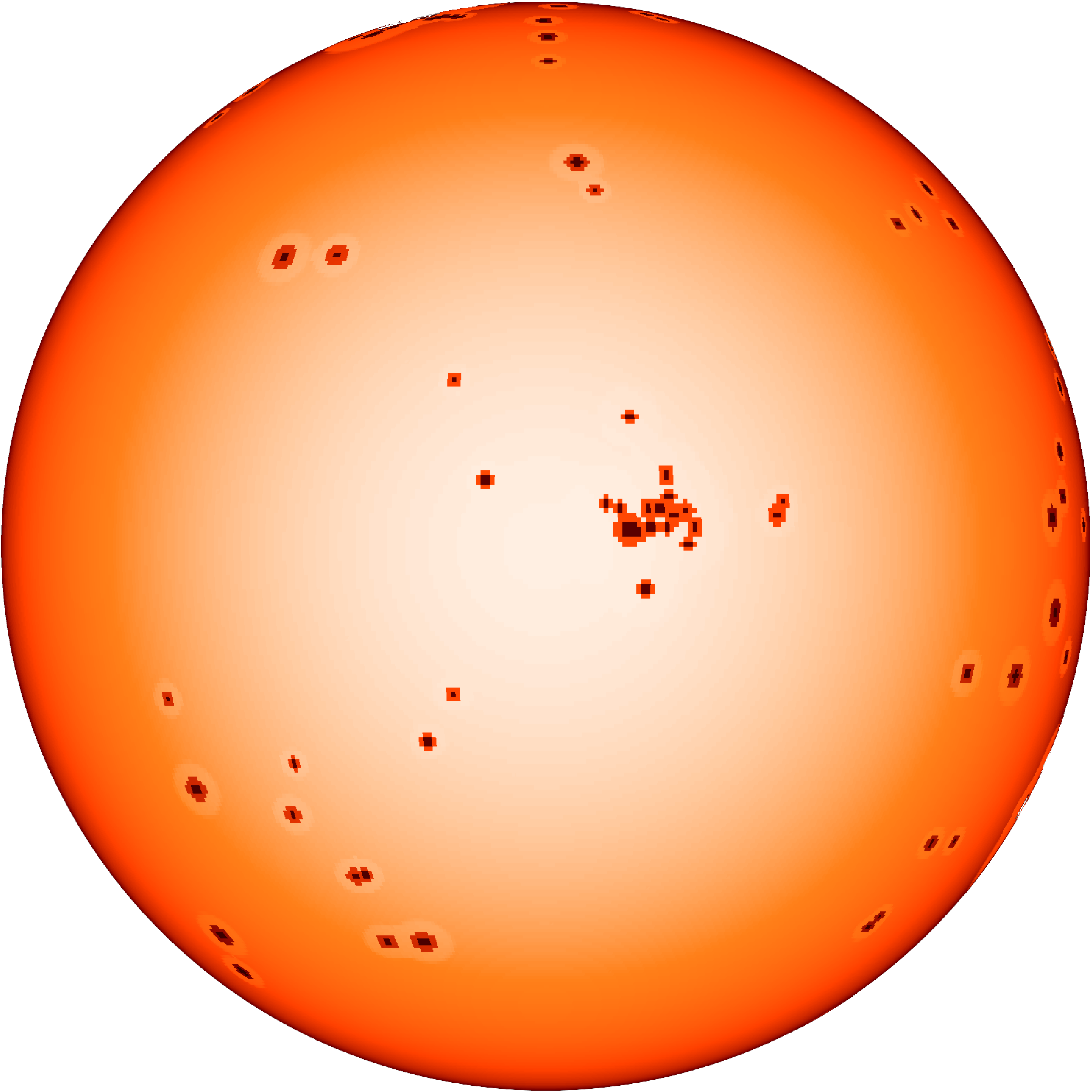} &
   \hspace{0mm} &
   \includegraphics[trim=0 -5mm 0 5mm, width=0.59\columnwidth]{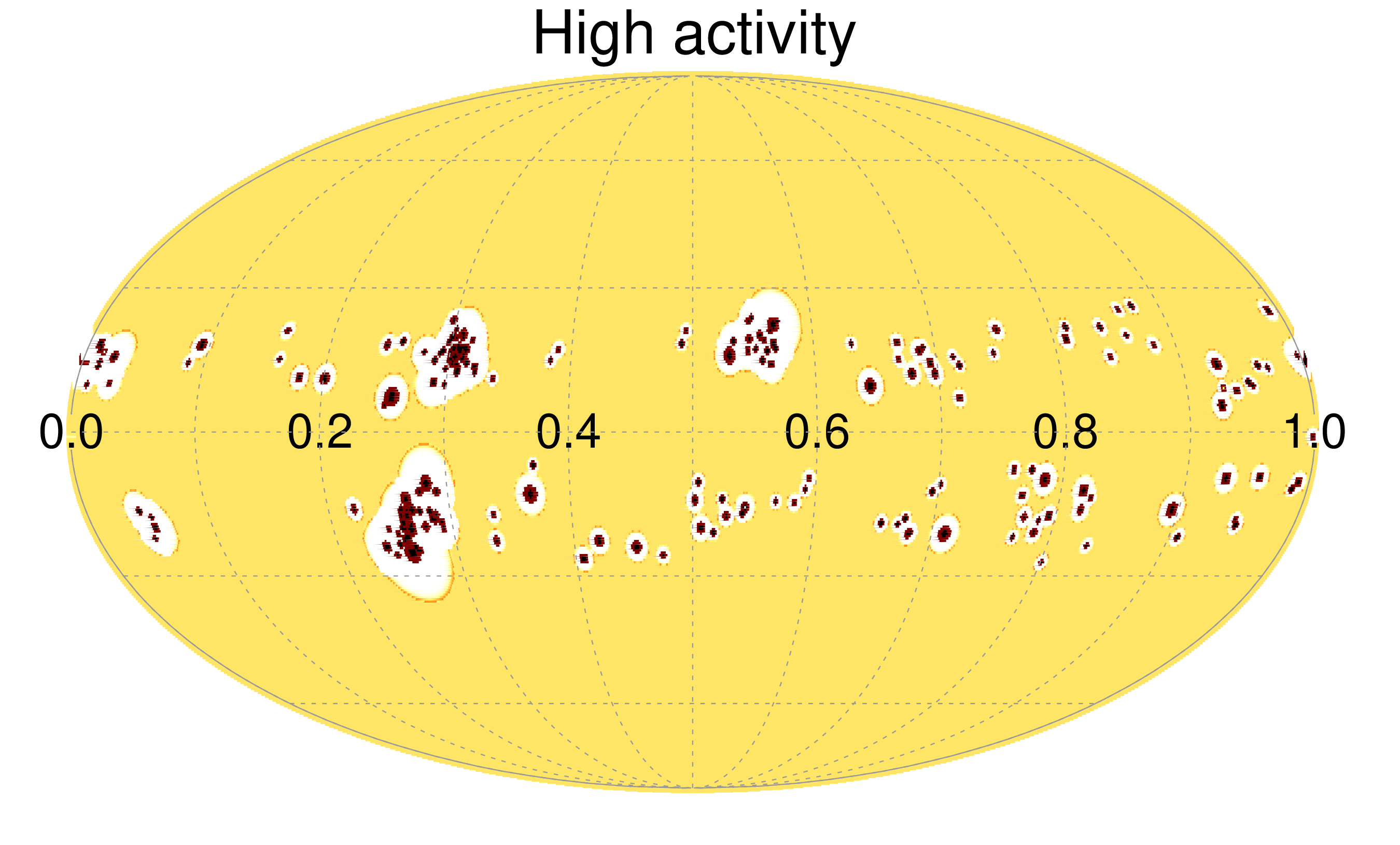} &
   \includegraphics[width=0.34\columnwidth]{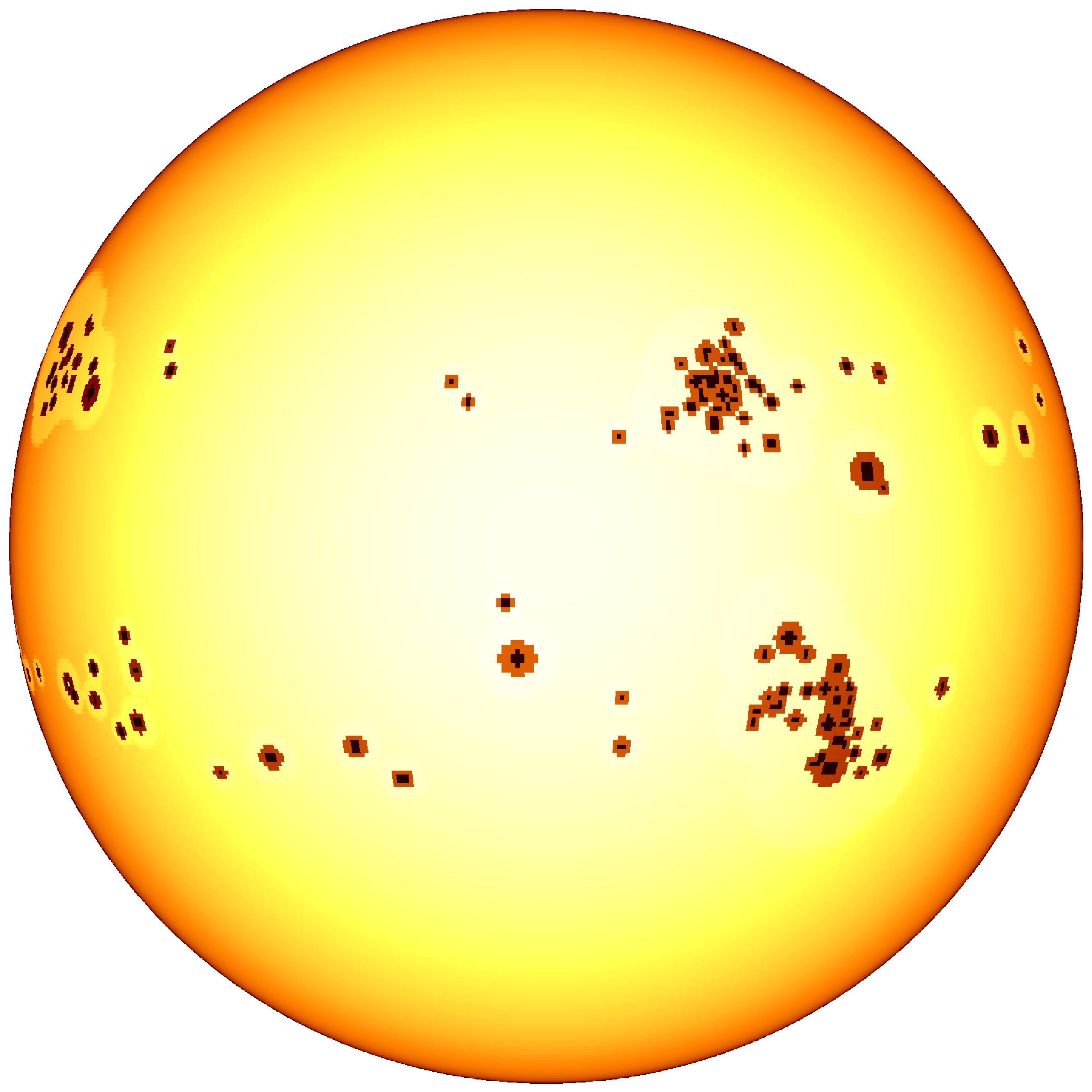} \\

   \includegraphics[trim=0 -5mm 0 0mm, width=0.59\columnwidth]{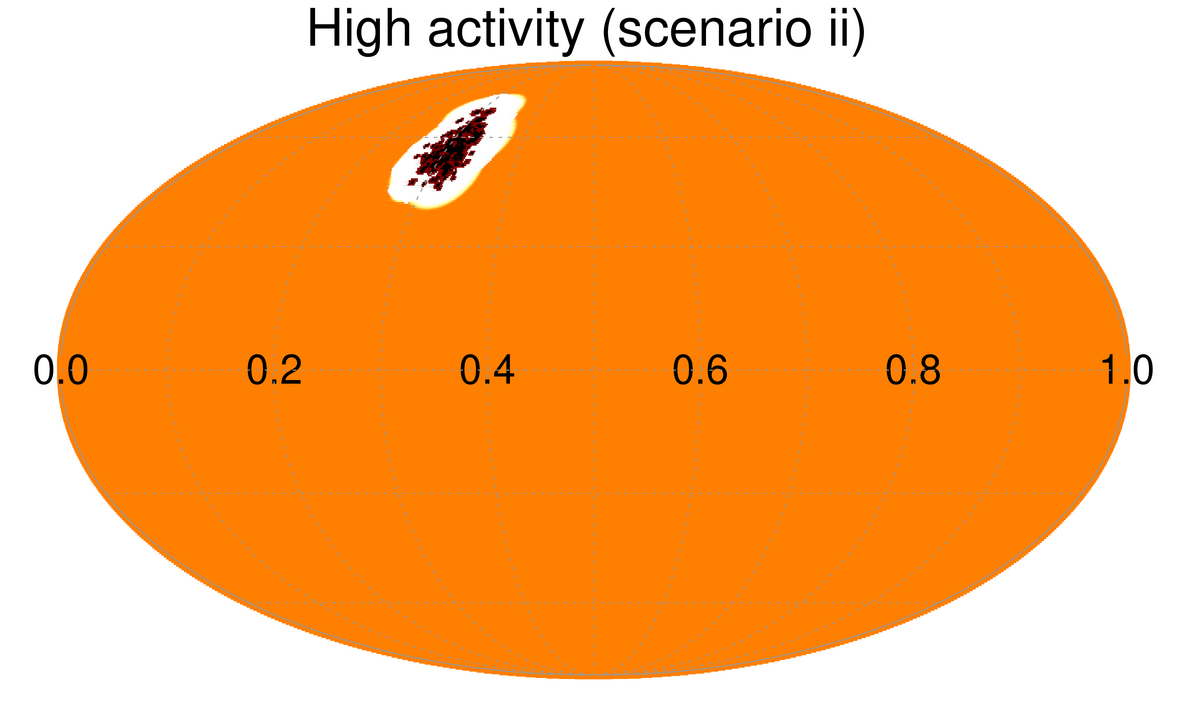} &
   \includegraphics[width=0.34\columnwidth]{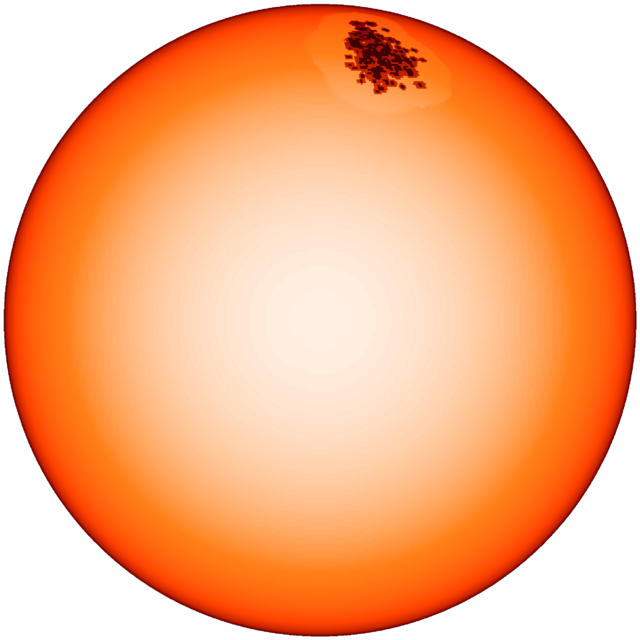} & & \\

   \end{tabular}
   \end{center}
\caption{Top: Starspot distribution Mollweide projection maps (showing rotation phase and latitude) and corresponding snapshot images at arbitrary phase of the M dwarf model (columns 1 and 2, orange) and G dwarf model (column 3 and 4, yellow). For the G dwarf, three spot models are shown: solar min, solar max and high activity models with respective cool spot coverage of 0.03\%, 0.3\% and 0.14\%. 
For the M dwarf high activity models, scenario i uses the same spot coverage as the G dwarf models, but with spots distributed at all latitudes. In scenario ii, for the high activity case only (0.14\% spot fraction), all spots are located within a single group at high latitude. Umbral (darkest), penumbral, photospheric (yellow/orange) and facular (white) regions are indicated.}
   \label{fig:spotmodels}
\end{figure*}

\section{Central Line Moments for scaled multi-spot distributions}
\label{section:spotmodels}

Based on solar observations, we don't expect a single spot to be present at most observation epochs, except during periods of lower activity.
Although simple quasi-sinusoidal variability is possible over a single stellar rotation, CLM variability is expected to be more complex when more than one spot or spot group is present at different stellar longitudes and latitudes. To investigate this further, we generated spot distributions for an M dwarf and G dwarf model.

\subsection{Multi-spot model construction}
\label{section:GMmodels}

We simulated three models with different activity levels based on solar observations. Spot coverage fractions and other statistics are given in Table \ref{tab:spotmodels}. Spot sizes were drawn from the log-normal distributions described by \cite{solanki04}. We assumed the same spot size distribution models for both G dwarfs and M dwarfs, with spot sizes defined via their angular radii (i.e. the absolute spot sizes scale with stellar radius).


We modelled spots in groups, with the first spot group as the largest, dominant group. To achieve this, the number of spots in each successive group was divided by an \textit{ad hoc} factor of 1.5 until a minimum spot group size of 2 spots. For the solar minimum, solar maximum and high activity models, we assumed the first group contained respectively 3, 24 and 36 spots (see Table \ref{tab:spotmodels} for further statistics). For each given spot group, a centroidal longitude and latitude was randomly allocated. Spots were then added to the group at Gaussian distributed random offsets with 1-sigma distance of 2\degs{} (solar minimum and maximum) and 3\degs{}(high activity). Spot sizes were randomly selected without replacement from the appropriate spot size distribution. We did not attempt to model individual spots as spot pairs. 

To account for any overlapping spots, the total spot count for each model was scaled to ensure the total cool spot area coverages of 0.03\%, 0.3\% and 1.4\% for the solar minimum, solar maximum and high activity models, as defined in \citep{solanki04}. The chances of spot overlap are higher for the larger facular regions, which were again painted as annuli around each cool spot. All facular regions were also scaled to maintain the appropriate $A_\textrm{f}$/($A_\textrm{u}$+$A_\textrm{p}$) ratios {used} in the one spot simulations (see {\S \ref{section:spotconstruction}}). A minimum spot size equivalent to the $0.5$\degs{} pixel resolution of our models was assumed. 

\begin{figure*}
    \begin{center}
    \includegraphics[width=0.99\textwidth]{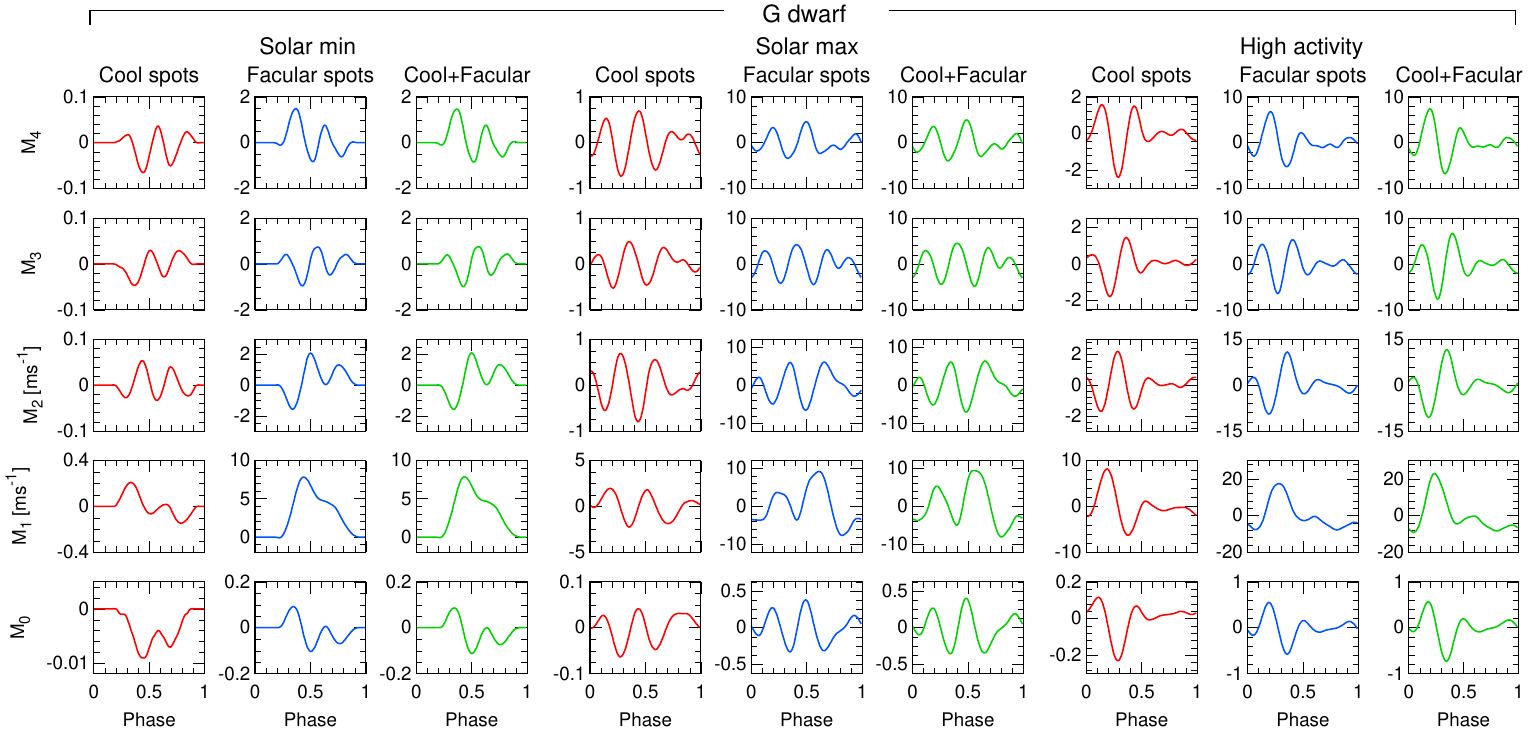}
    \vspace{5mm} \\
    \includegraphics[width=0.99\textwidth]{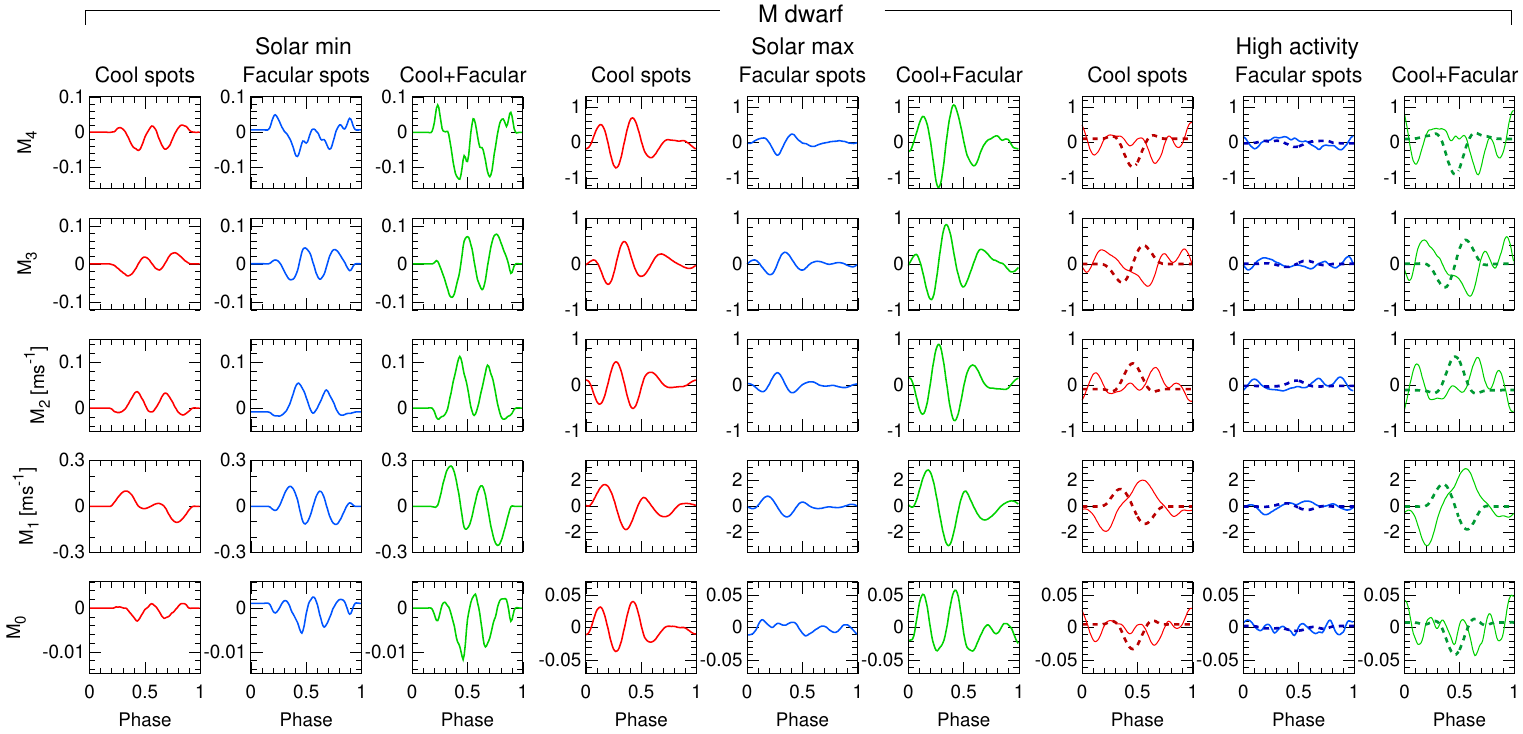}
    \end{center}
    \caption{
    Central Line Moments as a function of phase for spot solar min, solar max and high activity models for a star with $v$\,sin$i =2$~\kms. Bottom panels: M dwarf model CLMs for scenario i with uniformly distributed spot groups (solid lines) and for scenario ii, for the 
    high activity model only with a single high latitude spot group (dashed lines).
    Top panels: G dwarf models with spot group centroids restricted to latitudes $5$\degs~$ < l < 15$\degs~(solar min) and $25$\degs$ < l < 40$\degs~(solar max and high activity) and the two largest spot groups located $180$\degs~of longitude apart.}
    \label{fig:moments_spotmodels}
\end{figure*}

\subsubsection{{M dwarf spot distributions}}
\label{section:Mmodels}

The simulated M dwarf spot maps with cool spots and facular regions are shown on the left of Fig. \ref{fig:spotmodels} with snapshots of the corresponding stellar models at random phases (by definition, longitude on the models runs in the opposite sense to phase on the maps). There is evidence that supports both distributed dynamo activity in M dwarfs and also large-scale dynamo activity. While Zeeman Doppler imaging studies of early M dwarfs tend to favour toroidal and non-axisymmetric large scale field geometry, strong axisymmetric poloidal structure is more common in stars later than M3.5V, when stars are thought to become fully convective \citep{donati08mdwarfs, morin08mdwarfs, morin10mdwarfs}. Brightness Doppler maps of rapidly rotating M0\,-\,M9 M dwarfs reveal cool spot patterns that show spots both distributed on the stellar surface, but also in some cases, concentrated at higher or circumpolar latitudes (\citealt{barnes01mdwarfs, barnes04hkaqr, barnes15mdwarfs, barnes17mdwarfs, barnes17jitter}), in broad agreement with the magnetic field reconstructions. 
Discussion of dynamo models, numerical simulations and their various predictions can be found in \cite{morin10mdwarfs}. 

Although we simulated M2 dwarf spot contrasts, we investigated 
two 
M dwarf spot distribution patterns, which we subsequently refer to as scenario i and scenario ii, as follows:

\begin{enumerate}[wide, labelwidth=!,itemindent=!,labelindent=0em, leftmargin=0.5cm, label=\roman*), itemsep=0.1cm, parsep=0pt]
    \item random longitude and latitude spot group placement with group sizes defined as for the G dwarf (Fig. \ref{fig:spotmodels}) 
    \item  for the high activity model only, a single large group at latitude $60$\degs{}.
    The standard deviation of the spots within the group is $13$\degs   
\end{enumerate}

\subsubsection{{G dwarf spot distributions}}
\label{section:Gmodels}

The G dwarf model approximates Sp\"{o}rer's law for the Sun by restricting spot group centroids to latitudes $25^{\circ} < l < 40^{\circ}$ (solar min) and $5^{\circ} < l < 15^{\circ}$ (solar max and high activity. The simulated G dwarf spot maps are shown on the right of Fig. \ref{fig:spotmodels}. 



\begin{table}
	\begin{tabular}{lcccc}
		\hline
		Model           & Spot coverage & $N_\textrm{spots}$    & $r_\textrm{u,min}$    & $r_\textrm{u,max}$ \\
		& [per cent]    &                       & [degs]                & [degs] \\
		\hline
		Solar min      & 0.03           & 5                     & 0.31                  & 0.41 \\
		Solar max       & 0.3           & 65                    & 0.31                  & 0.61 \\
		High activity   & 1.4           & 206\,(305)            & 0.31                  & 0.87\,(0.92) \\
		\hline
	\end{tabular}
	\caption{Spot model statistics. The factional spot coverage of umbral regions, number of spots and maximum and minimum umbral radii are given. The high activity model for the single high latitude spot group (scenario ii) are given in parentheses.}
	\label{tab:spotmodels}
\end{table}

\subsection{Multi-spot model CLM signatures}
\label{section:GMsignatures}

Fig. \ref{fig:moments_spotmodels} illustrates the CLM signatures for the three activity models {of solar minimum, solar maximum and high activity with the three spot type models with cool, facular and cool\,+\,facular spot distributions for $v$\,sin\,$i = 2$~\kms. For each activity model, the facular spot models use the same annular regions as the cool\,+\,facular spot models, but with the cool spot umbral and penumbral regions replaced with photosphere. This more readily enables a comparison between the cool and facular signatures and their relative contributions to the cool\,+\,facular signatures. Overlap of randomly located spots can lead to modification of the facular to cool spot ratio. The mean ratio of $A_\textrm{f}/(A_\textrm{u}+A_\textrm{p})$ per spot was modified in each multi-spot model to ensure the global $A_\textrm{f}/(A_\textrm{u}+A_\textrm{p})$ values described in \S \ref{section:spotconstruction}.}
 
The morphology of the CLMs vs phase {in the multi-spot models} is more complex than for a single spot. 
For multi-spot models, since spots are more distributed, we expect continuous variation of CLMs to be common throughout a single rotation, unlike single spot models, where the CLMs may be constant for half the rotation phase. It is thus useful to describe a CLM's variability via its harmonic complexity, $\Psi$, {which simply describes the number of complete cycles during a single rotation spanning phases $0 < \phi < 1$. This relates directly to the previous simulations and discussion for single spot models: a harmonic complexity of $\Psi$ might reasonably be expected to lead to recovery of stellar rotation at $P_\textrm{rot}/\psi$. }

{Fig. \ref{fig:moments_spotmodels} shows that} the degree of harmonic complexity in the CLMs is often higher than for a single spot, as expected for more complex spot patterns. Nevertheless, despite the random longitude placement in both models, low-order harmonic complexities still prevail in the CLM signatures. {This is likely due to most} of the spots appearing in just a few spot groups.

\subsubsection{M dwarf {multi-spot} CLM signatures}
\label{section:Mbehaviour}

The M dwarf CLMs for spots distributed at all latitudes (scenario i) are shown in the lower panels of Figure \ref{fig:moments_spotmodels}. The solar minimum model, comprising 2 small spot groups, produces CLM signatures with relatively low $\Psi$, {but the semi-amplitudes are very small and do not exceed $\pm 0.3$\,\ms. It is also worth noting that because the two spot groups in the solar minimum model are separated by $\sim 90$\degs, no spots are visible for a significant portion of a rotation, resulting in little or no modulation of the CLMs. Since the cool and facular regions appear dark (at least for most facular limb angles), the CLM morphologies are similar. The much lower contrast of faculae on the solar minimum model is balanced by the high relative area ratio with cool spots, $A_\textrm{f}/(A_\textrm{u}+A_\textrm{p}) = 26.9$. These factors yield cool and facular signatures of similar amplitude that reinforce each other in the cool\,+\,facular spot model. The facular contribution introduces greater harmonic complexity at $M_4$, but at an extremely low amplitude.} 

There is {an order of magnitude} increase in CLM amplitude in the solar maximum models compared with the solar minimum models. {The harmonic complexity doesn't appear to change substantially, but shows greater amplitude in the signatures for phases in the approximate range $0 < \phi \ < 0.5$, where the dominant spot groups are located. As more spots are added in the high activity model, $\Psi$ appears to increase in the higher order moments, as a consequence of the more even distribution of spot groups (see Fig. \ref{fig:moments_spotmodels}). For $M_1$, on the other hand, $\Psi$ appears to show a decrease. As the relative ratio of  $A_\textrm{f}/(A_\textrm{u}+A_\textrm{p})$ decreases with increasing activity, the facular contribution can be seen to decrease relative to the cool spot contribution. The cool\,+\,facular, high activity, model CLMs are dominated by the cool spots.
 with the consequence that they have a smaller or smoothed-out contribution relative to the larger spot groups. In general, the amplitudes of the CLMs is relatively low for scenario i, suggesting that the signatures may be difficult to discern with realistic data and SNRs.


Scenario ii, simulated for the high activity model only with one large spot cluster, resembles a single spot at high latitude. In Fig. \ref{fig:moments_spotmodels}, the dashed curves in columns $7-9$ show the CLM amplitudes for scenario ii with amplitudes {divided by a factor of $5$ to aid visual comparison}. Clearly, $\Psi$ is much lower and exhibits behaviour that is closer to the single spot simulations with $l = 60$\degs{}.}

\begin{figure*}
\begin{tabular}{lll}
    \includegraphics[trim=2mm 0mm 2mm 0mm, height=9.05cm]{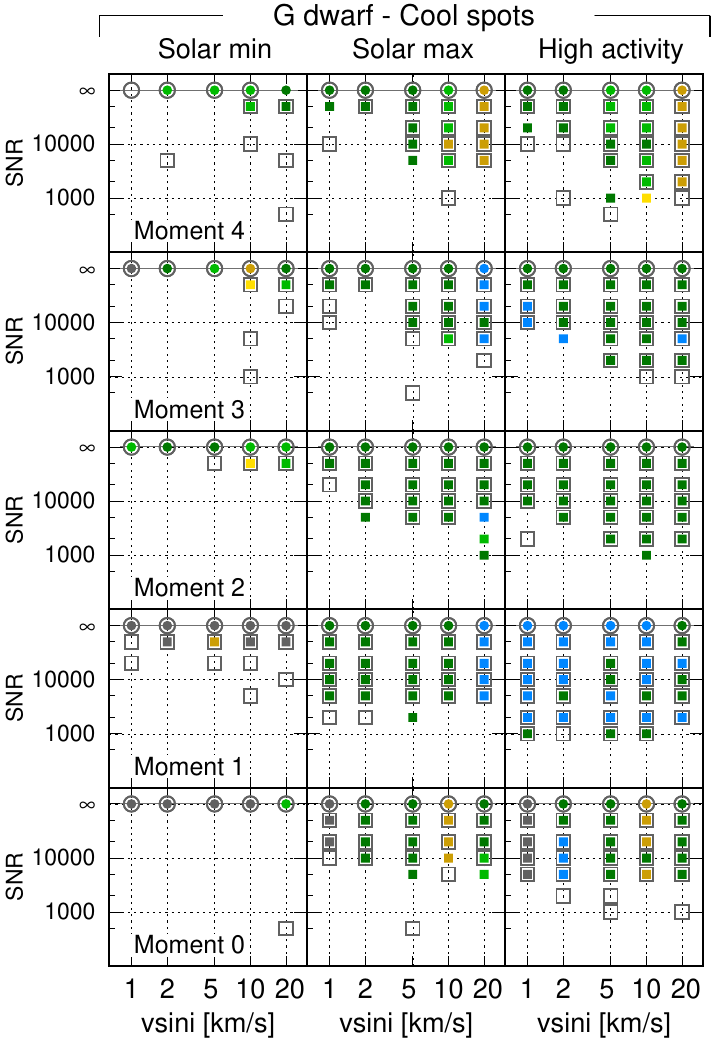} \hspace{0mm} &
\includegraphics[trim=2mm 0mm 2mm 0mm, height=9.05cm]{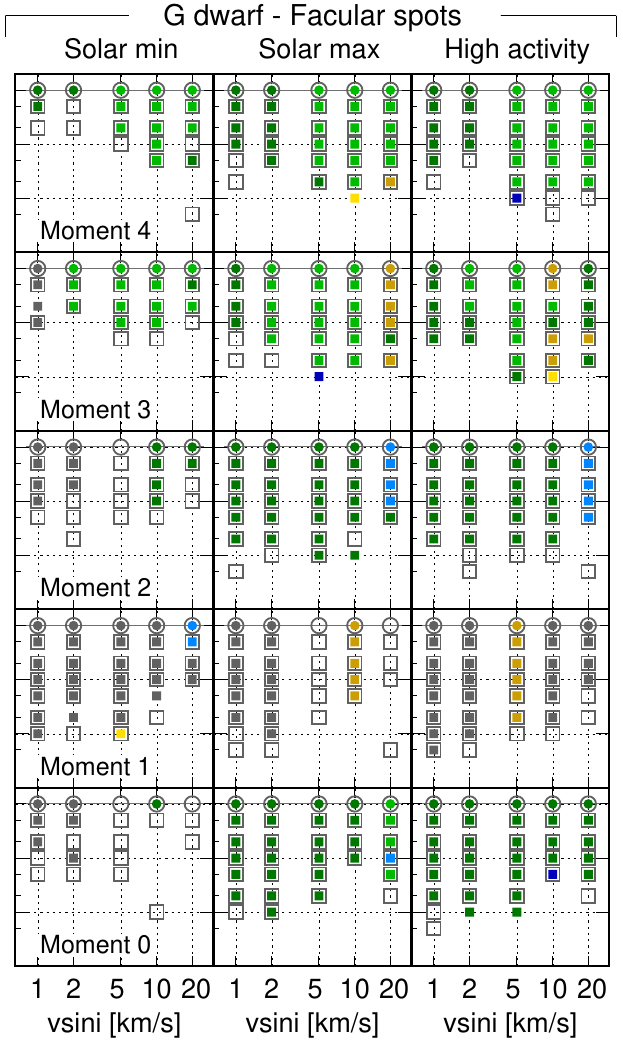} \hspace{0mm} &
\includegraphics[trim=2mm 0mm 2mm 0mm, height=9.05cm]{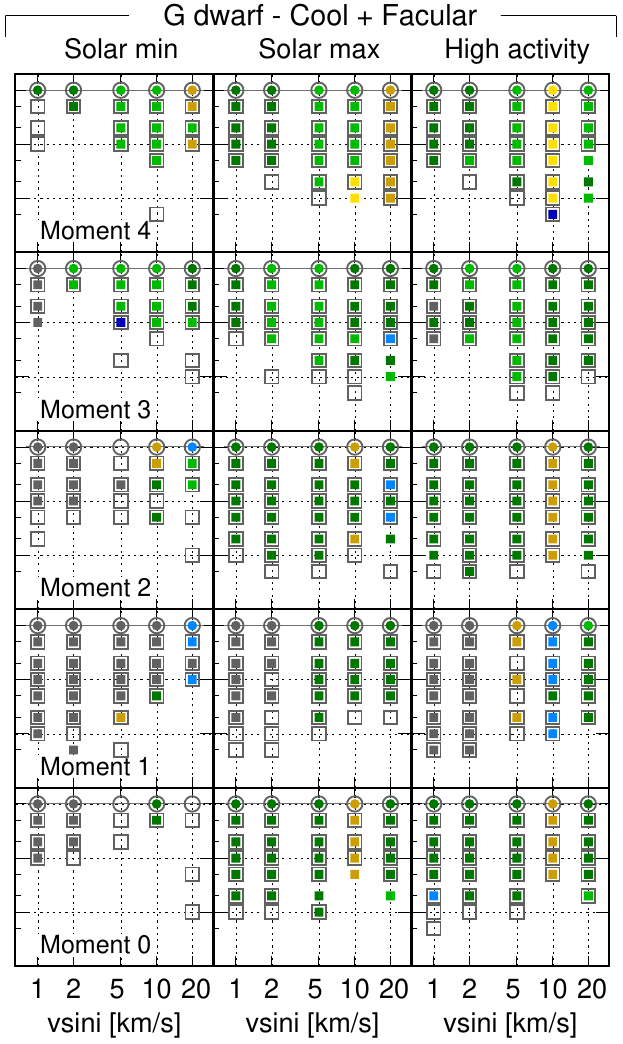}
    \end{tabular}
	\vspace{8mm} \\

\begin{tabular}{lll}
    \includegraphics[trim=2mm 0mm 2mm 0mm, height=9.05cm]{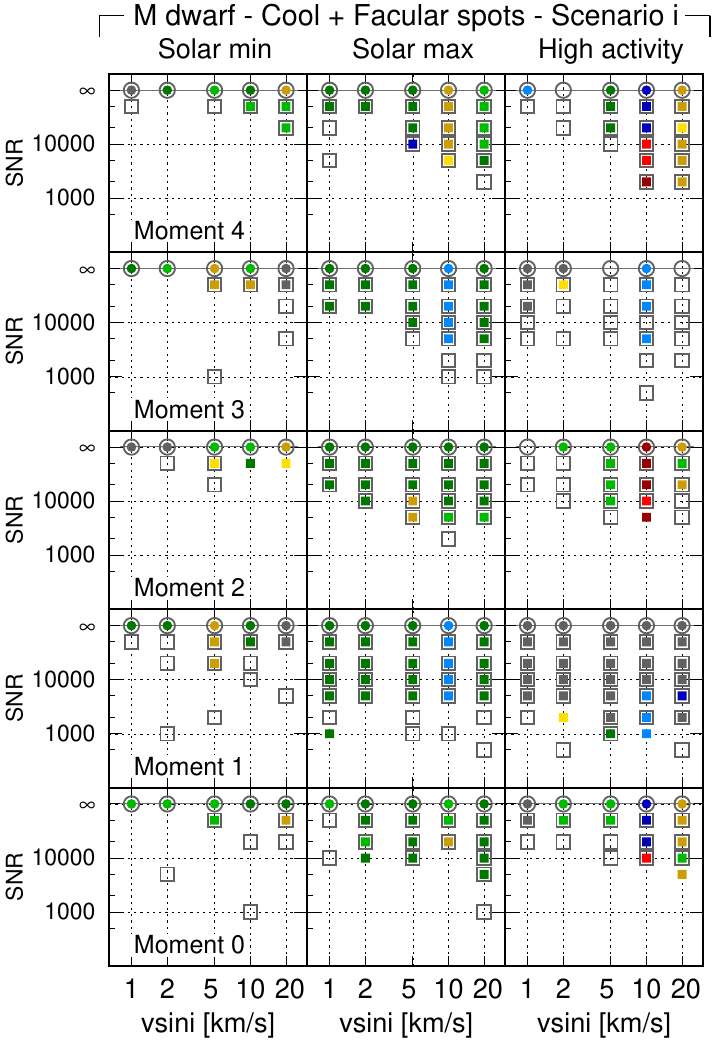} \hspace{1mm} &
	\hspace{10mm} 
	\includegraphics[trim=12mm 0mm 3mm 0mm, height=9.05cm]{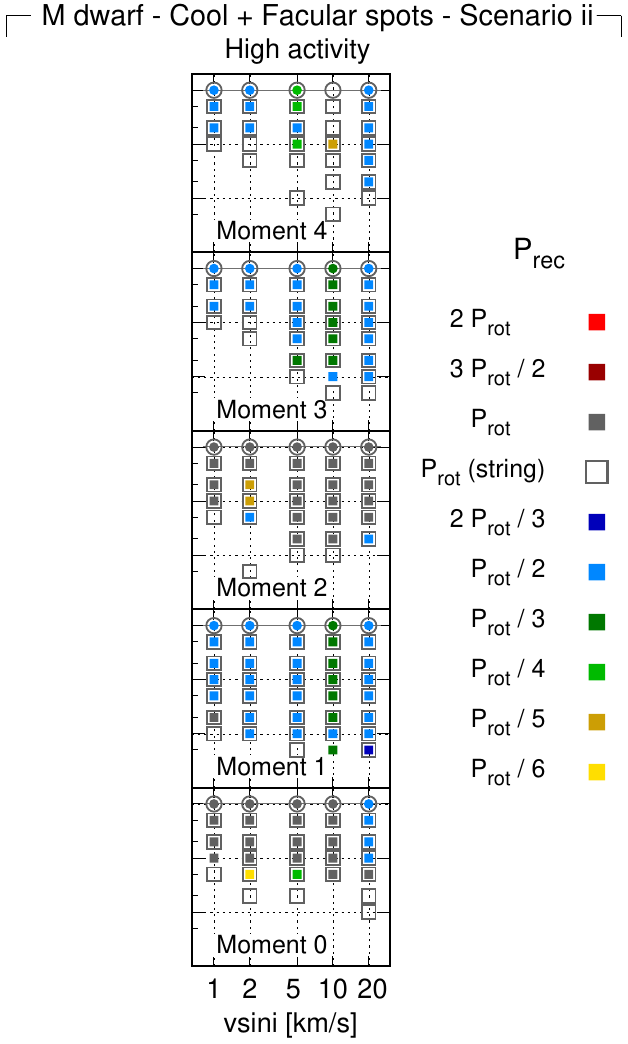} & \\
\end{tabular}
    \caption{Period recovery matrices for the simulated spot activity models using Central Line Moments. The matrices indicate the combinations of SNR and \vsini{} for which $P_\textrm{rec}$, the simulated rotation period or a harmonic of the rotation is recovered. For GLS periodogram searches, finite SNR recoveries are indicated with filled squares, and SNR = $\infty$ (noise-free) recoveries are shown with filled circles. The key shows the colours that represent the recovered period harmonics. Symbols with grey borders and open symbols plotted in grey, indicate SNR and \vsini{} combinations where the SL period analysis returns $P_\textrm{rec} = P_\textrm{rot}$. A stellar inclination of $i=90$\degs{} was simulated for all models. Bottom panels: Period recovery matrices for an M dwarf. The panels show results for scenario i with randomly placed cool\,+\,facular spot groups at solar minimum, solar maximum and high activity levels and scenario ii, with cool\,+\,facular spots for the high activity high latitude spot group. Top panels: Period recovery matrices for a G dwarf. The simulations are grouped into models with cool spots, facular spots and cool\,+\,facular spots, each showing the three simulated activity models: solar minimum, solar maximum and high activity.}
    \label{fig:sensitivity_spotmodels_GMdwarf}
\end{figure*}

\subsubsection{G dwarf {multi-spot} CLM signatures}
\label{section:Gbehaviour}

{The cool spot CLM amplitudes increase by an order of magnitude between the solar minimum and maximum models. This trend continues to a lesser degree between the solar maximum and high activity models, but is greater than was seen for the M dwarf. This is in part due to the spots being more clustered in lower and narrower latitude bands in the G dwarf model.}

The facular CLM amplitudes are somewhat larger when {compared with the corresponding cool spot CLMs.  The relative decrease in $A_\textrm{f}/(A_\textrm{u}+A_\textrm{p})$ with increasing activity means that the facular $M_1$ doesn't change amplitude between solar minimum and maximum models despite an increased total area. Further, the faculae are found in more groups and at relatively low contrast, so that their signatures are resolved at more longitudes or phases. In addition, since the facular signatures are greatest near the limb, when features are also foreshortened, the change in $M_1$ amplitude is fairly small as facular area increases. In these noise-free cases, the higher order moment amplitudes, $M_2$ - $M_4$, show comparatively larger changes. 
	
The morphology of the cool\,+\,facular model signature is dominated by the facular contribution, contrasting with the M dwarf models, where cool and facular CLM amplitudes were more closely matched. Even the high activity facular\,+\,cool G dwarf model with the lowest $A_\textrm{f}/(A_\textrm{u}+A_\textrm{p}) = 5.29$ more closely resembles the facular-only model for all the CLMs. The primary cause of facular dominance stems form a combination of the larger area, albeit at lower contrast than a cool spot, and the CB of $340$\,\ms, as was illustrated for the single spot models in Fig. \ref{fig:moments_single}.}
We note that in the high activity model, the CLM amplitude is greatest in the phase $0.0 < \phi < 0.5$ interval owing to the two spot groups at closely separated longitudes. The overall effect is reminiscent of the single spot simulations. Generally, the harmonic complexity, $\Psi$ does not appear to change considerably across the three activity models. {Arguably, it appears to decrease in the high activity model owing to the two dominant spot groups.}

Since $M_1$ is effectively equivalent to measuring spot induced RV (i.e. in our simulations with no dynamic Keplerian signals present), our findings are in good agreement with those of \cite{meunier10plage} for solar minimum and maximum models (e.g. see their Fig. 7) when comparing the cool spot and facular components with convective blueshift included. Because $M_1$ is effectively centre-weighted, the total amplitude is lower than for RV, for which it can be considered a proxy.

\subsubsection{CLM signatures summary}
\label{section:CLMsummary}

{Our simulations do not lead us to expect significant differences between cool spot and cool\,+\,facular model CLMs for an M dwarf because our models adopt the findings of \cite{beeck15} and \cite{johnson21}, who do not predict {strong facular contrasts} on M dwarfs, and also \cite{liebing21}, who {empirically} find that convective blueshift is absent. Because the low contrast faculae may appear dark on an M dwarf (except at the smallest $\mu$), the primary effect of adding them to the model is to fractionally increase the amplitude of cool spot signatures. There are however some further subtle changes to morphology of the CLMs when faculae are modelled in conjunction with cool spots. The more pronounced facular and convective contributions for earlier spectral types are the key ingredients that differentiate G dwarf CLM signatures from those of the M dwarf models.}

CLM signatures are lower in the M dwarf scenario i models compared with G dwarfs because (a) spots have lower contrast (b) spots may appear at higher latitudes, which means they are more foreshortened and do not traverse the full velocity extent of photospheric absorption lines and (c) {facular contrasts are low and} there is no convective blueshift effect. However, the relative amplitude of the CLMs {increases} considerably for the high activity M dwarf spot distribution scenario ii {with all spots in a single group, where the CLM amplitudes are similar to those of the high activity G dwarf model.}

\subsection{Recovered periodicities from the spot distribution models}
\protect\label{Section:models_periods}

We investigated period recovery for various combinations of SNR and \vsini\ using our spot distribution models. {Although evolution of activity features, including spot growth, decay and differential rotation will modify CLM signatures, for the purposes of this study we assumed fixed spot patterns with no spot evolution. Based on our findings in \S \ref{section:GMsignatures}, we consider M dwarf models with cool\,+\,facular spots only in this section and focus instead on differences between the M dwarf scenario i and scenario ii models.}

Determination of stellar rotation directly from CLMs is dependent on several factors apart from activity level. Activity is likely to be greater for more rapidly rotating stars with larger \vsini, making $P_\textrm{rot}$ easier to determine. For slow rotators, rotational modulation becomes difficult to resolve. The balance between activity signature amplitude, \vsini{} and data SNR is thus an important consideration, especially when we also consider that line equivalent width is conserved as \vsini\ increases (i.e. lines are shallower relative to the continuum).

In general, empirical data sets are obtained with unique sampling that is governed by various observing constraints, impacting on the ability to recover periodicities. For simplicity, {we simulated 60 observations} and assumed alternate-day sampling for 120 days with $3$ hr uncertainty on the time of each observation to reduce aliasing. As in \S \ref{section:single_periods}, the three spot activity models, solar minimum, solar maximum and high activity, were simulated with a range of \vsini{} and rotation periods appropriate for our G and M dwarf models. We assumed stellar axial rotation, $i = 90^{\circ}$ for these simulations. {Trial values for SNR = }500, 1000, 2000, 5000, 10000, 20000 and 50000, representative of {typical CCF profiles obtained in precision RV studies spanning several thousand absorption lines,}  were simulated along with the noise-free (SNR~$= \infty$) case. {We performed GLS and SL period searches to identify and compare the recovered period, $P_{\rm rec}$, with the simulated rotation period, $P_\textrm{rot}$ and its harmonics as outlined in \S\S \ref{section:single_periods}\,-\,\ref{section:single_period_sl}. Fig. \ref{fig:sensitivity_spotmodels_GMdwarf} shows the recovered GLS and SL periodicities for combinations of SNR and $v\,\textrm{sin}\,i$, colour coded according to the period harmonic of the dominant GLS peak.}

\subsubsection{M dwarf {multi-spot} CLM periodicities}
\protect\label{section:mdwarf_multispot_periodicities}

{The lower panels of Fig. \ref{fig:sensitivity_spotmodels_GMdwarf} show the periodicities recovered from CLMs for the simulated  cool\,+\,facular M dwarf spot model scenarios described in \S \ref{section:Mbehaviour}. Scenario i, with spots randomly located in longitude and latitude shows changes in GLS $P_\textrm{rec}$, as the spot coverage increases. For both SNR~$=\infty$ (filled circles) and finite SNR cases (filled squares), recovered GLS $P_\textrm{rec}$ are found at a range of $P_\textrm{rot}$ harmonics, though there is a preponderance of $P_\textrm{rot}/3$ in the solar maximum model and $P_\textrm{rot}$ for $M_1$ in the high activity model. It is also clear that recovery of periodicity in the solar minimum model, with only 5 small spots in 2 spot groups (Table \ref{tab:spotmodels}), will not be possible for feasible SNRs. For the high activity model, with many scattered spot groups, recovery sensitivities are similar to the solar maximum models, but the GLS $P_\textrm{rec}$ is more unpredictable in terms of which harmonic of $P_\textrm{rot}$ is recovered. Higher harmonics, such as $P_\textrm{rot}/5$ and $P_\textrm{rot}/6$ are recovered for \vsini{} =~$10$ and $20$~\kms, but the presence at low activity or detection borderlines is likely less secure.
		
Fig. \ref{fig:sensitivity_spotmodels_GMdwarf} also shows the recovered string length periodicity, SL $P_\textrm{rec}$, for instances where the true rotation period, $P_\textrm{rec}$ = $P_\textrm{rot}$, is recovered. As in \S \ref{section:single_period_sl}, we show the SL $P_\textrm{rec}$ as an open symbol where there is no GLS $P_\textrm{rec}$ at 0.1\% FAP and as a grey open symbol enclosing the filled GLS $P_\textrm{rec}$ symbols. For almost all cases, where we recover $P_\textrm{rot}$ or a harmonic of  $P_\textrm{rot}$ with GLS, we recover precisely the simulated $P_\textrm{rot}$ with SL. There are isolated borderline cases where the SL method was not able to recover a period or obtain $P_\textrm{rec}$ = $P_\textrm{rot}$. In some cases, with the $2.5-\sigma$ requirement for identifying SL minima (\S \ref{section:single_period_sl}), we are in fact more sensitive than the GLS method.
}



For the M dwarf {with scenario i spot distribution patterns}, it is clear that very high SNRs $\geq 5000$ - $10000$ are required to recover the stellar rotation signature. {But even then, this may only prove reliable when spots are clustered in groups with solar maximum activity numbers and not too distributed as in the high activity model. While $P_\textrm{rot}$ or a harmonic may be recovered with $M_1$ and a GLS period search in the scenario i high activity model, higher order moments that enable stellar activity and dynamically induced Keplerian signals to be distinguished do not confidently enable period recovery. The SL period method appears to be slightly more sensitive here and has the benefit of unambiguously recovering the true $P_\textrm{rot}$.
	
In scenario ii, CLMs from $M_1$ to $M_3$ allow rotation signatures to be recovered at lower SNRs of $1000 - 2000$. {But here, period recovery for the higher order moments requires \vsini $\geq 5$\,\kms. Although scenario ii resembles the single spot simulation with a spot at high latitude, we have used $i=90$\degs{} for the multi-spot simulation. Consequently, for the GLS $P_\textrm{rec}$, $P_\textrm{rot}/2$ is common in $M_1$ and $M_3$ since the high latitude spot group is not always visible. The string length minimisation method again effectively  recovers the true rotation period, $P_\textrm{rot}$, in contrast with GLS $P_\textrm{rec}$.}

}

\subsubsection{G dwarf {multi-spot} CLM periodicities}
\protect\label{section:gdwarf_multispot_periodicities}

{For the G dwarf model, GLS $P_\textrm{rec}$ is predominantly  $P_\textrm{rot}$ and the lower harmonics from $P_\textrm{rot}/2$ to $P_\textrm{rot}/4$. Again, $M_3$ and $M_4$ are most likely to return GLS $P_\textrm{rec}$ with higher harmonics of $P_\textrm{rot}$, especially for higher \vsini{} values. Inspection of the CLMs shows that considerable harmonic complexity may indeed be present within a single rotation, so GLS $P_\textrm{rec}$ = $P_\textrm{rot}/5$ and $P_\textrm{rot}/6$ should not be surprising. Given the additional higher order structure that we find for higher spectral resolutions in \S \ref{section:resolution} and Fig. \ref{fig:moments_single_resolution}, it isn't unreasonable to expect that higher order harmonics of $P_\textrm{rot}$ might be recovered at lower \vsini{} values when observing with instruments such as ESPRESSO.
	
The presence of faculae significantly improves the sensitivity at all activity levels, but particularly, makes recovering periodicity in the solar minimum model more feasible. Despite this, it is clear that for the solar minimum model, high CCF SNR $\geq 10,000$ is required to recover higher order moments $\geq$~$M_2$ that are typically useful as activity diagnostics. For solar maximum and high activity models, periodicities are clearly recovered at lower SNRs than for the solar minim models. Here, SNR $\sim 2000 - 5000$ is sufficient to recover periodicities in $M_2$ and $M_3$.
The increase in ability to recover a GLS $P_\textrm{rec}$ between solar minimum and solar maximum activities is more pronounced when compared with the M dwarf. The further slight increase in sensitivity seen in the G dwarf high activity model again contrasts with the M dwarf models, where the scattered nature of spots did not favour recovery of higher order moments.

As with the M dwarf, the SL periodograms return $P_\textrm{rec}$ = $P_\textrm{rot}$ when the GLS returns harmonics of $P_\textrm{rot}$. There are again some borderline sensitivity instances where the SL periodogram doesn't identify any peak. However, there are also instances where an SL $P_\textrm{rec}$ is found, but no GLS $P_\textrm{rec}$ is returned.
}

 \begin{figure*}
	\begin{center}
		\includegraphics[trim=0mm 0mm 0mm 0mm, width=0.49\textwidth]{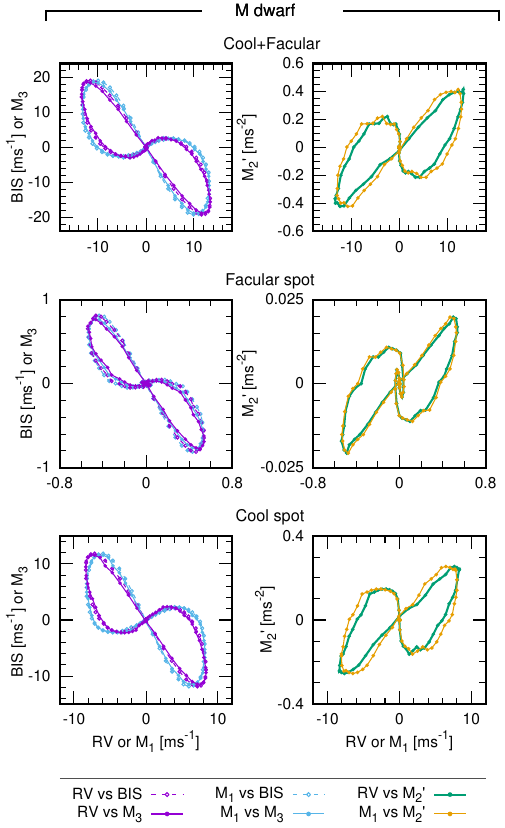}
		\hspace{2mm}
		\includegraphics[trim=0mm -11.25mm 0mm 0mm, width=0.49\textwidth]{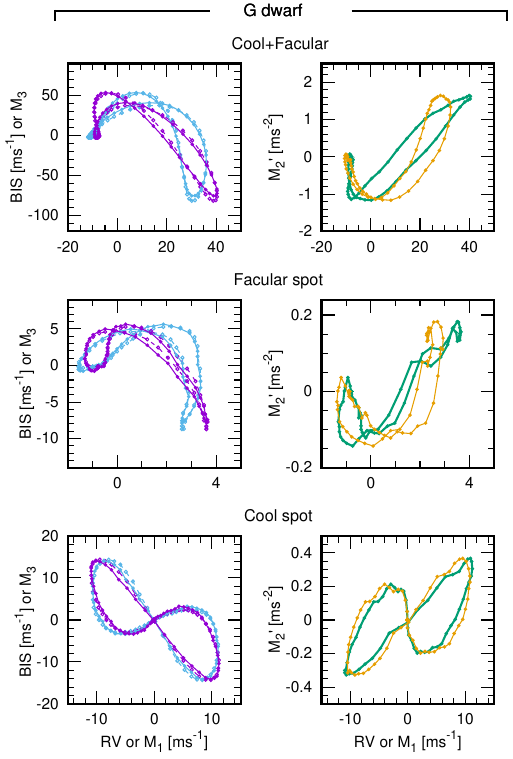}
	\end{center}
	\caption{The correlated behaviour of CLMs for M dwarf (left) and G dwarf (right) {models} with a cool spot (bottom), facular spot (middle) and cool\,+\,facular spot (top). The behaviour of BIS, $M_3$, and $M_2$ time derivative, $M_2^\prime$ (see key) as a function of RV or $M_1$ is shown for \vsini\ = 5 \kms{} for a spot located at latitude $l=0$\degs. The spot radii ($r_\textrm{u} = 2$\degs{}) are defined in \S \ref{section:spotconstruction}.}
	\label{fig:moments_correlation_signatures}
\end{figure*}

\subsubsection{CLM signatures and periodicities for a G dwarf with active longitudes}
\protect\label{section:gdwarf_multispot_periodicities_180degs}

We also investigated the CLM signatures and periodicities for a G dwarf for spots that appear predominantly at active longitudes. For the Sun, \cite{berdyugina03activelongs} showed that spots tend to appear in active regions separated by 180\degs{}. We simulated the two largest spot groups for each activity model to be separated by 180\degs{} in longitude in our simulations, rather than allocating their longitudes randomly. The CLM signature and period recovery matrices are shown in Figs \ref{fig:moments_spotmodels_activelongs} and \ref{fig:sensitivity_spotmodels_Gdwarf_activelongs}. The CLM signatures in \ref{fig:moments_spotmodels_activelongs} show a notable difference when compared with the spots placed randomly in longitude. Visually, there is an obvious preponderance of signatures that appear with harmonic complexity $\Psi = 2$. Higher order harmonic structure starts to assert itself for $M_3$ and $M_4$, but the amplitudes of the additional structure often appear smaller at the plotted \vsini~=~$2$\,\kms. These observations are born out in the period recovery matrices in Fig. \ref{fig:sensitivity_spotmodels_Gdwarf_activelongs}, {where there is a very clear dominance of GLS $P_\textrm{rec}$ = $P_\textrm{rot}/2$. The SL periodograms are again able to correctly recover the injected periods, so that SL $P_\textrm{rec}$ = $P_\textrm{rot}$.} The higher amplitude variability seen in the CLMs also increases the sensitivity slightly compared with the random spot group allocation. 

\protect\label{section:periods_models}

\begin{figure*}
    \begin{center}
    \includegraphics[width=0.9\textwidth]{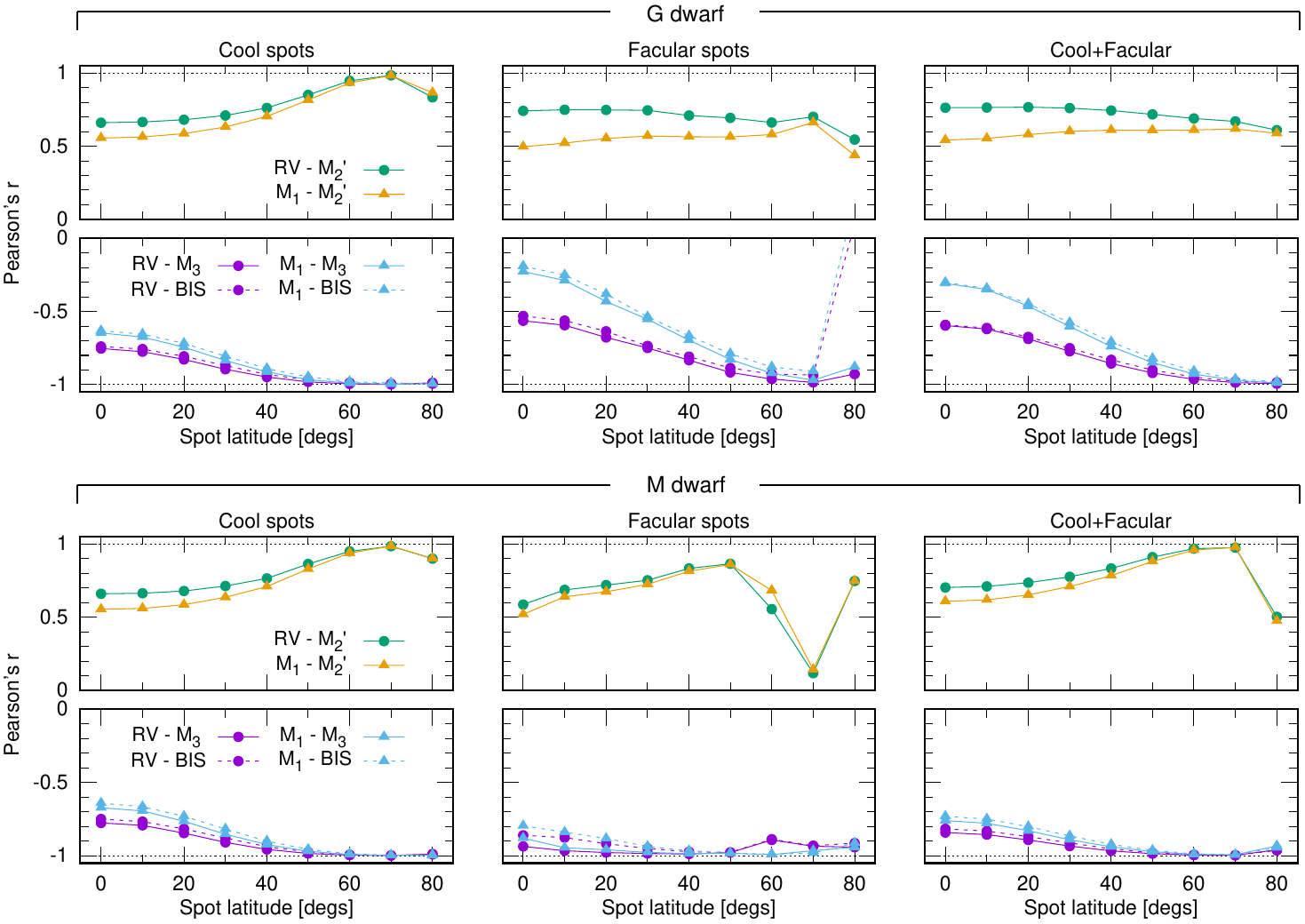}
    \caption{
    Illustrative Pearson's $r$ correlations between absorption line (or CCF) metrics for single spots with $r_\textrm{u} = 2$\degs and \vsini\ = 5\ \kms{} located at latitudes $0^\textrm{o}$ to $80^\textrm{o}$. The lower plots show Pearson's $r$ linear correlation for an M dwarf, while the upper panels show the same correlations for a G dwarf. The key indicates the line metric pairs that are considered in each case.}
    \label{fig:moments_correlations}
    \end{center}
\end{figure*}

\section{Central Line Moments as activity indicators}
\label{section:CLMactivity}

When stellar activity signatures are present in absorption lines, care is needed to correctly and efficiently  disentangle Keplerian signals from stellar astrophysical periodicities. It is common practice to use the {FWHM or BIS derived from the CCF} since they are able to describe line distortion periodicities due to stellar activity (e.g. \citealt{barragan22pyaneti2}, \citealt{zicher22}, \citealt{barnes23dmpp4}). In the following sections, we look at the most significant {CLM} correlation signatures of single spot models before investigating the signatures arising from our M dwarf and G dwarf spot models.

\subsection{Single spot CLM correlations}
\label{section:correlations}

The amplitudes and slopes of RV vs BIS variations for stars with various spectral types were first investigated by \cite{saar97} and {in detail by} \cite{desort07}. For a cool spot, the RV vs BIS shape resembles an inclined \textit{lemniscate} (i.e. figure-of-eight or $\infty$ shape) with negative slope correlation (\citealt{desort07}; \citealt{boisse11}). This behaviour is also seen when the RV correlation with chromatic index (CRX) \citep{zechmeister18serval} is measured for active M stars \citep{baroch20crx,jeffers22evlac}. The degeneracy of the BIS vs RV relationship implied by the lemniscate arises as a result of foreshortening and limb-darkening effects due to the velocity-resolved cool spot changing its {instantaneous} limb-angle on the stellar disc as the star rotates. As a result, the absolute value of the peak of the RV and the BIS deviations, and equivalently, $M_1$ and $M_3$ deviations, occur at different limb angles (\S \ref{section:Mdwarfcool} and Fig. \ref{fig:moments_single}). {It has already been noted that for BIS vs RV, the correlation length becomes shorter and the lemniscate collapses to a (near-)linear correlation as spot latitude increases \citep{boisse11}.}
The degree of linear correlation thus increases for high latitude cool spots, but in the presence of noise, the decreased amplitude of the negative correlation will likely hamper the measurement of a linear correlation. Rapidly rotating stars that only possess higher latitude cool spots might thus be expected to enable more efficient activity modelling via a linear correlation compared with lower activity stars{, provided that} good SNR data can be obtained.

Fig. \ref{fig:moments_correlation_signatures} illustrates the correlations of line moments for cases with a single $r_\textrm{u} = 2$\degs{}, $r_\textrm{p} = 4.47$\degs{} spot. For {purely} facular spots, {we used} $r_\textrm{f} = 4.47$\degs{} and for the combined cool\,+\,facular spot cases, we assumed the high activity area ratio,  $A_\textrm{f}/(A_\textrm{u} + A_\textrm{p}) = 5.29$ (see \S \ref{section:spotconstruction}). The spot was located on the equator at $l=0$\degs{} of an M~dwarf and G~dwarf rotating with \vsini\ $= 5$\ \kms. CLMs were derived from absorption lines with $3$\degs{} rotation resolution.

The lower left panel for the cool spot cases illustrates the behaviour of RV and $M_1$ with BIS and $M_3$. The lemniscate with negative slope is clear and is comparable for both M dwarf and G dwarf models.  In all sub-panels, the extent of $M_1$ is scaled {by a constant} to {optimally} match RV. {Similarly,} $M_3$ is scaled up to match BIS to enable a visual comparison of the correlations. Because RV and BIS have respectively higher corresponding amplitudes than $M_1$ and $M_3$, we might expect a higher degree of correlation, suggesting that BIS is a more sensitive indicator of activity. However, as Fig. \ref{fig:moments_correlation_signatures} {columns~1 and 3 reveal}, RV vs $M_3$ {marginally} shows the highest degree of (negative) linear correlation. {The degree of correlation of RV vs $M_3$ compared with RV vs BIS is around $\sim 3$\% greater in both the M dwarf and G dwarf cool and cool\,+\,facular models for spots at low-mid latitudes. Corresponding improvements of respectively 6\% and 10\% are found for facular spots for the M dwarf and G dwarf models.}



{
Although inspection of Fig. \ref{fig:moments_single} leads us to expect negative linear correlations of $M_1$ and $M_3$ in many cases, it can also be seen that a correlation between $M_1$ and $M_2$ is less obvious. The FF$^\prime$ method for predicting RVs from photometric data \citep{aigrain12ff} involves taking the (time) derivative of the photometric signal. In terms of morphology, this has the equivalent effect of transforming the {\em even} CLM signatures ($M_0$, $M_2$ and $M_4$) for a cool single spot into a signature more closely resembling the {\em odd} ($M_1$ and $M_3$) CLM signatures. Fig.~\ref{fig:moments_correlation_signatures} columns~2 and 4} show the correlation of RV and $M_1$ with the time derivative of $M_2$, which we denote $M_2^\prime$. While lemniscate figures with negative correlation are seen for all M dwarf $M_3$ signatures as well as the cool single spot G dwarf signature, the corresponding $M_2^\prime$ signatures show lemniscates with positive correlation, though visual inspection suggests the degree of correlation is lower compared with $M_3$. {For models with both facular and cool\,+\,facular spots, the signatures for the G dwarf model are somewhat different in both morphology and amplitude where a skewed figure is seen. Facular spots have higher intensity nearer to the limb, when the velocity amplitude is greatest, leading to a more linear correlation between RV and $M_2^\prime$ for the G dwarf where the asymmetry of the correlation over the rotation cycle is a consequence of the CB effect. The lower spot contrasts and lack of CB effect in the M dwarf leads to low-amplitude facular signals.} The M dwarf facular spot signatures are an order of magnitude smaller than the cool spot signatures. For the G dwarf, the cool\,+\,facular signatures appear to be much higher than in the individual cool and facular spots cases because the area of the facular annulus (i.e. we used the high activity ratio $A_\textrm{f}/(A_\textrm{u} + A_\textrm{p}) = 5.29$) is considerably greater than for a single $r_\textrm{f} = 4.47$\degs{} facular spot. Obviously, for solar minimum and solar maximum with $A_\textrm{f}/(A_\textrm{u} + A_\textrm{p}) = 26.9$ and $10.3$, the facular contribution will dominate the G dwarf signatures even further.

Practically, taking time derivatives is only possible with dense time-sampling over a stellar rotation cycle. Instead, \cite{barragan22pyaneti2} incorporated the time derivative of the GP when performing a multidimensional approach to analysing spectroscopic timeseries to account for the difference in behaviour of RV with FWHM as compared with RV and BIS. A better correspondence between RV and FWHM variability due to cool spots is obtained this way, {though good time sampling will still be desired to reliably inform the modelling process.} In the following sections we investigate the effectiveness of using {$M_3$ and $M_2^\prime$ for our M dwarf and G dwarf {single spot} models via the Pearson's~$r$ linear correlation statistic.}

\subsection{CLMs single spot correlations as a function of spot latitude}
\label{section:CLMonespot}

To further investigate the effective correlations between CLMs, we first performed simulations with single spot models for a $2$\degs\ spot located in the latitude range $0$\degs\ $\leq l \leq$ $80$\degs\ at $10$\degs\ intervals. We obtained RV and BIS in addition to the CLMs and their time derivatives. 
We again simulated 120 measurements over a single rotation with no noise for the case with axial inclination $i=90$\degs. Fig. \ref{fig:moments_correlations} shows Pearson's $r$ for single cool spots, facular spots and cool\,+\,facular spots for both the M dwarf and G dwarf models {with rotational broadening of \vsini~=~$5$\,\kms.} {The most significant correlations are plotted for RV vs BIS, RV vs $M_3$, $M_1$ vs BIS and $M_1$ vs $M_3$ (purple and cyan) and RV vs $M_2^\prime$ and $M_1$ vs $M_2^\prime$ (green and yellow).

The Pearson's $r$ correlations} can be summarised as follows: \\

{
\noindent
M dwarf and G dwarf - cool spots:
\begin{enumerate}[wide, labelwidth=!,itemindent=!,labelindent=0em, leftmargin=0.5cm, label=(\roman*), itemsep=0.1cm, parsep=0pt]
	\item $M_3$ and BIS show negative linear correlation while $M_2^\prime$ shows positive correlation with RV or $M_1$
	\item $M_3$ shows a greater degree of linear correlation than $M_2^\prime$
	\item $M_3$ and $M_2^\prime$ show a greater degree of correlation with RV than BIS
    \item The degree of correlation is greatest for high latitude spots
\end{enumerate}

\noindent
M dwarf - facular and cool\,+\,facular spots:
\begin{enumerate}[wide, labelwidth=!,itemindent=!,labelindent=0em, leftmargin=0.5cm, label=(\roman*), itemsep=0.1cm, parsep=0pt]
    \item The $M_3$ and BIS correlations, follow the same trend as cool spots, but with improved correlation at low-intermediate latitudes
    \item $M_2^\prime$ is {moderately} correlated with RV or $M_1$ with greatest correlation for a spot at intermediate latitudes and a fall in correlation for high latitude spots
\end{enumerate}

\noindent
G dwarf - facular and cool\,+\,facular spots:
\begin{enumerate}[wide, labelwidth=!,itemindent=!,labelindent=0em, leftmargin=0.5cm, label=(\roman*), itemsep=0.1cm, parsep=0pt]
    \item $M_1$ vs $M_3$ and $M_1$ vs BIS correlations for low latitude spots are significantly lower compared with cool spots
    \item  $M_2^\prime$ shows fairly uniform moderate-high correlation at all latitudes
\end{enumerate}
}

If only cool spots are found on M dwarfs, we can expect both $M_3$ and $M_2^\prime$ to act as good linear activity correlators with RV. {For the M dwarf model, the drop in correlation of RV with $M_2^\prime$ for high latitude spots with a facular contribution is likely due to the switch in contrast behaviour of our models near to the limb. For the G dwarf models, cool spots also induce high linear correlations in $M_3$ and $M_2^\prime$. Here, the negative correlation between RV and $M_3$ is high, with $r \leq -0.75$ for low-intermediate latitude spots (Fig. \ref{fig:moments_correlations}, solid purple line in lower left panel); the range over which spots are typically found on the Sun. The inclusion of {a facular contribution} in the cool\,+\,facular model reduces the RV vs $M_3$ correlation for intermediate latitudes to $r \leq -0.60$. RV vs $M_2^\prime$ also performs well for G dwarf model cool spots, with $r = 0.66$ for an equatorial $l=0$\degs{} spot and $r = 0.83$ for $l=80$\degs{}. RV vs $M_2^\prime$ exhibits less variation when cool\,+\,facular regions are present on the G dwarf model, with $r = 0.76$ ($l=0$\degs{}) and $r = 0.61$ ($l=80$\degs{}). 

Increasing the spectral resolution leads to very small improvements in correlation between RV and $M_3$. At \vsini{} = $5$\,\kms{}, for resolutions of $R=140,000$ and $190,000$, we find respective (negative) linearity increases in Pearson's r of $1.3$\% and $2.3$\% for an equatorial spot. These improvements in linearity decrease with spot latitude with essentially no improvement for a spot with $l=80$\degs{}. The gains are smaller for RV and $M_2^\prime$, with $0.7$\% and $1.2$\% increases in Pearson's r for low latitude spots. 

For rotational broadening of \vsini{} = $2$\,\kms, Fig.~\ref{fig:moments_correlations_vsini2} shows that correlations are reduced. In particular, high latitude spots become harder to discern, leading to poorer correlations compared with \vsini{} = $5$\,\kms. The correlations with BIS and $M_3$ in Fig.~\ref{fig:moments_correlations_vsini2} are plotted for $-1 < r < 1$ since the correlations clearly either switch sign, or become positive when faculae are present. This is particularly pronounced for the G dwarf model and likely arises because the 340\ms{} convective blueshift is a significant fraction of the stellar \vsini{} and because the shape of the bisectors is a function of limb angle. Spots at low latitude traverse wider ranges of limb angles compared with spots at higher latitudes. The exact correlation coefficient is therefore dependent on combinations of these factors.

For RV vs $M_2^\prime$, moderate correlations persist for spots at most latitudes. In addition, Fig.~\ref{fig:moments_correlations_vsini2} shows that despite a reduction in Pearson's r at the lower \vsini{} = $2$\,\kms, a moderate correlation remains for spots at lower latitudes. For solar-like activity levels, with solar-like facular levels, it may thus be more useful to use the time derivative of $M_2$ or a similar metric. 
}

{We find that $M_1$ generally correlates less well than RV}, but since regular precision radial velocity work {uses} RVs from template-matched spectra of cross-correlation profiles (as in our simulations), $M_1$ is potentially most useful as an additional monitoring diagnostic parameter only. 
Unlike the BIS, which measures the skewness of the line or CCF via the difference of the mean of two sub-regions of the profile, $M_3$ {can potentially make use of} the entire profile, leading to increased correlation sensitivity of $M_3$ over BIS. {We note that in practice, it is necessary to reject the unstable profile wing regions when measuring line shapes. This is routine procedure when measuring BIS in observed data and should also be applied to CLM analysis.}


\begin{figure*}
    \begin{center}
    \begin{tabular}{c|c}

      \includegraphics[trim=0mm -53mm 0mm 0mm, width=0.475\textwidth]{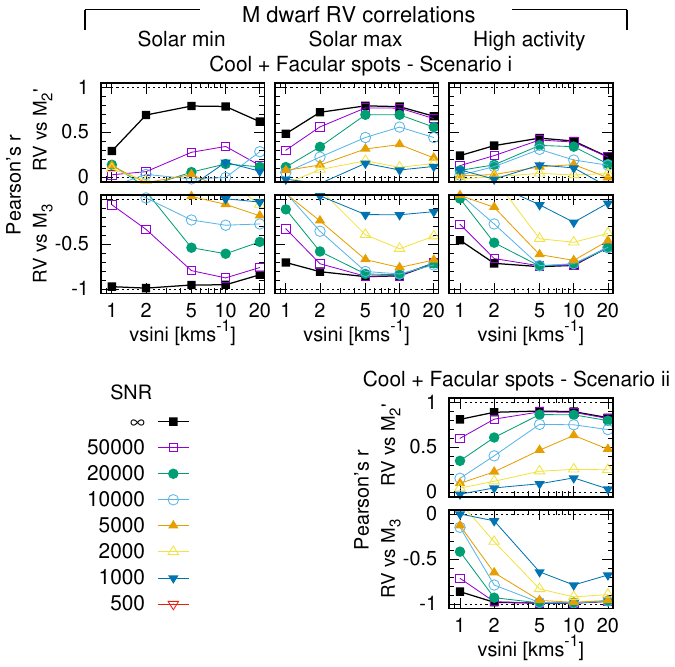} &
      \includegraphics[width=0.475\textwidth]{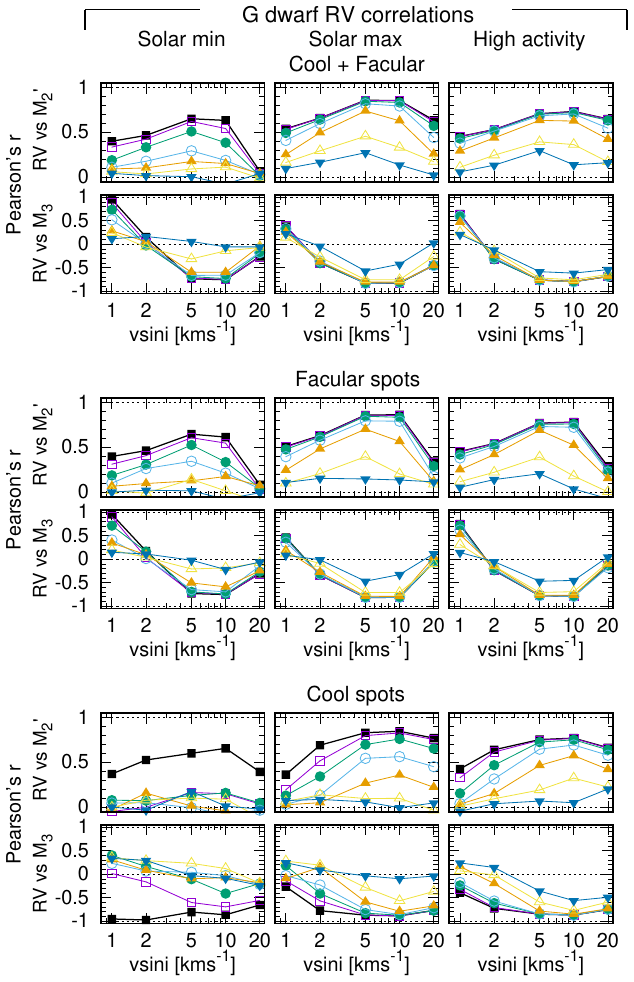} \\
    \end{tabular}
    \end{center}
    \caption{
    Pearson's $r$ correlations for M dwarf (left panels) and G dwarf (right panels) for solar minimum, solar maximum and high activity models. The correlations for RV vs $M_3$ and RV vs $M_2^\prime$ are shown for a range of \vsini{} values and SNR combinations in each sub-panel (see key). For the M dwarf model, the two spot distributions, scenario i and scenario ii, are shown for cool spots. For the G dwarf model, the cool, facular and cool\,+\,facular spot model correlations are shown.}
    \label{fig:moments_correlations_spotmodels}
\end{figure*}

\subsection{CLM correlations for spot models}
\label{section:CLMspotmodels}

We computed the RV and CLM correlations using the solar minimum, solar maximum and high activity spot models. For the M dwarf, we show correlations for the {cool\,+\,facular spot models only} using scenario i and also scenario ii for the high activity case only. For the G dwarf, correlations are shown for cool, facular and cool\,+\,facular spot models. Pearson's $r$ values are plotted in Fig.~\ref{fig:moments_correlations_spotmodels} for RV vs $M_3$ and RV vs $M_2^\prime$ for \vsini\ $= 1, 2, 5, 10$ and $20$~\kms{} and are colour coded for $500 \leq \textrm{SNR} \leq \infty$. The SNR = $\infty$ correlations (black {curves} in Fig. \ref{fig:moments_correlations_spotmodels}) give an indication of the maximum attainable correlation {and indicate the true sense of a correlation (i.e. either negative, with $r<0$, or positive, with $r>0$).} Typically, {a lower SNR} reduces the degree of negative or positive correlation, and in some cases, results in a misleading outcome, where a true negative correlation may appear as a positive correlation {or} vice-versa. 


{
Even rotation velocities of \vsini{}~$=1$~\kms{} enable significant solar minimum RV vs $M_3$ activity correlations to be detected, but only for very high SNR or SNR $= \infty$. For RV vs $M_3$, when faculae are present, the correlations change sign at \vsini{}~=~$1$ and $2$~\kms{}, as we found for single spots in \S \ref{section:CLMonespot}; hence, for the appropriate panels in Fig.~\ref{fig:moments_correlations_spotmodels}, we again plot $-1 < r < 1$. This suggests that odd moments such as $M_3$ alone can not reliably distinguish between activity induced RV shits and dynamically induced shifts.
Although the corresponding RV vs $M_2^\prime$ correlations may be lower at a given finite SNR, they are more stable in the presence of facular regions. $M_2^\prime$ may thus be a more reliable activity indicator at low \vsini{} values, though the necessary SNR and low correlations may prove challenging for solar minimum activity levels. Further details for the correlations with RV from the three activity models are discussed briefly below.}

\subsubsection{Solar minimum}

{Reliable correlations are not discernible for low \vsini{}~and activity levels for the M dwarf model with SNRs that could feasibly be obtained with empirical data. A positive RV vs $M_3$ correlation of $r=0.5$ is found for the G dwarf cool\,+\,facular model at \vsini{}~=~$1$~\kms{} and SNR = 10,000. Here, it would likely be more advisable to use $M_2^\prime$, but clearly, the degree of correlation is much lower and very high CCF SNR $\geq 20,000$ would be needed. For the G dwarf models, \vsini{} $= 5$\,\kms and $10$\,\kms, SNR $\geq 5000$ yields RV vs $M_3$ with Pearson's $r \leq -0.5$. Low-moderate degrees of correlation are recovered for RV vs $M_2^\prime$ for SNR $> 10000$ when facular regions are present.}

\subsubsection{Solar maximum}

{The degree of correlation at a given SNR improves notably for both M dwarf and G dwarf models compared with the solar minimum models.} For the M dwarf, SNR $\sim 5000 - 10000$ and \vsini{}~$\geq 5$ \kms{} yields Pearson's $r < -0.5$ for RV vs $M_3$, but SNR $\sim 20000$ is needed to yield similar correlations for $M_2^\prime$. For the cool\,+\,facular G dwarf model; a star with rotation \vsini{}~$\geq 5$\,\kms, {requires SNR $\sim 1000$ for strong RV vs $M_3$ correlation to be recovered. Here, moderate to strong correlations are also found in RV vs $M_2^\prime$ for SNR $\geq 5000$. The degree of linear correlation decreases significantly by \vsini{}~$=20$\,\kms\ for models with facular spots.}

\subsubsection{High activity}

{For the M dwarf, the degree of correlation in the high activity scenario i models decreases compared with the solar maximum model. This is a consequence of the uniform distribution of spots. As expected, for scenario ii with all spots are located in a single high latitude spot group, more significant correlations are found. For scenario ii, SNR $\geq 5000$ enables $r < -0.5$  to be recovered at \vsini{} $\geq 2\,$\kms{} for RV vs $M_3$, though lower SNRs are needed when \vsini{} $\geq 5$ \kms. Correlations increase significantly in the G dwarf high activity cool spot models. Here, despite a larger number of spots, the decline in relative facular contribution in the high activity models yields very similar correlation levels to the solar maximum models. Again, strong correlations are found for \vsini{} $= 5$~\kms and $10$~\kms with high SNRs generally required for $M_2^\prime$ when compared with the $M_3$ at a given \vsini.} \\

{We also investigated the effect of increasing the instrumental resolution. Figure \ref{fig:moments_correlations_spotmodels2} shows Pearson's $r$ correlations for the G dwarf models at instrumental resolutions of $R=140,000$ and $R=190,000$. Subtle increases in correlation are seen at a given SNR as resolution increases. While low activity and low \vsini{} values appear to preclude effective recovery of activity signatures at all simulated resolutions, it is also important to realise that, within the specified bounds of the models, the simulated spot distributions are just a single realisation. As such, the spot model correlations described above should only be used as a basic guideline for discerning trends and correlation amplitudes.}

\section{Summary and Discussion}
\label{section:summary}

Our primary goal has been to demonstrate the use of CLMs derived from simulated high SNR absorption lines or cross-correlation functions to identify and recover activity related periodicities and trends. We have focused on stars with activity levels similar to the Sun to enable us to determine the effective limits at which CLMs can provide useful information. 

\subsection{Summary of main results}
\label{section:summary_results}

For a single spot, we find the following behaviour of CLM signatures

\begin{enumerate}[wide, labelwidth=!,itemindent=!,labelindent=0em, leftmargin=0.5cm, label=(\roman*), itemsep=0.1cm, parsep=0pt]
    \item Cool spot CLM signatures are similar for both M dwarfs and G dwarfs; higher intrinsic spot-photosphere contrasts in G dwarfs (and also V band simulations vs R band for M dwarfs) yield larger signatures
    \item Facular regions show a negative correlation with cool spots when convective blueshift is not present - i.e. in M dwarfs 
    \item Convective blueshift modifies the signature of CLMs in facular regions (and hence ii above) since a large asymmetric velocity shift affects the CCF wings where facular signatures have highest amplitude (i.e. for G dwarfs)
    \item A consequence of (iii) is that the facular component dominates combined cool\,+\,facular signatures (i.e. for G dwarfs)
\end{enumerate}

\medskip
\noindent
The recovered GLS periodicities, $P_\textrm{rec}$, that appear in the CLMs for a \textit{single spot} are as follows:

\begin{enumerate}[wide, labelwidth=!,itemindent=!,labelindent=0em, leftmargin=0.5cm, label=(\roman*), itemsep=0.1cm, parsep=0pt]
\item Low and intermediate latitude spot: a mixture of $P_\textrm{rot}$ and harmonics of $P_\textrm{rot}$ 
\item High latitude spot: predominantly, $P_\textrm{rot}$
\item Higher harmonics of $P_\textrm{rot}$ appear in stars with higher \vsini\ where more phase dependent CLM structure is resolved
\end{enumerate}

{Period searches that assume sinusoidal variability, such as the (Generalised) Lomb-Scargle method, are not optimally matched to recover stellar activity signals and often lead to recovery of a harmonic of the true stellar rotation period. Instead, string length minimisation is particularly effective at recovering the true $P_\textrm{rot}$ for spots at any latitude for the CCF SNRs that are readily achieved with precision radial velocity observations.} 

\medskip
\noindent
For the three simulated activity models with multiple spots: solar minimum, solar maximum and high activity, we similarly find

\begin{enumerate}[wide, labelwidth=!,itemindent=!,labelindent=0em, leftmargin=0.5cm, label=(\roman*), itemsep=0.1cm, parsep=0pt]
\item For G dwarfs, the GLS period or harmonic of the period can be recovered for moderate to high SNRs and is generally easier at moderate and high \vsini{} values
\item Recovery of periods with GLS analysis for M dwarfs is challenging, {often} requiring SNR $\geq 10000$ and \vsini\ $\ge 10$ \kms{} and is most successful for high activity when cool spots are located at high latitudes in a single group
\item {String length period searches reliably recover the true rotation period in most cases for the simulated range of \vsini{} and SNRs where GLS periodogram searches recover harmonics of the true rotation period}
\end{enumerate}

\smallskip
\noindent
{When CLM periodicities are present, correlations between RV and higher order CLMs or their time derivatives show moderate-high degrees of linear correlation.}

\begin{enumerate}[wide, labelwidth=!,itemindent=!,labelindent=0em, leftmargin=0.5cm, label=(\roman*), itemsep=0.1cm, parsep=0pt]
\item line moments $M_3$ and the time derivative, $M_2^\prime$, show the strongest correlations with activity induced RV variability.
\item RV vs $M_3$ shows a stronger linear trend than the traditionally adopted RV vs BIS
\item {RV vs $M_2^\prime$ shows a moderate but more consistent linear trend across the range of simulated \vsini{} values, particularly for the G dwarf models with a facular component}
\item {Spot groups confined to high latitudes yield much higher degrees of linear correlation}
	
\end{enumerate}


Our simulations have restricted spot latitude distributions for the G dwarf model to those seen on the Sun and expected from more distributed dynamo activity in M dwarfs (scenario i). However, based on the simulations of \cite{granzer00}, we might expect spots to appear at intermediate latitudes for a G dwarf and high latitudes for an M dwarf (as per our M dwarf scenario ii model) for the fastest rotation rates we considered. {While this may improve the degree of linear correlation, restricting spot latitudes in this way will also to lead to} smaller CLM signatures for a fixed spot size or ensemble of spots, making recovery more challenging at any fixed SNR. Similarly, we expect a preference for periodicities recovered as lower harmonics of $P_\textrm{rot}$ {with GLS periodogram analysis}, as evidenced by the M dwarf scenario ii model we considered. However, simulating changes in spot distributions with longitude and spot clustering within spot groups is generally more difficult.

\subsection{Application of CLM analysis to observations: AU Mic}
\label{section:summary_discussion}

\begin{figure}
    \centering
    \includegraphics[width=0.45\textwidth]{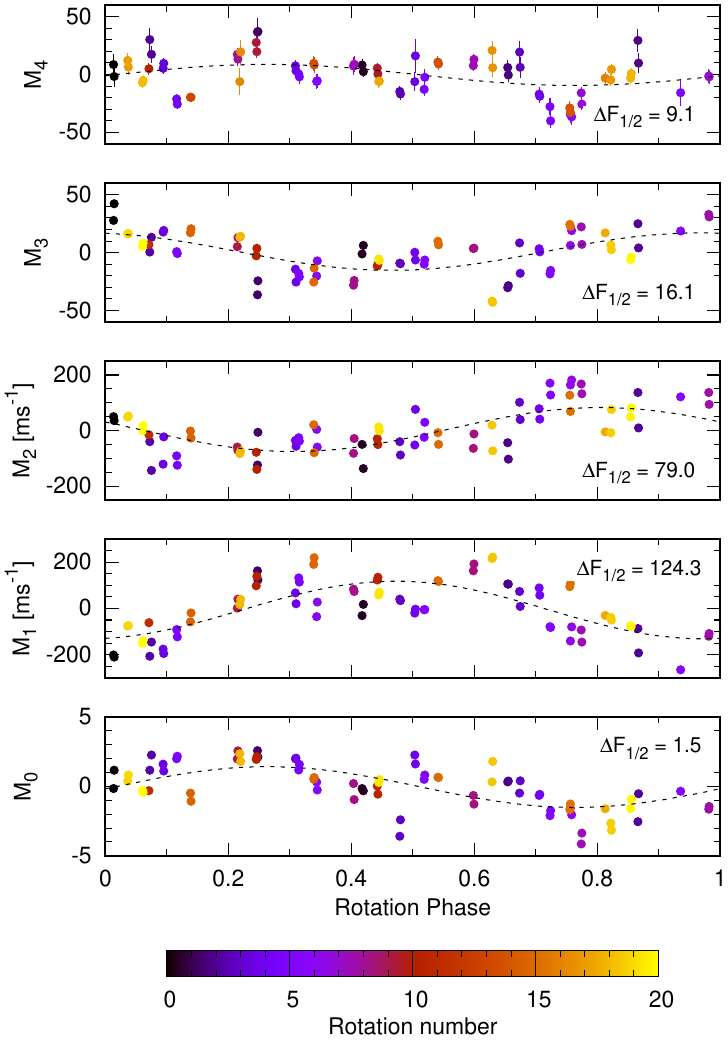}
     \caption{AU Mic line moments, $M_0$ - $M_4$, phased on the rotation period. The data are colour coded according to rotation number and phase relative to the first observation for the 18 rotations of AU Mic in the time interval of the observations. The dashed lines show a sinusoidal fit to the phased data.}
     
     \label{fig:AUMic_moments_phased}
\end{figure}

\begin{figure*}
	\centering
	\includegraphics[width=0.92\textwidth]{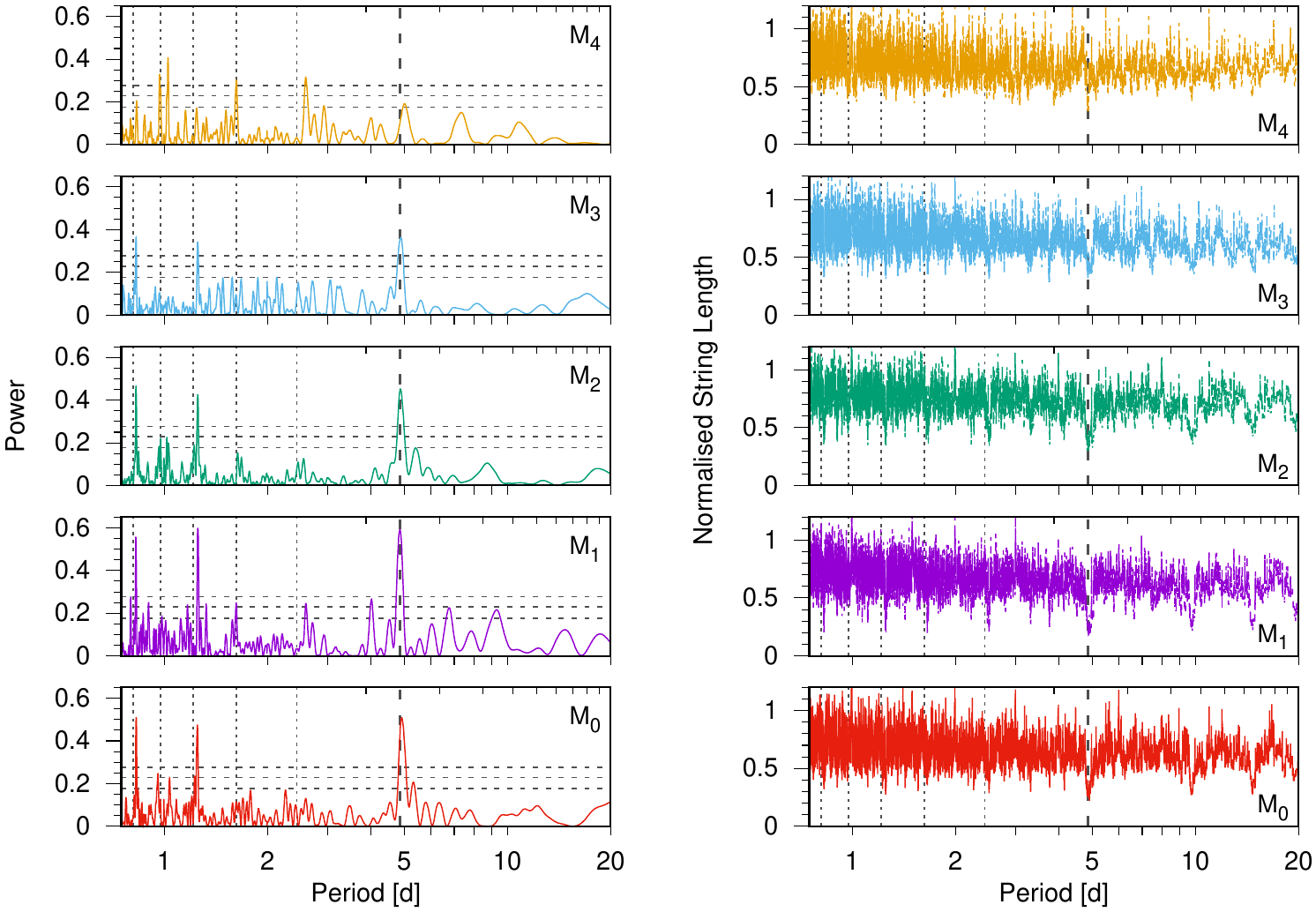}
	\caption{{Periodograms of the Central Line Moments for AU Mic. Left: The Generalised Lomb Scargle periodograms with false alarm probabilities at FAP = 0.1, 1 and 10\%. Right: normalised SL periodograms. The known period of $P_\textrm{rot} = 9.863$~d is indicated with the vertical dashed line. The fundamental harmonics at $P_\textrm{rot}/2$ to $P_\textrm{rot}/6$ are indicated by the dotted vertical lines. See Table \ref{tab:aumicp} for further details of recovered periodicities.}}
	\label{fig:AUMic_moments_periodograms}
\end{figure*}

\begin{figure}
    \centering
    \includegraphics[width=0.35\textwidth]{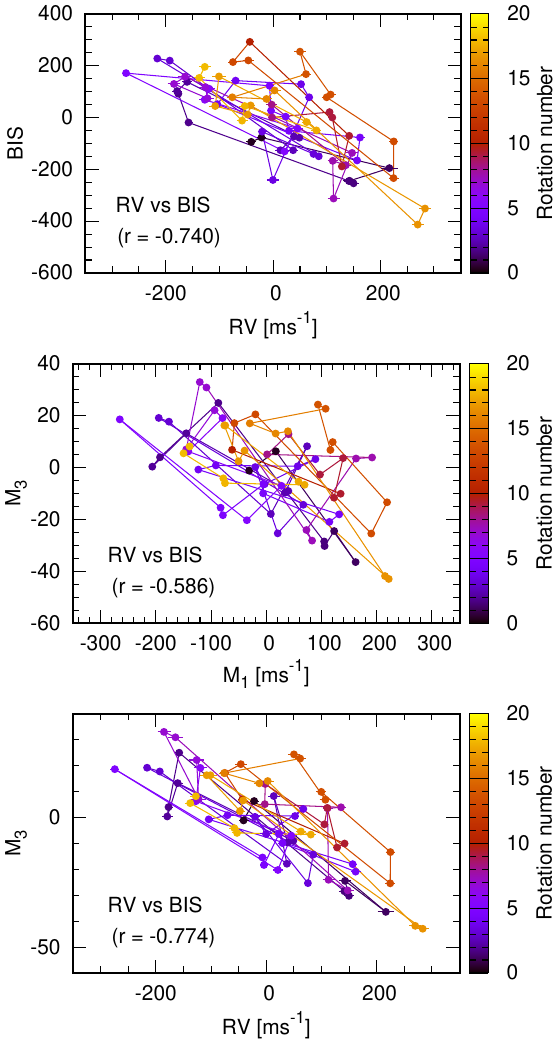}
     \caption{The trends of RV vs BIS, $M_1$ vs $M_3$ and RV vs $M_3$ over 18 rotations of AU Mic. Error bars are plotted on the points. The {lines connect the points in observing sequence with the colour coding used} in Fig. \ref{fig:AUMic_moments_phased}.}
     \label{fig:AUMic_moments_comparison}
\end{figure}

To illustrate the recovery of periodicities from CLMs, {we used observations of} the young active M1Ve star, AU Mic {(HD~197481 / GJ~803), with photometric $P_\textrm{rot} =$ $4.863 \pm 0.010$~d \citep{plavchan20aumic}}. 
$P_\textrm{rot}$ (or very close to $P_\textrm{rot}$) has been recovered {by Doppler imaging methods, RVs and activity indicators, including chromospheric indicators, FWHM  and BIS (\citealt{klein21aumic}, \citealt{klein22b})}. Most importantly, the Zeeman Doppler map presented in \cite{klein21aumic} reveals a decentered polar region of radial magnetic field suggesting a highly dipolar global magnetic field. The brightness Doppler images however suggest very weak cool spot structure close to the equator. In the HARPS images presented in \cite{klein22b} at the later epoch, a weak high latitude cool spot at latitude $60$\degs{} is seen in the 2020 image. Two weak spots at equatorial latitudes are also seen in the same rotation hemisphere. The images also map weak, warm structure, at high and equatorial latitudes. The 2021 image reveals very weak, cool and warm, structure. It is likely that AU Mic is more active than our high activity spot simulations and the recovered radial magnetic field structure is probably more akin to a hybrid of scenarios i and ii for the high activity model. We note that the \vsini\ $= 7.8$ \kms{} is very low for brightness Doppler imaging and that reliable spot recovery, and particularly, latitude recovery of features, is difficult in this regime, especially considering the observations were made with a spectral resolution equivalent to $\sim 4$\ \kms. There is also a degeneracy between the signatures of facular and cool regions in Doppler images, which makes it difficult to recover hot spots in addition to cool spots. Further, \cite{klein21aumic} note that the variability seen in the line profiles may well be the result of magnetic field effects (i.e. from Zeeman broadening) rather than true brightness variations. 

We posit that cool high latitude spots are the most likely cause of measurable line profile and CCF distortions, particularly since the facular contribution decreases with active stars. Given the simulations of \cite{beeck15}, {we do not expect bright facular signatures to be discerned} on M dwarfs.  We note that purely high latitude cool spots do lead us to expect recovery of $P_\textrm{rot}$ in all CLMs as shown in Fig. \ref{fig:sensitivity_onespot} for a single spot, exactly as the GLS periodogram of \cite{klein22b} suggests. The distributed spot group in the high activity scenario ii model predicts a mixture of $P_\textrm{rot}$ and harmonics of $P_\textrm{rot}$, likely because our simulated spots occur in a group with a larger effective radius. In addition, our scenario ii was simulated for an $i=90$\degs{} case with mean latitude of $l=60$\degs{}. The inclination of AU Mic's rotation axis is very close to $i=90$\degs{}, so a decentered or asymmetric polar spot would be sufficient to yield periodicities at $P_\textrm{rot}$ in the RVs and CLMs.

Observations of AU Mic were made with CARMENES in 2019 and 2020 \citep{ribas23carmenes}. We used the more densely sampled observations in July to October 2019 to search for the signature of periodicity in the CLMs. We applied our implementation \citep{barnes98aper} of least squares deconvolution \citep{donati97zdi} to derive line profiles with SNR $= 3700$ with the effects of line blending removed. 
{The mean deconvolved profile velocity increment of $1.25$~\kms\ is large compared with the \vsini~=~$9.6$~\kms \citep{han23mdwarfs}. Since the line distortions due to active regions can be small, before calculating the CLMs, we re-sample to a finer velocity increment via bicubic spline interpolation to minimise systematics. This also enables the degree of linear correlation between RV and the CLMs to be fine-tuned empirically by rejecting the wing regions of the profiles that tend to degrade the correlation. We found the greatest degree of anti-correlation in Pearson's $r$ by rejecting profile pixels at a fixed level corresponding to the 28.7 per cent of the mean deconvolved profile in the wings (i.e. the pixels nearest the effective continuum level). Pixels within the velocity range of $\pm 10.6$\,\kms{} were thus used to calculate the CLMs. This range is likely determined by the \vsini~=~$9.6$~\kms\ combined with the stellar line broadening and instrumental broadening. Determination of CLM uncertainties is important if they are to be used for subsequent analysis. The least squares deconvolved profile uncertainties are fully propagated for each CLM according to Eqn. \ref{eqn:moments} and account for the re-interpolation described above.}

The CLMs derived from 75 spectra are shown in Fig. \ref{fig:AUMic_moments_phased} phased on the {known rotation period, $P_\textrm{rot} =$ $4.863$~d}. The relative amplitude of the weighted moments $M_1$ to $M_4$ decreases in accordance with the simulations shown in Figs \ref{fig:moments_single} and \ref{fig:moments_spotmodels}. $M_0$ shows much lower amplitude variability as expected for activity features covering a small fraction of the stellar surface. To first order, the CLMs show sinusoidal variability on the rotation timescale, though there are indications of {either higher order harmonic structure or} low-level evolution over the $88$~d timespan of the observations.

{Fig.~\ref{fig:AUMic_moments_periodograms} shows the GLS and SL periodograms for each line moment. The most significant periods, at the maximum GLS power and minimum string length, are tabulated in Table \ref{tab:aumicp} for each CLM. While the GLS method preferentially ($P_1$ column) recovers a harmonic of $P_\textrm{rot}$, the second strongest peak ($P_2$ column) generally recovers $P_\textrm{rot}$ with almost the same significance as indicated by the fractional $P_1$ - $P_2$ change in power (Table~\ref{tab:aumicp}, column 4). For $M_4$, we note that $P_\textrm{rot}$ is only just recovered above the 10\% false alarm level. The SL periodograms on the other hand recover $P_\textrm{rot}$ for all CLMs, except for $M_3$, where $P_\textrm{rot}$ yields the 4th shortest string length after harmonics close to $3P_\textrm{rot}/4$, $P_\textrm{rot}/2$ and $P_\textrm{rot}/4$. Given the simulations in the previous sections, it should be pretty clear that taken together, the GLS and SL periodograms complement each other and are able to convincingly identify $P_\textrm{rot}$. In brief, the folded CLM signatures show evidence for higher order harmonics, which is supported by the presence of periodogram peaks at harmonics of $P_\textrm{rot}$. This appears to confirm the findings of \cite{klein21aumic} that high latitude spot structure is the most satisfactory explanation for the signatures. 
}

\begin{table}
 \begin{center}
  \begin{tabular}{c|c|c|c}
\hline
\multicolumn{4}{c}{GLS} \\
\hline
CLM		&	$P_1$ [d]					& $P_2$ [d]	                   & Fractional Power \\
        &                               &                              & ratio decrease @ $P_2$       \\
\hline
M0		&	0.829 ($P_\textrm{rot}/6$)	& 4.920 ($P_\textrm{rot}$)~~~~ & 0.00~ \\
M1		&	1.254 ($P_\textrm{rot}/4$)	& 4.866 ($P_\textrm{rot}$)~~~~ & 0.02~ \\
M2		&	0.828 ($P_\textrm{rot}/6$)	& 4.882 ($P_\textrm{rot}$)~~~~ & 0.03~ \\
M3		&	0.828 ($P_\textrm{rot}/6$)	& 4.893 ($P_\textrm{rot}$)~~~~ & 0.00~ \\
M4		&	2.586 ($P_\textrm{rot}/2$)	& 1.624 ($P_\textrm{rot}/3$)   & 0.05~ \\
\multicolumn{4}{c}{\vspace{0mm}} \\
\hline
\multicolumn{4}{c}{String Length} \\
\hline
CLM		&	$P_1$ [d]					& $P_2$ [d]	                     & Fractional string \\
        &                               &                                & length increase @ $P_2$     \\
\hline
M0		&	4.912 ($P_\textrm{rot}/6$)~~	& 9.726 ($P_\textrm{rot}$)~~~~~~   & 0.22~ \\
M1		&	4.903 ($P_\textrm{rot}/4$)~~	& 1.656 ($P_\textrm{rot}$)~~~~~~   & 0.16~ \\
M2		&	4.881 ($P_\textrm{rot}/6$)~~	& 1.253 ($P_\textrm{rot}$)~~~~~~   & 0.18~ \\
M3		&	3.763 ($3P_\textrm{rot}/4$)	& 4.860 ($P_\textrm{rot}$)~~~~~~   & 0.15$^{\,1}$ \\
M4		&	4.855 ($P_\textrm{rot}/2$)~~	& 3.871 ($3P_\textrm{rot}/4$)     & 0.07~ \\
		
  \end{tabular}
 \end{center}
\caption{{Recovered AU Mic GLS and SL periodicities showing peaks with the highest and second highest power (Note 1: $P_\textrm{rot}$ is recovered as the 4th strongest peak in the String Length $M_3$ case).}}
\protect\label{tab:aumicp}
\end{table}


We investigated the correlations between RV and BIS, $M_1$ and $M_3$ and RV and $M_3$. Figure \ref{fig:AUMic_moments_comparison} shows the results for these three correlations. The points {and connecting lines} are colour-coded in time. We find exactly the same behaviour as our simulations predicted, with the tightest negative correlation between RV vs $M_3$, {where Pearson's $r = -0.775$, a modest improvement over the more usual RV vs BIS of 4.6\%}.

{Finally, we also applied the same analysis to the young F8V Hyades cluster star, HD 30589. \cite{ross24hd30589} have presented evidence for both Keplerian RV variability and have used BIS to simultaneously model significant stellar activity RV modulation. \cite{ross24hd30589} also presented an analysis of the CLMs, finding that the RV vs $M_3$ correlation of $r = -0.659$ yields an 8.0\% improvement over the RV vs BIS correlation of $r = -0.610$. 
}


\section{Conclusions and Future work}
\protect\label{section:future}

As we attempt to retrieve low amplitude dynamical RV signals, it becomes increasingly important to be able to discern and correct for the effects of purely stellar signals in RV timeseries. We have explored the first few central line moments and have shown that distinct signatures of $P_\textrm{rot}$ and its harmonics are to be expected. {The String Length minimisation method is particularly effective at recovering $P_\textrm{rot}$, where more widely adopted periodogram methods tend to return harmonics of $P_\textrm{rot}$. We advocate the use of both methods in tandem to obtain a more robust assessment of activity induced modulations.} The behaviour of the CLMs differs between M dwarfs and solar type stars due to the differing character of stellar activity in the two cases. For low activity levels, we have shown that recovering activity signatures is challenging. Sensitivity improvements may be obtained by considering the most activity-sensitive lines \citep{dumusque18}, though a balance between loss of SNR in the resulting CCF against the potential gain in activity induced amplitude, would need careful consideration. The calculation of CLMs is straightforward to implement, with $M_3$ {offering a potential improvement over BIS as an activity correlator}. We similarly advocate the use of methods that exploit the time derivative of second order line moment indicators, such as $M_2$.

In this work, we have limited simulations to a single realisation (e.g. random spot pattern) within the bounds of our models: with further runs, uncertainties could be determined on each detection periodicity, though the CPU time to achieve this would be considerable. Similarly, we have assumed static spot patterns and a fixed number of observations to estimate recovery of signal periodicities at fixed SNRs. {We have not considered the effects of differential rotation, which complicates the recovery of periodicities and long term stability of active features, particularly for earlier spectral types \citep{reiners03diffrot,reiners03diffrot2,barnes05diffrot,barnes17mdwarfs}}. Active regions are known to be relatively stable on both M dwarfs and G dwarfs, which may exhibit active longitudes on timescales of months to many years. Further exploration of the latitude dependence of activity as a function of rotation rate could provide another refinement of the simulations. Ultimately, customised simulations, informed by observations and tailored to individual stars that include time evolution of features would likely shed more light on further factors that could impact on recovery of periodicities and activity signals.
With the increasing number of facilities pushing to 10 c\ms{} stability, simulations that include stellar variability arising from lower activity levels in addition to more stochastic processes may provide further useful insights.

\section*{Acknowledgements}
JRB and CAH were funded by STFC under consolidated grant ST/T000295/1 and ST/X001164/1. 
SVJ and FL acknowledge the support of the DFG priority program SPP 1992 "Exploring the Diversity of Extrasolar Planets (FL, SVJ: JE 701/5-1).
MP and GAE acknowledge support by spanish grants PID2021-125627OB-C31
funded by MCIU/AEI/10.13039/501100011033 and by “ERDF A way of making
Europe”, PID2020-120375GB-I00 funded by MCIU/AEI, by the programme
Unidad de Excelencia María de Maeztu CEX2020-001058-M, and by the
Generalitat de Catalunya/CERCA programme.
This work makes use of data from the CARMENES data archive at CAB (CSIC-INTA). The CARMENES archive is  part of the Spanish Virtual Observatory project
(http://svo.cab.inta-csic.es), funded by MCIN/AEI/10.13039/501100011033/ through grant PID2020-112949GB-I00.
We thank the anonymous referee for providing constructive suggestions that led to an improved manuscipt.

\section*{Data Availability}
The data underlying this article will be shared on reasonable request to the corresponding author.




\bibliographystyle{mnras}
\bibliography{master,ownrefs}



\appendix

\section{Additional Figures}

\begin{figure*}
	\centering
	\includegraphics[trim=5mm 0mm 5mm 0mm, width=2.0\columnwidth]{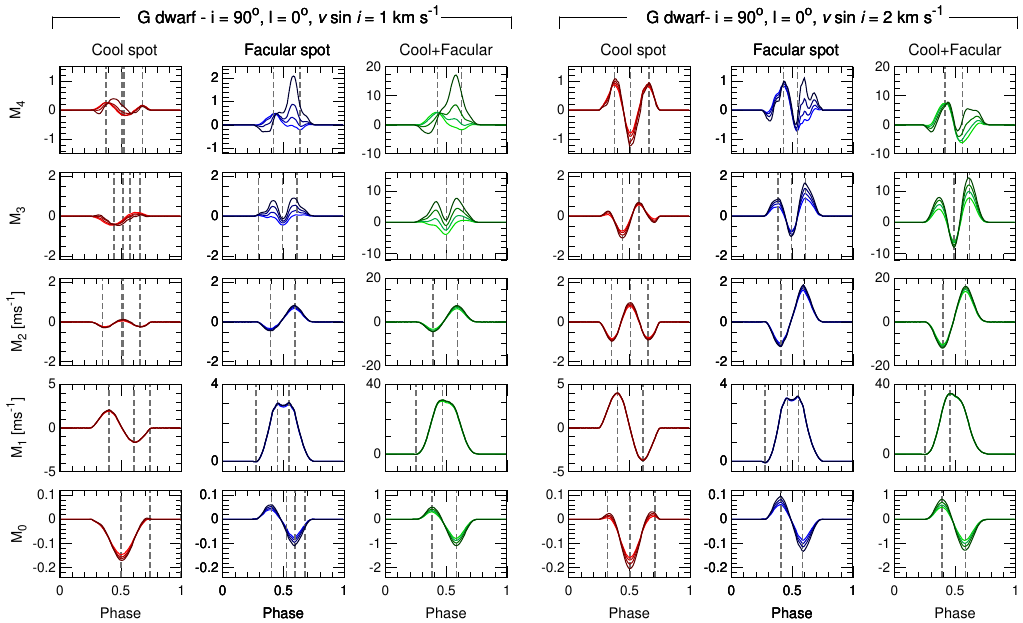}
	\caption{{The effect on CLMs of increasing resolution from $R=115,000$ (lightest colour curves with smallest amplitudes), through resolutions of $R=140,000$ and $R=190,000$, to $R=500,000$ (darkest curves with highest amplitude or degree of complexity).}}
	\label{fig:moments_single_resolution}
\end{figure*}

\begin{figure*}
    \begin{center}
    \includegraphics[width=0.99\textwidth]{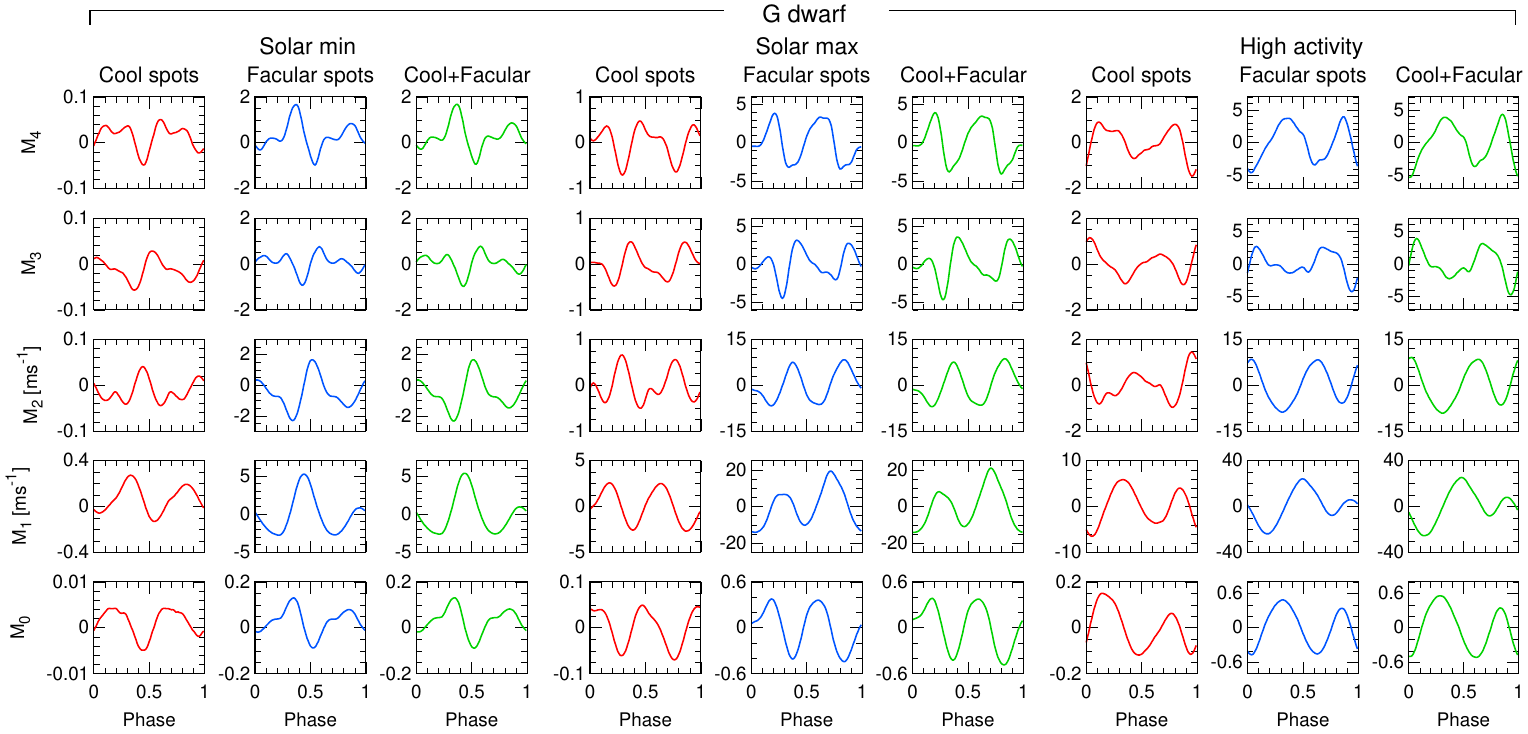}
    \end{center}
    \caption{Central Line Moments as a function of phase for a G dwarf model with the two largest spot groups separated by 180\degs{}. For further details, see main text and Fig. \ref{fig:moments_spotmodels}.}
    \label{fig:moments_spotmodels_activelongs}
\end{figure*}

\begin{figure*}
\begin{tabular}{lll}
    \includegraphics[trim=0mm 0mm 2mm 0mm, height=9cm]{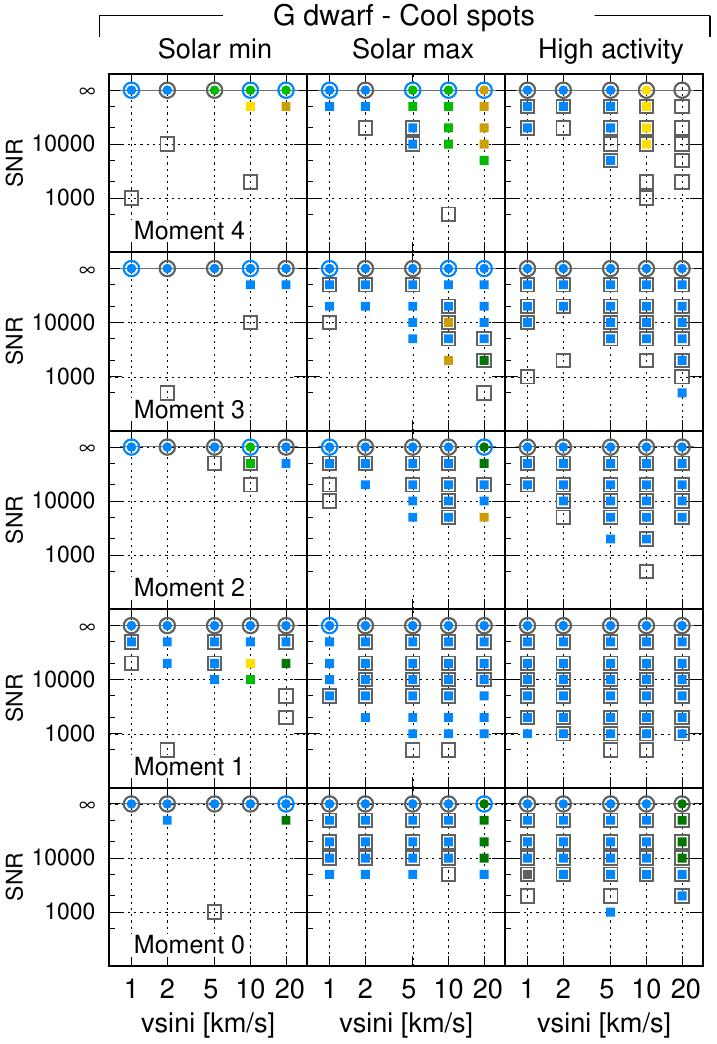} \hspace{1mm} &
    \includegraphics[trim=0mm 0mm 2mm 0mm, height=9cm]{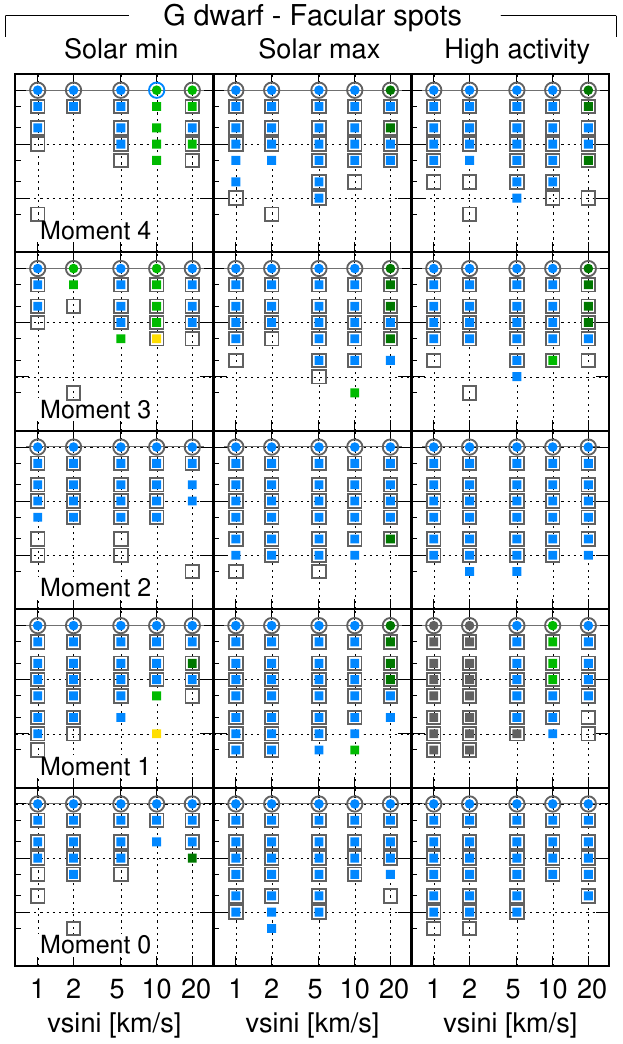} \hspace{1mm} &
    \includegraphics[trim=0mm 0mm 2mm 0mm, height=9cm]{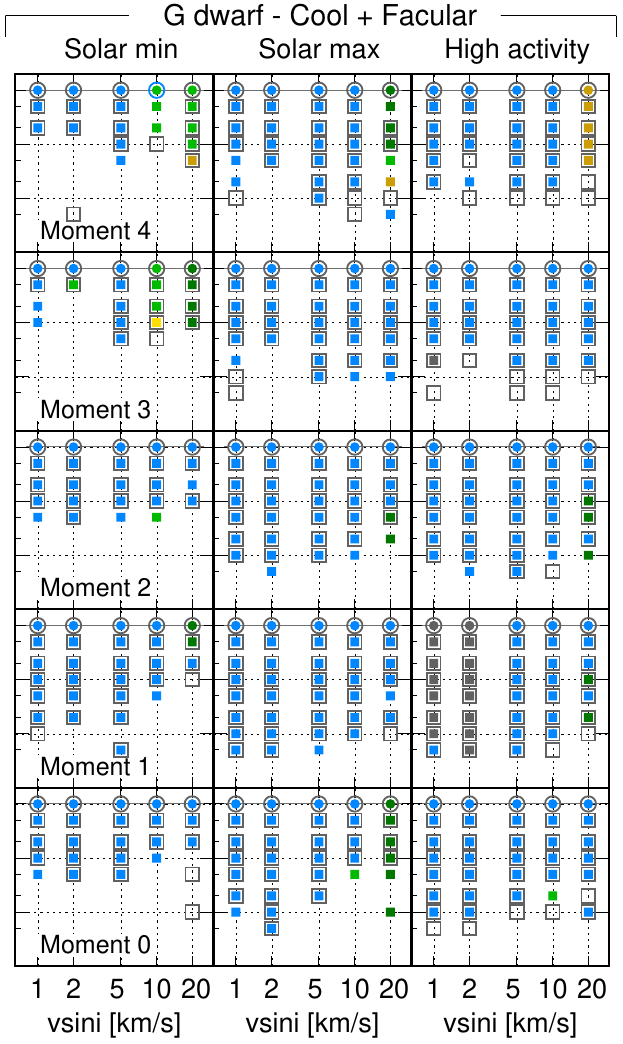}
    \end{tabular}
    \caption{Period recovery matrices for a G dwarf with the two largest spot groups separated by 180\degs{} in longitude. For further details, see main text and Fig. \ref{fig:sensitivity_spotmodels_GMdwarf}.}
    \label{fig:sensitivity_spotmodels_Gdwarf_activelongs}
\end{figure*}

\begin{figure*}
    \begin{center}
    \includegraphics[width=0.9\textwidth]{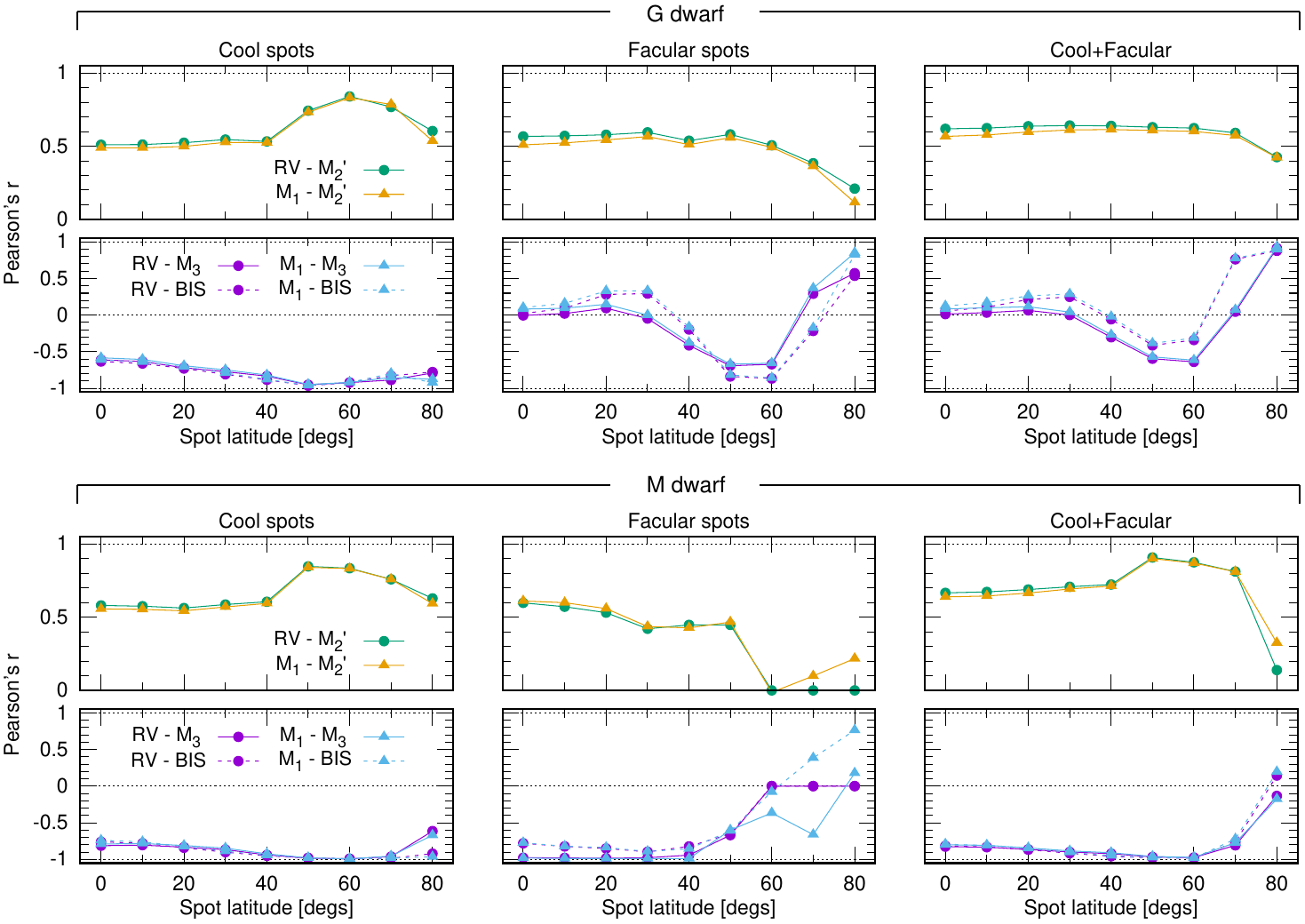}
    \caption{
    {Companion to Fig. \ref{fig:moments_correlations} in the main text for \vsini\ = 2\ \kms.  Pearson's $r$ correlations between absorption line (or CCF) metrics for single spots with $r_\textrm{u} = 2$\degs and \vsini\ = 2\ \kms{} located at latitudes $0^\textrm{o}$ to $80^\textrm{o}$. The upper plots show Pearson's $r$ linear correlation for an M dwarf, while the lower panels show the same correlations for a G dwarf. The key indicates the line metric pairs that are plotted in each case.}}
    \label{fig:moments_correlations_vsini2}
    \end{center}
\end{figure*}

\begin{figure*}
	\begin{center}
		\begin{tabular}{c|c}
			
			\includegraphics[trim=0mm 0mm 0mm 0mm, width=0.475\textwidth]{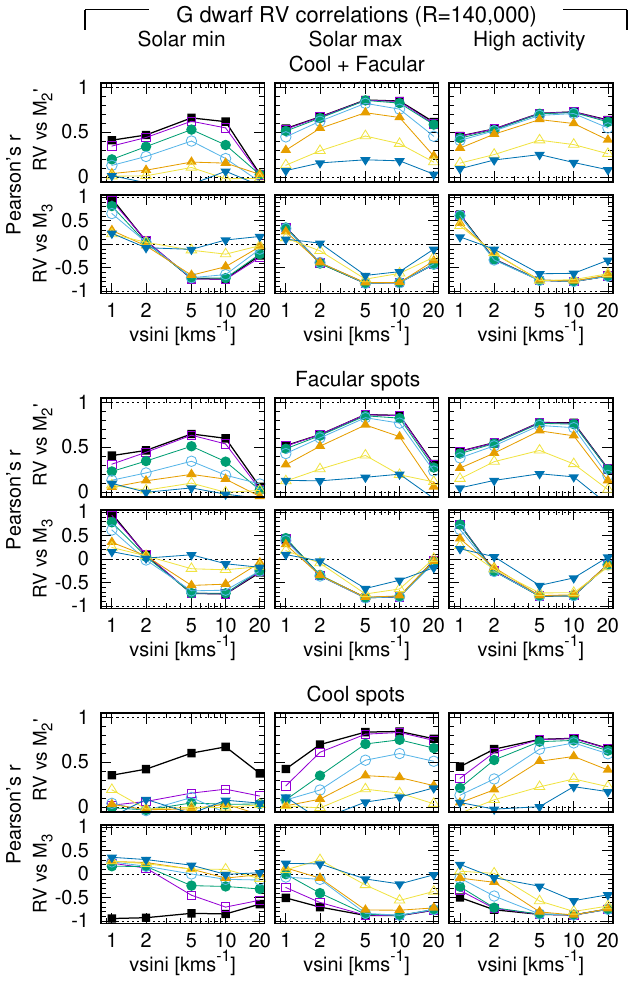} &
			\includegraphics[width=0.475\textwidth]{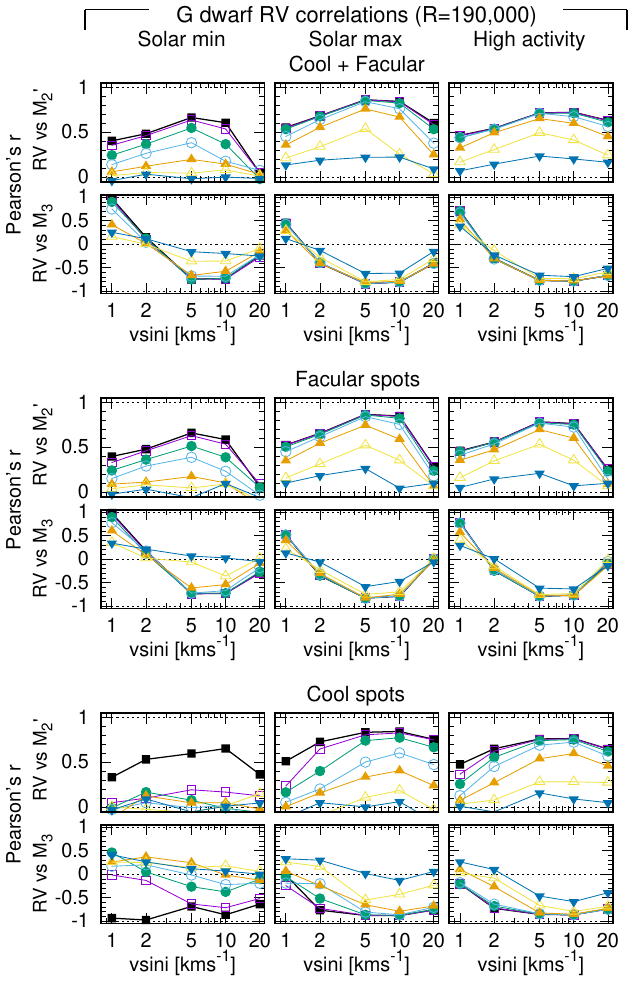} \\
		\end{tabular}
	\end{center}
	\caption{
		{Pearson's $r$ correlations for G dwarf models at spectral resolutions of $R=140,000$ (left panels) and $R=190,000$ (right panels). The can be compared directly with the G dwarf correlations for $R=115,000$ in Fig. \ref{fig:moments_correlations_spotmodels} (right panels).}}
	\label{fig:moments_correlations_spotmodels2}
\end{figure*}

%



\bsp	
\label{lastpage}
\end{document}